\documentclass{article}

\usepackage{iclr2027_conference,times}

\usepackage{amsmath,amssymb}
\usepackage{booktabs}
\usepackage{graphicx}
\usepackage{flafter}
\usepackage[below]{placeins}
\usepackage{xcolor}
\usepackage{hyperref}
\usepackage{url}
\iclrfinalcopy

\AddToHook{env/table/begin}{\setlength{\abovecaptionskip}{5pt}}
\AddToHook{env/table*/begin}{\setlength{\abovecaptionskip}{5pt}}

\definecolor{revcolor}{RGB}{0,70,180}
\DeclareRobustCommand{\rev}[1]{{\color{revcolor}#1}}
\newenvironment{revblock}{\color{revcolor}}{}
\newcommand{\revtable}[1]{\color{revcolor}\addtocounter{table}{-1}%
  \renewcommand{\thetable}{#1}\def\theHtable{#1}}
\newcommand{\revfit}[1]{\resizebox{\ifdim\width>\linewidth\linewidth\else\width\fi}{!}{#1}}

\title{Image Watermarking with Memory\\across Reissuance}

\author{
Xinyuan Li\textsuperscript{*} \quad
Yi Nian\textsuperscript{*} \quad
Xiaoqin Feng \quad
Tiankai Yang\quad
Yue Zhao\\
University of Southern California \\
\texttt{\{xli02281,yinian,xiaoqinf,tiankaiy,yzhao010\}@usc.edu}, \\ 
\vspace{1ex} 
    \textit{\small $^*$Co-first authors}
}

\begin{document}

\maketitle
\lhead{Preprint}
\begin{abstract}
Publishers often revise and reissue images, yet valid watermarks may persist
on superseded releases or unapproved edits.  Even a current release can
contain regions last authorized at different revisions.  We introduce and
formalize \emph{reissuable provenance}: determining an image's current,
superseded, or unissued status and recovering each region's last authorized
revision (its generation), with abstention when evidence is insufficient.
\emph{Generational Stratigraphy Fields} (GSF) combine watermark evidence with
an authenticated live reference to the current release, enabling replacement
or retirement without rewriting circulated copies.  Each reissue freshly
embeds new generation labels in approved regions and inherited labels
elsewhere, preserving history without requiring old watermarks to survive
successive edits.  Our complete-output reissuance evaluation follows 480
histories through four releases, retaining each AI editor's entire output.
GSF achieves 99.67\% accuracy in classifying issued images as current or
superseded and 94.09\% recall of unissued edits.  After three reissues, one
generation map provides time and location: it recovers regional labels with
93.13\% median accuracy and localizes the latest approved revision with
92.86\% IoU.  Regional labels remain readable under tested delivery
transformations that impair issued-state recognition.  Watermarks can thus
express who issued an image, which release is authoritative now, and when
each region was last authorized.
\end{abstract}

\section{Introduction}
\label{sec:introduction}

\begin{figure}[!ht]
    \centering
    \includegraphics[width=1\textwidth]{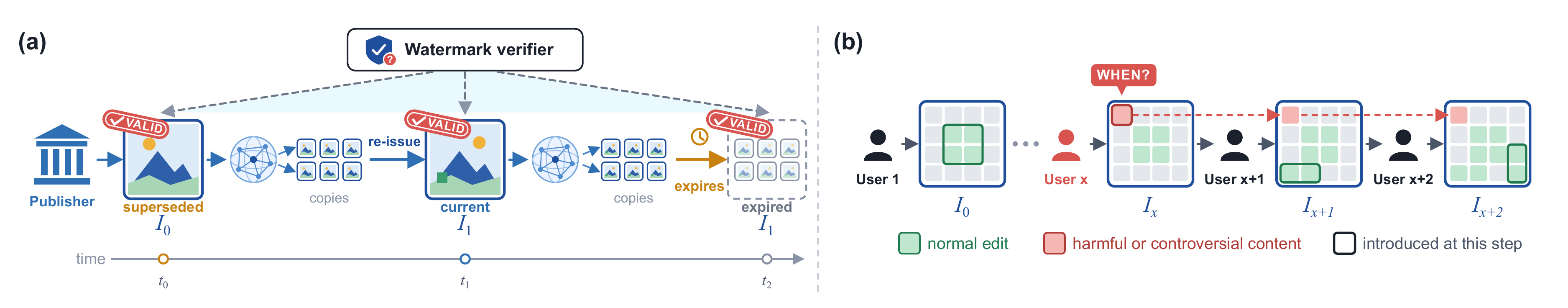}
    \caption{\textbf{Why present status and regional history are both
    necessary.} (a) \textbf{Authentic does not mean current.}  A
    publisher issues $I_0$, corrects and reissues $I_1$, and $I_1$ later
    expires.  The
    superseded issue, the current issue, and the retired copies all carry
    valid marks and receive the same \textsc{Valid} answer from a static
    verifier.  (b) \textbf{A disputed edit propagates.}  A harmful or controversial local change
    introduced at revision $x$ persists through otherwise normal later
    revisions.}
    \label{fig:intro-problem}
\end{figure}

Watermark verification answers whether an image carries a publisher's mark,
and a verified image may be treated as authentic \citep{gowal2025synthid,nemecek2026integrity,c2pa2026softbinding,nian2026only}.  Publication, however, is not always a one-time event.
A publisher may revise and reissue an image to protect privacy, address safety risks, correct errors, or meet compliance requirements.  We call each authorized release an \emph{issue}.  Others may edit an issued image without the
publisher's approval, yet enough of the mark may survive for the result to verify \citep{bui2025trustmark,lu2025vine,nemecek2026integrity}.  The superseded issue, the current issue, and the unissued edit
therefore all point to the same trusted source while demanding different
decisions.  The publisher cannot simply clean up, either: once copies have
spread across websites, group chats, and caches, correction is not recall; the publisher can replace or withdraw
the image but cannot rewrite circulated copies \citep{melley2024kate,gowal2025synthid}, whose marks may remain valid (Figure~\ref{fig:intro-problem}a).  Nothing
has failed: the mark is
telling the truth about the past.  But the verifier now needs an answer
about the present.  Is this the current issue, a genuine but superseded
one, an edit that was never issued, or an image without sufficient
evidence?

This is only half of the question: whole-image currency alone is insufficient because an issue need not change every region.  The sky of a current issue may date from the first issue, the
subject from the second, and the caption from the third; a version number
says when the image was issued, not when each region entered authorized history. This distinction matters for auditing \citep{moreira2018provenance,nian2026auditable}.  A disputed object or background change
introduced at one revision persists through later, otherwise normal
revisions (Figure~\ref{fig:intro-problem}b); to retrieve the record that
approved it, an auditor needs each region's last authorizing revision, not
a change mask for the latest edit.  Figure~\ref{fig:intro-problem} shows
both halves: authentic does not mean current, and a disputed region keeps
its origin through later issues. Both halves are becoming routine: instruction-guided AI editors put a revision one prompt away \citep{brooks2023instructpix2pix,blackforestlabs2025kontext}, and return a complete re-rendering rather than a patch \citep{meng2022sdedit,gu2026denoising,qi2026editing}, so no part of the previous mark can be assumed to survive.

\FloatBarrier

We call this joint task \emph{reissuable provenance}.  Its inputs are a query
image, the issuer's key, and a small trusted reference to the lineage's current authorization state.  A \emph{lineage} is the set of issues descending from one source.
Its first output is the query's present status: \textsc{Current} for the
active issue; \textsc{Superseded} for an issue displaced by a successor or
explicitly retired; \textsc{Open} for an edit of the lineage that was never
issued---a fact of its history, not a verdict; and \textsc{Unknown} when
the evidence is insufficient---an abstention, since a destroyed mark is not
by itself an accusation.  Its second output is a \emph{field}: for each
region, the authorized revision that last changed it.  The two are coupled:
one reissuance both moves the image's status and labels the approved regions. The two answers belong in different places.  Present authority can change while the image does not, so no fixed image, key, and verifier can determine it from the pixels alone; it lives in the trusted reference, one entry per lineage.  Regional history is fixed at issuance and travels inside every circulated copy: pixels persist when metadata is stripped, and any party entrusted with the issuer's key and the lineage's list of issues reads the history from the copy, without the trusted reference or the issuer's edit records.  No existing watermark answers either half, for a structural
reason.

Existing watermark pipelines pose one round (embed, edit, detect), so time
and authorization never enter. 
A robust mark is expected to survive \citep{bui2025trustmark,lu2025vine}; a fragile mark is expected to break at the edit
\citep{wong2001secret,wu2023sepmark,zhang2024editguard,zhang2025omniguard}; re-embedding is
an overwrite attack \citep{bulychev2026watermarks} or a removal problem
\citep{bui2025trustmark,liu2025waterup}; and an editor's output is treated as adversarial, never as a draft.  Each reports whether and where a mark is intact
at one moment; none can say whether an issue is still current (an answer
that changes after the image is fixed) or which approval a region came
from.  Common localization benchmarks hide the gap by restoring the premise AI editors removed, an unchanged protected background: localization is scored by inpainting a
region and pasting it back into the protected image, so the background and
its old mark survive by construction
\citep{zhang2024editguard,zhang2025omniguard}, and supersession, approval,
and inheritance never occur.  Manifest standards such as C2PA do record
revisions and revocation in signed metadata, which a watermark can help
rediscover \citep{c2pa2026spec,c2pa2026softbinding}, but the record
describes one asset's history, not which issue of a lineage is current or
which approval each region came from. 


Where a one-round pipeline sees a credential that survives or breaks, our design reads the mark as evidence at three moments (when an image becomes \textsc{Open}, \textsc{Current}, or \textsc{Superseded}), so that time and authorization enter. After an edit (\textsc{Open}), surviving source evidence ties the output to its lineage and a disrupted field marks an unresolved change: a broken mark is not the end of provenance, but the edit remains unauthorized until the issuer approves it. At approval (\textsc{Current}), the issuer writes the successor a fresh, complete field---the new revision on the approved regions, inherited labels elsewhere---so continuity comes from reissuance, not survival: no predecessor mark needs to survive beyond the next reissuance.  A region approved twice carries the later approval; the earlier approval remains in that issue's record. The same edit is untrusted before issuance and official after it; what separates the two is authorization, not pixels. After a later reissue (\textsc{Superseded}), the predecessor remains unchanged but is no longer current. We therefore keep this mutable fact in the trusted reference introduced above, an authenticated \emph{live head}, like OCSP for certificates \citep{santesson2013ocsp}.  The publisher updates it without touching any circulated copy. Everything inheritable is written into each reissued image; only what cannot be frozen lives in the head.


\emph{Generational Stratigraphy Fields} (GSF) realize this division.  The
field records each region's last authorized revision (its
\emph{generation}) as a keyed two-bit fingerprint, so its rate does not
grow with the history; the price is that fingerprints of different
generations collide in some regions as the history deepens.  A spatial
reader resolves such collisions using unambiguous neighboring regions and the coherence of revisions, which tend to occupy connected areas.  The
head says which issue, if any, is current; the field says which issue last
shaped each region.

A lifecycle question needs a lifecycle evaluation.  We retain the complete
outputs of AI editors, in which essentially every block changes, and follow
480 histories of four issued generations through review, authorization,
reissuance, branching, delivery, and later queries, scoring both outputs on
the same trajectories (Section~\ref{sec:evaluation}).  Three findings
follow.  Supersession works without rewriting copies: GSF classifies issued
images as current or superseded with 99.67\% accuracy, and recalls 94.09\%
of unissued edits as \textsc{Open}, so surviving marks do not make drafts
official.  Spatial reading is necessary: on cells with ambiguous two-bit fingerprints, independent decoding accuracy falls to 25.6\% at the deepest tested
history while the spatial reader stays near 89\%.  One field answers when
and where: it recovers each region's last authorizing revision with 93.13\%
accuracy, and its newest slice overlaps the approved revision with 92.86\%
IoU.  Continuity thus survives legitimate revision and
unresolved modification when survival, disruption, authorization, and
reissuance are treated as moments of one lifecycle.
Our contributions are:
\begin{itemize}
    \item \textbf{A reissuance problem.}  We formalize reissuable provenance as a four-way status decision coupled with each region's last authorized revision.  Present authority is stored in a trusted mutable reference because no static in-image credential can express it.
    \item \textbf{Generational Stratigraphy Fields (GSF)}.  We introduce a reissuance protocol whose authenticated live head exposes replacement and retirement without rewriting circulated copies.  In its field instantiation, a constant-rate keyed field carries complete inherited regional history, and a candidate-equivariant spatial reader recovers that compressed history.
    \item \textbf{A complete-output reissuance evaluation.}  We evaluate the
    joint task in realistic AI-editing workflows, retaining full outputs
    through review, authorization, reissuance, branching, delivery transformations,
    and later verification. The results show that GSF sustains issue-level authority and regional history
    together.
\end{itemize}

\section{Related Work}
\label{sec:related_work}

Prior work supplies the ingredients of reissuable provenance separately but does not natively couple live authorization state with in-image regional revision history.

\paragraph{Re-watermarking.}
Repeated watermarking predates this work \citep{mintzer1999multiple}: Multiple Re-Watermarking
traces trading paths through accumulated receiver marks, WaterUp \mbox{updates} a
distribution-stage identity, and TrustMark's remover supports removal and
re-embedding \citep{mascherkampfer2006multiple,liu2025waterup,bui2025trustmark};
a later mark can suppress an earlier one \citep{bulychev2026watermarks}.
Each embedding changes the message an image carries; these operations do not themselves define lineage-wide currentness or an inherited field of last-authorized regional revisions.  GSF
re-purposes the same interaction as a controlled revision within one
lineage: an authorized reissue, through the live head, supersedes every
unchanged predecessor (pixels unchanged, status changed), while an authenticated head update can retire the lineage. Each successor receives a freshly written field of last-authorized regional revisions, including inherited labels.

\paragraph{Manifest provenance.}
C2PA records provenance in signed, updatable manifests, checks signer
revocation through OCSP, and rediscovers stripped manifests through soft
bindings \citep{c2pa2026spec,c2pa2026softbinding,collomosse2024authenticity}; it already has temporal
records and mutable resolution.  Its validation, however, remains asset-scoped: an active manifest establishes validity for the queried asset, not which issue of a lineage is currently authorized, while revocation concerns the signer's credential rather than one issue's authority. An
application could assemble a
\textsc{Current}/\textsc{Superseded}/\textsc{Open}/\textsc{Unknown} policy
from repository and lineage state and replay regions along the selected branch for each
region's last authorized revision, but neither is native and both live
outside the pixels. GSF instead separates the placement: regional history travels inside every reissued image, while an authenticated live head supplies the mutable information needed for lineage currentness, naming the current issue or declaring the lineage inactive (Section~\ref{sec:problem}).
 
\paragraph{Spatial and forensic watermarks.}
Spatial watermarks recover localized or multiple messages
\citep{sander2025wam,hu2025maskwm}, and forensic watermarks pair global
attribution with local tamper evidence
\citep{zhang2024editguard,zhang2025omniguard,chen2025tagwm,yang2025stableguard,gan2026genptw,liu2026pai}.
Both output a snapshot: where a message remains readable or tampering left
evidence.  Their native snapshot outputs do not specify how authorization state and regional revision labels evolve through reissuance.  GSF
reads the same local evidence in time: in an unissued edit, surviving and disrupted evidence supports recognition of an unresolved transition; after reissue, the generation map records the last authorized revision of each region.  Each slice marks regions by their last authorized revision; the newest localizes the issue's latest revision. One field thus provides time and location without a separate localization output.
 
\paragraph{Robustness and evaluation under generative editing.}
Robust watermarks are tested under regeneration \citep{zhao2024removable,an2024waves} and instruction-guided editing \citep{hu2024robustwide,lu2025vine}, and two concurrent analyses model diffusion editing as an information-contracting channel \citep{gu2026denoising,qi2026editing}; both focus on whether a static payload remains recoverable after editing. GSF asks instead what preserved and disrupted evidence means before and after issuer commitment, for generative and conventional edits alike. Evaluation also depends on which image is retained: paste-back preserves
the carrier outside the edited region, whereas a complete output may
disrupt it throughout. Forensic schemes include complete generative edits
\citep{zhang2025omniguard,gan2026genptw,liu2026pai}, but evaluate post-edit
recovery or tamper evidence rather than authorization transitions
through reissuance. Our evaluation carries the complete editor output through review, reissuance, branching, and later queries; payload robustness thereby becomes lifecycle robustness.
 
\section{Reissuable Provenance}
\label{sec:problem}
 
This section formalizes the two outputs introduced above: a status relative
to the live head and a field of last authorized revisions.  Throughout, the
issuer holds the watermark key $K$ and an authenticated \emph{registry} that
records, for each lineage $L$, its issued labels $\mathcal G_L$ and its live
head $h$.
 
\paragraph{Objects and authorization.}
Let $\mathcal B=\{B_i\}_{i=1}^{N}$ partition the image into cells.  A
lineage $L$ has a finite set of issued labels $\mathcal G_L$ and a live head
$h\in\mathcal G_L\cup\{\dagger\}$: $h\in\mathcal G_L$ names the current
issue, and $h=\dagger$ declares none active.  A proposed successor has a
fresh label $g\notin\mathcal G_L$ and an issued parent
$\operatorname{pa}(g)\in\mathcal G_L$; an editor produces an unissued output
$U$ from the parent's issued image.  If the issuer approves $U$ as the
content $X_g$ of a new issue, it also commits (records in the
registry) an \emph{authorized revision scope} $A_g\subseteq\{1,\ldots,N\}$,
the record of what it approves as revision $g$.  Pixel difference, watermark
disruption, or an authenticated edit record may propose the scope, but only
the issuer's commitment makes it $A_g$.  Authorization then updates the
last-authorized-generation field
\begin{equation}
\tau_0(i)=0,\qquad
\tau_g(i)=
\begin{cases}
g, & i\in A_g,\\
\tau_{\operatorname{pa}(g)}(i), & i\notin A_g.
\end{cases}
\label{eq:field-update}
\end{equation}
Cells in $A_g$ advance to $g$; all others inherit the parent's last
authorized revision, so the parent map covers branches as well as linear
histories.  The field $\tau_g$ is logical; Section~\ref{sec:writer} gives its
encoding.
 
\paragraph{Verification semantics.}
The \emph{event graph} of issuance and retirement events defines an
image's true status; the verifier must infer it from evidence.  A query $Y$
yields lineage evidence $\lambda_L(Y)\in\{0,1\}$, whether the evidence
supports association with $L$, and issue evidence
$\mu_L(Y)\in\mathcal G_L\cup\{\bot\}$, a supported match to an issued label,
if any; the decoded lineage identifier selects $L$ in the registry, which
returns $\mathcal G_L$ and $h$.  Both are evidence-supported conclusions,
and the verifier's evidence-conditioned prediction is
\begin{equation}
\widehat s_L(Y;h)=
\begin{cases}
\mathsf{UNKNOWN}, & \lambda_L(Y)=0,\\
\mathsf{CURRENT}, & \lambda_L(Y)=1,\quad
  h\in\mathcal G_L,\ \mu_L(Y)=h,\\
\mathsf{SUPERSEDED}, & \lambda_L(Y)=1,\quad
  \mu_L(Y)\in\mathcal G_L,\ \mu_L(Y)\neq h,\\
\mathsf{OPEN}, & \lambda_L(Y)=1,\ \mu_L(Y)=\bot.
\end{cases}
\label{eq:lifecycle-semantics}
\end{equation}
With $h=\dagger$, every matched issue is \textsc{Superseded}, whether
displaced or retired, so updating $h$ changes the meaning of an unchanged
image.  \textsc{Open} is predicted when lineage evidence remains but no
issued-label match is supported; such an output may come from an editor or
an attack, and the prediction authorizes nothing.  The verifier returns
$\mathcal V(Y;L,K,h)=(\widehat s,\widehat\tau)$ with
$\widehat\tau(i)\in\mathcal G_L\cup\{\bot\}$, where $\bot$ marks
insufficient regional evidence.  For an unissued output derived from parent
$r$, the target field is $\tau_r$: an unissued edit does not advance the
field, and only issuance of $g$ assigns $g$ to $A_g$.
 
\paragraph{Necessity of the live head.}
Supersession and retirement change the status of an image that itself does
not change.  A verifier whose every trusted input is fixed at
issuance (the image, its key, any static reference) receives identical
inputs before and after such an event and returns the same answer, so it
cannot move an image from \textsc{Current} to \textsc{Superseded}; that
status must come from a trusted input the issuer updates, and telling the
$|\mathcal G_L|$ possible current issues and the inactive value apart takes
at least $\lceil\log_2(|\mathcal G_L|+1)\rceil$ bits of it
(Appendix~\ref{app:formal-semantics}).
 
\section{Method}
\label{sec:method}
 
\subsection{Reissuance Protocol}
\label{sec:reissuance-protocol}
 
Our reissuance protocol connects authorization, regional inheritance, and
head-relative verification.  For a realization supporting both task
outputs, each issued image carries
\begin{equation}
\mathcal P_g=
\bigl(P^{\mathrm{lin}}_L,
      P^{\mathrm{issue}}_{g,X_g},
      C_g\bigr),
\qquad
\widetilde X_g=\operatorname{Embed}_K(X_g;\mathcal P_g),
\label{eq:gsf-state}
\end{equation}
where $P^{\mathrm{lin}}_L$ is lineage evidence, $P^{\mathrm{issue}}_{g,X_g}$
is issue evidence for label $g$ and its issued content $X_g$, and $C_g$
encodes the field $\tau_g$.  The encoding and embedding operator are
implementation choices (Section~\ref{sec:writer}).  Lineage evidence may
remain decodable when no issue match is supported; once an issued label is
matched, the live head determines whether it is \textsc{Current} or
\textsc{Superseded}.
 
\begin{figure}[t]
    \centering
    \includegraphics[width=1\textwidth]{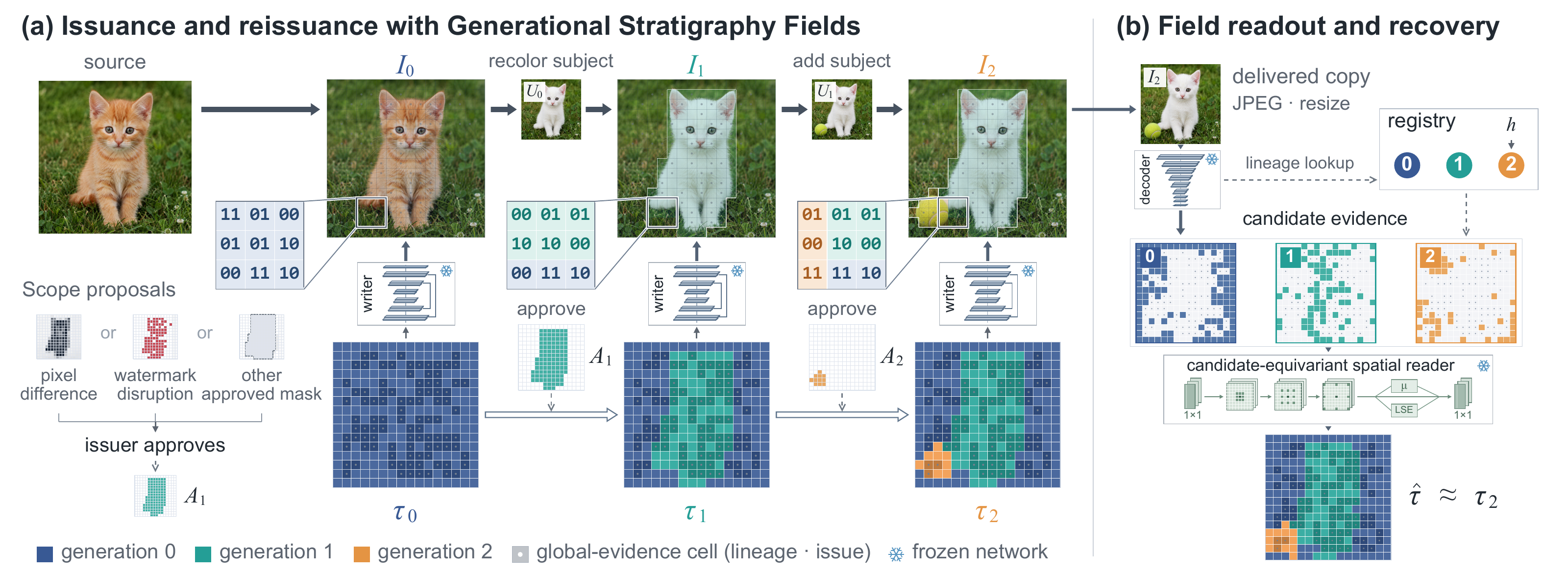}
    \caption{\textbf{Overview of reissuable provenance with GSF.}
\textbf{(a)} Every issued image $I_g$ carries its regional history as a
$16\times16$ field $\tau_g$.  A reissue starts from an editor's unissued
output $U_g$; the issuer approves a scope $A_g$ (from pixel difference,
watermark disruption, or another approved mask), and the writer rewrites the
whole field: cells in $A_g$ advance to generation $g$, the rest inherit
theirs.  Each cell holds a two-bit keyed fingerprint of its generation
(insets: the same window across $I_0$, $I_1$, $I_2$); dotted cells carry
lineage and issue evidence.
\textbf{(b)} From a delivered copy, a frozen decoder reads the field bits
and the lineage identifier; the registry supplies the candidate generations
and the live head $h$, and the candidate-equivariant spatial reader resolves
fingerprint collisions among the candidates to recover $\widehat\tau$.}
    \label{fig:overview}
\end{figure}

Editing and reissuance form one transition (Figure~\ref{fig:overview}).  An
unissued output with lineage support but no supported issue match is
predicted \textsc{Open}.  If the issuer approves the output, it commits a
revision scope $A_g$ (Section~\ref{sec:problem}).  The issuer then applies
Eq.~\eqref{eq:field-update}, and the writer freshly embeds the complete
payload $\mathcal P_g$ into the approved content $X_g$, including both new
and inherited regional labels.  Once the successor is embedded, the issuer
atomically records $g$ with its parent link in the registry and sets
$h\leftarrow g$, making all earlier issues \textsc{Superseded} without
rewriting their copies.  Retirement sets $h\leftarrow\dagger$ without
touching any image or payload; a later issuance reactivates the lineage.
 
\subsection{Generational Stratigraphy Fields: A Constant-Rate Keyed Two-Bit Encoding}
\label{sec:writer}
 
GSF instantiates this protocol with a shared watermark carrier for the three
terms in Eq.~\eqref{eq:gsf-state}.  Both global terms are keyed and
error-corrected, with issue evidence additionally bound to a descriptor of
the issued content (Appendix~\ref{app:frozen-instantiation}).  The field
admits two encodings at the same two-bit rate: direct generation labels and
keyed candidate fingerprints
(Appendix~\ref{app:same-rate-counterfactual}).  Direct two-bit labels are
collision-free for a fixed alphabet of at most four generations, but a label for an unbounded lineage needs
$\lceil\log_2|\mathcal G_L|\rceil$ bits per cell, so its rate grows with the
history.  GSF adopts keyed fingerprints for an extensible candidate set,
trading local ambiguity for position-dependent compatibility and explicit
no-match or ambiguous readouts.  With a pseudorandom function $F_K$,
\begin{equation}
c_{L,a,i}=\operatorname{Trunc}_{2}
F_K(L\mathbin\Vert a\mathbin\Vert i)\in\{0,1\}^{2},
\qquad
C_g(i)=c_{L,\tau_g(i),i}.
\label{eq:keyed-code}
\end{equation}
The two bits are not a version number but a compatibility test against the
registry's candidate set $\mathcal G_L$, so one interface serves any finite
candidate set.
 
At constant rate, collisions are unavoidable: under an ideal PRF a wrong
candidate matches the true two-bit value with probability $1/4$, so against
$d$ other candidates a cell identifies its generation uniquely with
probability
\begin{equation}
p^{(2)}_{\mathrm{uniq}}(d)=\left(\frac34\right)^d,
\qquad
p^{(2)}_{\mathrm{amb}}(d)=1-\left(\frac34\right)^d.
\label{eq:collision-law}
\end{equation}
Wider fingerprints collide less but, under a fixed slot budget, yield fewer
and coarser cells.  Under the same ideal-PRF model (keyed-iid) and a fixed
slot budget, two bits maximize the worst-case expected number of uniquely
identifying cells over $d\in\{1,2,3\}$, normalized by the best width at
each depth; deeper histories gradually favor wider codes
(Appendix~\ref{app:bitwidth-ladder} gives the law, its proof, and depth
sensitivity).
 
At reissuance the writer embeds all three terms of Eq.~\eqref{eq:gsf-state}
and rewrites every field cell: cells in $A_g$ receive the fingerprint of
$g$, all others the fingerprints of their inherited labels.  Because
$c_{L,a,i}$ is deterministic, a reissue refreshes the physical embedding
without changing inherited values, so a successor carries its history
itself rather than in residual pixels of its parent (embedding network and
channel simulation: Appendix~\ref{app:frozen-instantiation}).
 
\subsection{Candidate-Equivariant Generation-Field Reader}
\label{sec:spatial-inference}
 
Several generations can share a cell's fingerprint, so an independent
two-bit decision cannot always identify the generation.  The frozen decoder
returns bit logits $\ell_i\in\mathbb R^2$ at cell $i$; for each issued
generation $a\in\mathcal G_L$, $u_{i,a}$ sums the two bitwise
log-likelihoods under $c_{L,a,i}$
(Eq.~\eqref{eq:local-likelihood}, Appendix~\ref{app:spatial-reader-details}).
A hard read $z_i$ admits the candidates
$\mathcal C_i(z_i)=\{a\in\mathcal G_L:c_{L,a,i}=z_i\}$: a singleton is an
unambiguous temporal anchor, a larger set cannot be resolved from that cell
alone, but neighboring cells can help because authorized revisions occupy
coherent regions.
 
The reader forms a cell--candidate evidence field $Z_{i,a}$ from $u_{i,a}$,
bitwise match evidence, and decoder reliability.  A shared pointwise
operator recalibrates likelihood, confidence, and margin evidence at each
cell; a shared multiscale spatial operator then propagates reliable anchors
across coherent revision regions.  With $\operatorname{Pool}$ the
concatenated mean and log-sum-exp over candidate planes,
\[
\begin{aligned}
E_a
  &=\Phi^{\mathrm{sp}}_\psi
    \!\left(\Phi^{\mathrm{pt}}_\psi(Z_a)\right),\\
p_i(a)
  &=\operatorname{softmax}_{a}
    \Psi_\psi\!\left(
      E_a,\operatorname{Pool}_{a'\in\mathcal G_L}(E_{a'})
    \right)_i,
\end{aligned}
\]
where $\Phi^{\mathrm{pt}}_\psi$, $\Phi^{\mathrm{sp}}_\psi$, and $\Psi_\psi$
share weights across candidate planes and $\operatorname{Pool}$ is
permutation-symmetric.  Hence
$\mathbf p_\psi(\pi Z)=\pi\mathbf p_\psi(Z)$ for any permutation $\pi$ of
the candidate order (Eq.~\eqref{eq:reader-equivariance}), and the label map
$\widehat\tau(i)=\arg\max_{a\in\mathcal G_L}p_i(a)$ carries no candidate-order
semantics.
 
The reader returns $\widehat\tau$ where field evidence is readable and
$\bot$ elsewhere; the predicted status still follows
Eq.~\eqref{eq:lifecycle-semantics} from lineage and issue evidence.
\section{Experiments and Results}
\label{sec:results}
\setcounter{topnumber}{3}
\setcounter{bottomnumber}{2}
\setcounter{totalnumber}{4}

\subsection{A Complete-Output Reissuance Evaluation}
\label{sec:evaluation}
 
We follow 480 lineages, each rooted in one COCO {\citep{lin2014coco}} image, through four issues
$I_0$--$I_3$.  Between consecutive issues,
one of four editors (FLUX.1 Kontext \citep{blackforestlabs2025kontext}, InstructPix2Pix \citep{brooks2023instructpix2pix}, Qwen Image Edit \citep{wu2025qwenimage},
SDXL-Turbo \citep{sauer2024adversarial}) returns a complete image, which the issuer approves and reissues;
each root also keeps two outputs that are never issued, the editor's output
after $I_3$ and one branch from $I_1$ or $I_2$ (2,400 editor calls).  Every
return is kept whole (the median fraction of field cells with any pixel
change is 1.00 for all four editors), so the evaluation does not rely on
preserved background pixels.  Thresholded blockwise pixel difference
proposes each scope, a median of about 150 of the 256 cells; the issuer commits it as $A_g$, the
writer rewrites the whole field, and the head advances
(Section~\ref{sec:method}).  Here watermark disruption plays no part in the
scope; it independently provides evidence to the verifier.  True status is
fixed by the event graph before verification; queries cover each issued image
under its own head ($4\times480$, \textsc{Current}) and every later head
($6\times480$, \textsc{Superseded}), and each unissued output under its
parent's head and, for branches, later heads (3,129 queries, \textsc{Open}).
Every query supplies the lineage $L$, its key $K$, and the head $h$, so the
verifier decides among the four statuses for an image attributed to the
lineage; negative queries, unmarked or of another lineage, are tested in
Appendix~\ref{app:negative-queries}.
One frozen verifier answers every query; metrics are status accuracy,
\textsc{Open} recall, false-issue acceptance (an unissued output predicted
\textsc{Current} or \textsc{Superseded}), and for the field cell accuracy,
mIoU over generations, and the IoU of the newest slice
$\{i:\widehat\tau(i)=g\}$ with $A_g$; the root is the statistical unit.
Construction details, the paste-back controls, and the statistical
treatment are in Appendices~\ref{app:evaluation-instantiation},
\ref{app:complete-output-controls}, and~\ref{app:fidelity-statistics}.

\subsection{GSF Keeps Authority Current across Reissuance}
\label{sec:results-lifecycle}

Queried as issued, status follows the live head without touching any issued
image, and unissued outputs are read as \textsc{Open}
(Table~\ref{tab:lifecycle-main}).  Issued images are classified correctly in
99.67\% of 4,800 queries, and no query reverses the head relation; of the
unissued outputs, 94.09\% are read as \textsc{Open} and 2.24\% as an issue.
Issue evidence separates the two: issued images keep a decodable issue
match, while three quarters of unissued outputs lose it and a disrupted
cluster catches most of the rest
(Appendix~\ref{app:lifecycle-results}).  Similar recall for
outputs later reissued and never reissued (Table~\ref{tab:lifecycle-main}b)
is consistent with a decision made on the commitment at query time rather
than on subsequent reissuance.

\begin{table*}[!ht]
\caption{Lifecycle decisions (\%).  \textbf{(a)} Issued
images under their own head (\textsc{Current}) and under later heads
(\textsc{Superseded}).  \textbf{(b)} Unissued outputs.  The 16
misclassified issued queries are read as \textsc{Open} (13) or
\textsc{Unknown} (3).}
\label{tab:lifecycle-main}
\centering
\scriptsize
\begin{minipage}[t]{0.53\textwidth}
\vspace{0pt}
\centering
\textbf{(a) Issued images: status accuracy (\%) $\uparrow$}\par\smallskip
\setlength{\tabcolsep}{2.0pt}
\begin{tabular}{@{}lccccc@{}}
\toprule
True status & $h=0$ & $h=1$ & $h=2$ & $h=3$ & Overall \\
\midrule
\textsc{Current}    & 99.58 & 99.79 & 99.79 & 98.96 & \textbf{99.53} \\
\textsc{Superseded} & --    & 99.58 & 99.79 & 99.79 & \textbf{99.76} \\
\bottomrule
\end{tabular}
\end{minipage}\hfill
\begin{minipage}[t]{0.45\textwidth}
\vspace{0pt}
\centering
\textbf{(b) Unissued outputs (\%)}\par\smallskip
\setlength{\tabcolsep}{1.5pt}
\begin{tabular}{@{}lrrrr@{}}
\toprule
Output group & Queries & \textsc{Open} $\uparrow$ & False issue $\downarrow$ &
\textsc{Unknown} $\downarrow$ \\
\midrule
Later reissued & $3\!\times\!480$ & 94.38 & 2.08 & 3.54 \\
Never reissued & 1,689              & 93.84 & 2.37 & 3.79 \\
All            & 3,129              & \textbf{94.09} & \textbf{2.24} & 3.68 \\
\bottomrule
\end{tabular}
\end{minipage}
\end{table*}

\subsection{One Generation Map Provides Time and Location}
\label{sec:results-generation-map}

\begin{figure*}[!ht]
\centering
\includegraphics[width=0.88\textwidth]{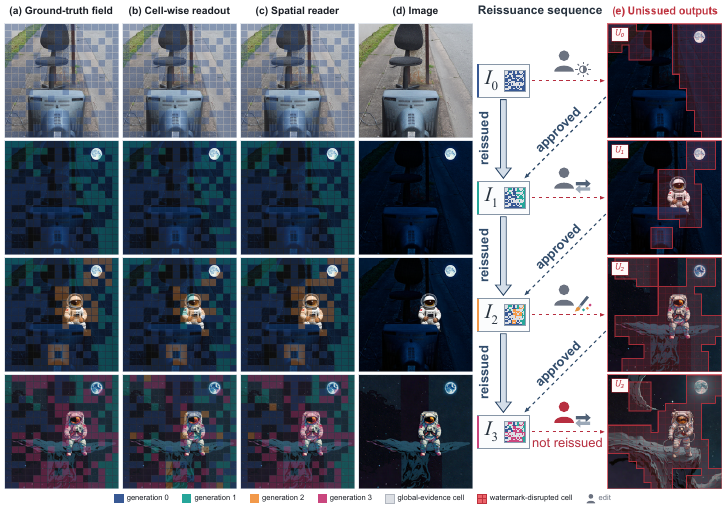}
\caption{One field follows an authorized revision sequence
$I_0\!\rightarrow\!I_3$ and its unissued outputs.  Colors give each region's
last authorized generation, so an older change stays attributable after later
revisions.  Red cells mark disrupted field evidence in unissued outputs; gray
cells carry lineage and issue evidence.}
\label{fig:lifecycle-sequence}
\end{figure*}

Once a revision is approved, the field records where and when it entered
history.  Repeated reissuance leaves a mosaic: 463 of 480 final fields carry
all four generations, since cells outside each committed scope inherit their
labels; a current issue thus holds more history than a whole-image version
number can express.  Table~\ref{tab:generation-map-results} reads that
mosaic on the current issue at depths 1--3; at depth 3 the field reaches
93.13\% map accuracy and 78.96\% mIoU.  mIoU falls from 93.87\% at depth 1
because older slices become small and fragmented.  On the 58.13\% of cells
whose fingerprint is ambiguous, the spatial reader reaches 88.27\% accuracy
against 25.56\% for independent decoding\rev{, and 84.84\% at depth 7, after four
further reissues of every root, where map accuracy and mIoU are 86.88\% and 59.03\%
(Appendix~\ref{app:generation-map-depth7})}.
Figure~\ref{fig:lifecycle-sequence} shows the audit path: a change that
entered the lineage at generation $a$, whether an object, an occlusion, or
a background, keeps label $a$ while later committed scopes leave it
untouched, so its region still points to the record that approved it.
The newest slice of the same map overlaps the committed scope $A_g$ with
92.86\% IoU, so the field that dates each region also localizes the latest
approved revision, without a separate localization output.  Time and
location are two readings of one field, and both hold on the same
successors as Table~\ref{tab:lifecycle-main}.

\begin{table*}[!ht]
\caption{Regional history read from the current issue at depths 1--3 (\%,
medians over roots).  Ambig.: share of the 160 field cells carrying
generation fingerprints (the other 96 carry lineage and issue evidence)
whose fingerprint is consistent
with more than one issued generation; Cell/Spatial acc.: independent decoding
and the spatial reader on those cells; Map acc./mIoU: all 160 cells; Newest
IoU/F1: the newest slice against $A_g$.}
\label{tab:generation-map-results}
\centering
\scriptsize
\setlength{\tabcolsep}{3.0pt}
\begin{tabular*}{\textwidth}{@{\extracolsep{\fill}}crrrrrrr@{}}
\toprule
& \multicolumn{3}{c}{Ambiguous cells} &
  \multicolumn{4}{c}{Whole field and newest slice} \\
\cmidrule(lr){2-4}\cmidrule(lr){5-8}
Depth & Ambig. & Cell acc. $\uparrow$ & Spatial acc. $\uparrow$ &
Map acc. $\uparrow$ & Map mIoU $\uparrow$ & Newest IoU $\uparrow$ &
Newest F1 $\uparrow$ \\
\midrule
1 & 24.38 & 46.67 & \textbf{88.64} & 97.19 & 93.87 & 94.92 & 97.39 \\
2 & 43.75 & 35.14 & \textbf{89.23} & 95.00 & 87.27 & 94.09 & 96.95 \\
3 & 58.13 & 25.56 & \textbf{88.27} & 93.13 & 78.96 & 92.86 & 96.30 \\
\bottomrule
\end{tabular*}
\end{table*}

\subsection{Status and Regional History under Delivery Transformations}
\label{sec:results-robustness}

\begin{figure*}[!htb]
\centering
\includegraphics[width=\textwidth]{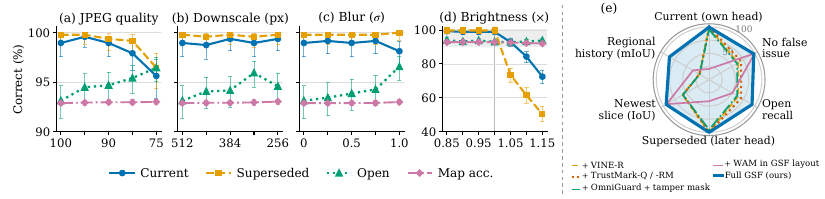}
\caption{\textbf{(a--d)} Status decisions under ordinary delivery at head 3:
\textsc{Current}, the current issue $I_3$; \textsc{Superseded}, the initial
issue $I_0$; \textsc{Open}, the never-issued output after $I_3$ and the
branch, pooled; Map acc., generation-map accuracy of the same $I_3$ at
depth 3, averaged over all 480 transformed images whatever their predicted
status.  \textbf{(e)} One polygon per system: \textsc{Current} and
\textsc{Superseded} accuracy, \textsc{Open} recall, the no-false-issue
rate, and map mIoU and newest-slice IoU averaged over depths 1--3;
TrustMark-RM coincides with TrustMark-Q.}
\label{fig:delivery-robustness}
\end{figure*}

Regional labels remain readable under tested transformations that impair
status recognition.  Applying one delivery transformation at a time to the
final-head queries repeats the three status decisions and the map reading
on the same current issue (Figure~\ref{fig:delivery-robustness}a--d).
Under JPEG down to quality 75, downscaling to 256 px, blur up to
$\sigma=1.0$, and darkening, all three statuses stay above 93\% without
per-condition tuning, and \textsc{Open} recall does not fall.  Brightening
is the boundary for status: at $\times1.15$ the current issue is still
\textsc{Current} in only 72.5\% of cases, while the map read from the same
images stays within 0.8 points of its untransformed value, as it does under
every condition shown.
Over the 31 transformations of Appendix~\ref{app:rendition-robustness},
map accuracy falls below 90\% only under JPEG 30 and 192-px resizing, where
at most 1\% of issued images keep their status.  Across the tested
issued-image queries, status errors are \textsc{Open} or \textsc{Unknown}
rather than a \textsc{Current}--\textsc{Superseded} reversal, and false
acceptance of unissued outputs stays at or below 2.6\%.  Certifying an
issue and reading its history rest on different evidence: a status needs a
supported issue match for the whole image, which local field disruption or
a failed issue-level check removes, whereas regional labels are read from
the field itself and remain recoverable after that whole-image check fails.

\subsection{The Reissuance Protocol Transfers across Watermark Carriers}
\label{sec:results-cross-method}

The evaluated watermarks do not natively expose the combined live-head
status, unissued-output, and regional-history interface.  We therefore equip
each with the reissuance protocol without retraining---a
generation-conditioned issue message rewritten at every reissue, the live
head, and the registry---plus OmniGuard's tamper mask and, for WAM, GSF's
field layout (Appendix~\ref{app:cross-method-comparison}).
Table~\ref{tab:cross-method-lifecycle} thus tests support for the joint
task, not the methods' native objectives.  Every global-message watermark
separates \textsc{Current} from \textsc{Superseded} at 94.7--100\%.  This
is a contribution of our protocol itself.  Unissued recognition and
inherited regional history, by contrast, need image evidence the protocol
does not carry.  The only such signal a global message carries is its
destruction, which reads as \textsc{Open}, whereas a surviving message
decodes to the old issue and is accepted.  VINE-R and both TrustMark
variants therefore recall 63.5--70.5\% of unissued outputs and accept the
remaining third as issues; OmniGuard reaches 61.3\% recall, 26.3 points
from its tamper mask, but its message survives more edits than any other,
so its false-issue acceptance is the highest, 38.7\%, against GSF's 94.0\%
and 2.1\%.  With the field layout, WAM localizes the newest revision as
well as GSF (newest-slice IoU 93.44\% against 92.86\% at depth 3) but keeps
little inherited history (24.30\% mIoU against 78.96\%\rev{; 34.22 and 43.15\% when every field
cell is rewritten at each reissue, at 46.7 and 37.8~dB, Appendix~\ref{app:cross-method-refresh}}): locating the
latest change is not keeping the history behind it.  GSF couples the live head with
image evidence of disruption and history, and among the evaluated systems it
alone closes all three parts of the task
(Figure~\ref{fig:delivery-robustness}e).
\providecommand{\nomech}[1]{\textcolor{gray}{#1}}
\begin{table*}[!htb]
\caption{Evaluation on the same 480 roots and 2,400 edits (\%).  ``$+$'': a
released watermark carrying our reissuance protocol.  Issued columns: status
accuracy under the own and later heads; unissued columns: the
$6\!\times\!480$ unissued queries shared by all systems; mIoU: medians at
depths 1--3.  One rule for all systems: destroyed messages read as
\textsc{Open}; carriers without a field never abstain (--); global messages
are scored as fields labeling every cell with the head (gray).
$^\dagger$WAM's generation block is readable on 442/480 roots; the rest
count as errors.  \rev{$^\ddagger$Every field cell rewritten at each reissue; initial-issue
PSNR as stated (Appendix~\ref{app:cross-method-refresh}).}}
\label{tab:cross-method-lifecycle}
\centering
\scriptsize
\setlength{\tabcolsep}{2.8pt}
\begin{tabular*}{\textwidth}{@{\extracolsep{\fill}}lrrrrrrrr@{}}
\toprule
& \multicolumn{2}{c}{Issued images $\uparrow$} &
  \multicolumn{3}{c}{Unissued outputs} &
  \multicolumn{3}{c}{Generation-map mIoU $\uparrow$} \\
\cmidrule(lr){2-3}\cmidrule(lr){4-6}\cmidrule(lr){7-9}
Existing carrier $+$ our reissuance protocol & \textsc{Current} & \textsc{Superseded} &
\textsc{Open} $\uparrow$ & False issue $\downarrow$ &
\textsc{Unknown} $\downarrow$ & $d{=}1$ & $d{=}2$ & $d{=}3$ \\
\midrule
$+$ VINE-R                          & 98.2 & 98.9 & 63.5 & 36.5 & -- & \nomech{28.91} & \nomech{19.40} & \nomech{14.70} \\
$+$ TrustMark-Q                     & \textbf{99.9} & \textbf{100.0} & 70.1 & 29.9 & -- & \nomech{28.52} & \nomech{19.27} & \nomech{14.40} \\
$+$ TrustMark-RM                    & 99.6 & 99.8 & 70.5 & 29.5 & -- & \nomech{28.52} & \nomech{19.21} & \nomech{14.65} \\
$+$ OmniGuard $+$ tamper mask       & 95.3 & 94.7 & 61.3 & 38.7 & -- & \nomech{28.32} & \nomech{19.08} & \nomech{14.65} \\
$+$ WAM in GSF layout$^\dagger$     & 20.4 & 40.8 & 50.8 & 0.4 & 48.8 & 49.05 & 32.38 & 24.30 \\
\rev{$+$ WAM in GSF layout, full refresh, 46.7~dB$^\ddagger$} & \rev{90.7} & \rev{95.7} & \rev{19.1} & \rev{23.2} & \rev{57.7} & \rev{63.61} & \rev{43.56} & \rev{34.22} \\
\rev{$+$ WAM in GSF layout, full refresh, 37.8~dB$^\ddagger$} & \rev{90.9} & \rev{95.7} & \rev{58.5} & \rev{3.8} & \rev{37.7} & \rev{80.40} & \rev{55.68} & \rev{43.15} \\
\midrule
\textbf{Full GSF (ours)}            & 99.5 & 99.8 & \textbf{94.0} & 2.1 & 3.9 & \textbf{93.87} & \textbf{87.27} & \textbf{78.96} \\
\bottomrule
\end{tabular*}
\end{table*}

\FloatBarrier
\setcounter{topnumber}{2}
\setcounter{bottomnumber}{1}
\setcounter{totalnumber}{3}

\section{Conclusion}
\label{sec:conclusion}

We introduced reissuable provenance, a joint task beyond watermark validity:
tracking current authorization, which changes with the live head, and each
region's last authorized revision, updated only at issuance.  The reissuance
protocol connects authorization, reissuance, inheritance, and later
verification, and transfers across watermark carriers.  The GSF
instantiation answers this task end to end: the live head and image
evidence distinguish current issues, genuine but superseded issues, and
unissued edits without rewriting circulated copies.  Its generation map
recovers inherited regional history rather than only locating the latest
revision.  Delivery experiments show that this history can remain readable
when issue certification fails.  The task, protocol, and complete-output
evaluation provide a common basis for alternative codes, readers, and
deployments.


\section*{Ethics Statement}
This work involves no human subjects and collects no personal data.  Source
images are drawn from the public COCO dataset and used under its terms; the
edited images are produced by released image editors, and the compared
watermarks are released models, all used under their respective licenses,
and no image is redistributed beyond what those terms permit.  The reissuance protocol lets an
issuer certify its own revisions; it judges authorization, not how content
was produced, and authorizes nothing by itself: an \textsc{Open} decision
marks an unresolved
change and is not an accusation against whoever holds the image
(Appendix~\ref{app:rendition-robustness}), and a \textsc{Superseded} decision
changes only which issue is current.  The live head is under the issuer's
control and the watermark key is secret, so the mechanism confers no authority
on third parties, and the paper reports the conditions under which its
decisions fail or abstain (Appendices~\ref{app:rendition-robustness}
and~\ref{app:negative-queries}).  We foresee no other ethical concerns.

\section*{Reproducibility Statement}
The task and verifier semantics are stated in Section~\ref{sec:problem} and
formalized in Appendix~\ref{app:formal-semantics}, which also proves the
live-reference bound.  The frozen instantiation is specified in
Appendix~\ref{app:frozen-instantiation}: payload coding and cell layout
(Table~\ref{tab:app-coding}), writer and decoder architectures
(Tables~\ref{tab:app-writer-architecture}
and~\ref{tab:app-decoder-architecture}), the training procedure and the
physical-channel check (Table~\ref{tab:app-carrier-gate}), blind alignment,
and the evidence-routing constants used for every reported decision
(Table~\ref{tab:app-verifier-parameters}).  The evaluation is specified in
Appendix~\ref{app:evaluation-instantiation}: root sampling, the prompt, seed,
parent, and branch schedule fixed before generation
(Tables~\ref{tab:app-trajectory-composition}
and~\ref{tab:app-editor-config}), and the event-defined query schedule
(Table~\ref{tab:app-query-schedule}).  The statistical unit and interval
procedure are given in Appendix~\ref{app:fidelity-statistics}, whose
Table~\ref{tab:app-scope-of-evidence} maps each claim to the experiment that
supports it; the cross-method adapters and their reproducibility notes are in
Appendix~\ref{app:cross-method-comparison}.  All results come from one frozen
verifier without per-condition tuning.  Code, the frozen checkpoints, the
trajectory assignment, and every query row will be released upon
publication.

\section*{AI Use Statement}
In this work, we used generative AI tools for the following tasks that
require disclosure.  Generating synthetic data sets: every edited image in the
evaluation was produced by the released AI image editors named in
Section~\ref{sec:evaluation} and Appendix~\ref{app:evaluation-instantiation}
under a prompt, seed, parent, and branch schedule fixed before generation;
these editors are the object of study, and no output was selected or
regenerated by hand.  Research ideation, conceptual framework, methodology,
and experiment design:
the core designs were provided by the authors; large language models
assisted in refining their details and in writing the corresponding code,
which the authors reviewed and tested.  We have not used generative AI tools to
formulate mathematical claims, to provide ingredients for or write proofs, or
to interpret results; translation, dataset cleaning, and qualitative data
analysis are not applicable to this work.  Additionally, we used generative
AI tools for recommended-disclosure tasks: drafting parts of the paper
(portions of the appendices, later revised by the authors), editing the text
for readability, identifying and summarizing related literature, and checking
the consistency of tables and text against the experiment records.  We have
reviewed all AI-assisted work: every number, table, and claim was checked by
the authors against the experiment records; model-drafted text was rewritten
and verified by the authors; works suggested by the model were read by the
authors before being cited; and model-assisted code was reviewed and its
outputs checked against independent computations by the authors.  We take
responsibility for the final content of this work, including text, claims,
and artifacts produced with the aid of generative AI.

\bibliography{references}
\bibliographystyle{iclr2027_conference}

\clearpage
\appendix
\renewcommand{\floatpagefraction}{0.85}
\renewcommand{\topfraction}{0.9}
\renewcommand{\bottomfraction}{0.5}
\renewcommand{\textfraction}{0.1}
\setcounter{topnumber}{3}
\setcounter{bottomnumber}{2}
\setcounter{totalnumber}{4}
\makeatletter
\setlength{\@fptop}{0pt}
\setlength{\@fpsep}{8pt plus 2pt}
\setlength{\@fpbot}{0pt plus 1fil}
\makeatother
\section{Verification Semantics and Atomic Reissuance}
\label{app:formal-semantics}

The central semantic change in reissuable provenance occurs without changing
the queried bitmap.  Once a successor becomes live, every earlier issue must
be interpreted as \textsc{Superseded}; when the lineage is retired, no issue
remains current.  A small authenticated live head makes this new meaning
available to every distributed copy at verification time, without rewriting
the copy or its watermark.  We formalize that behavior by separating
event-defined truth, image evidence, and the verifier output supported by the
evidence.

\subsection{Truth, Evidence, and Regional Targets}

Let $r$ denote an issued parent, $g$ a proposed successor label, and
$h\in\mathcal G_L\cup\{\dagger\}$ the live head, with $\dagger$ denoting an
authenticated inactive lineage.  Table~\ref{tab:app-truth-evidence} gives the
complete distinction.
For an unissued editor output, the edit event does not advance the regional
field: the last-authorized-generation target remains $\tau_r$ wherever the
field is readable and is $\bot$ where evidence is insufficient.  Only an
issuer commit assigns $g$ to the committed scope.

\begin{table}[!htbp]
\caption{Event truth, observable evidence, and verifier outputs.  Truth is
defined by the authorization graph; predictions depend on evidence.}
\label{tab:app-truth-evidence}
\centering
\small
\resizebox{\linewidth}{!}{%
\begin{tabular}{lllll}
\toprule
Query event & Event-defined truth & Supported issue evidence & Head role & Regional target \\
\midrule
Issue $a=h\in\mathcal G_L$ & \textsc{Current} & matches $a$ & selects $a$ as live & $\tau_a$ \\
Issue $a\neq h$, $h\in\mathcal G_L$ & \textsc{Superseded} & matches $a$ & establishes $a\neq h$ & $\tau_a$ \\
Any issue, $h=\dagger$ & \textsc{Superseded} & matches its issue & declares no issue active & $\tau_a$ \\
Unissued output from $r$ & \textsc{Open} & no committed issue match & none until issuance & $\tau_r$ or $\bot$ \\
Insufficient/out-of-lineage evidence & \textsc{Unknown} & unsupported & cannot be applied & $\bot$ \\
\bottomrule
\end{tabular}}
\end{table}

These evidence channels have complementary roles.  Lineage evidence
supports association with $L$; issue evidence supports a match to one
issue; the live head interprets that match as \textsc{Current} or
\textsc{Superseded}; physical disruption can expose an unresolved change; and
issuer authentication alone turns a proposed spatial support into an
authorized revision.  Thus disruption may help predict \textsc{Open}, but it
does not itself authorize content.

\subsection{Authorization as an Atomic Reissuance Transaction}
\label{app:issue-transaction}

The implementation follows the following transaction, expressed without
code-specific function names.

\begin{enumerate}
    \item \textbf{Verify the parent.}  Resolve an issued parent
    $r\in\mathcal G_L$ and interpret it relative to the current head $h$.
    The proposed label satisfies $g\notin\mathcal G_L$.
    \item \textbf{Produce and review an unissued output.}  An editor returns a
    complete image $U$.  The event graph assigns $s^\star(U;h)=\mathsf{OPEN}$,
    while its regional last-authorized target remains $\tau_r$.
    \item \textbf{Propose a revision scope.}  Physical image evidence or an
    authenticated edit record proposes $R_g$.  The proposal has no authority
    by itself.
    \item \textbf{Authorize and construct the successor.}  The issuer accepts
    or revises the proposal to obtain $A_g$, applies
    Eq.~\eqref{eq:field-update}, and embeds a complete issue payload
    $\mathcal P_g$ into the approved content.  This step freshly embeds both
    labels introduced by $g$ and labels inherited from $r$.
    \item \textbf{Validate, record, and advance.}  The completed successor is
    checked before visibility.  The registry then atomically performs
    \begin{equation}
       \mathcal G_L\leftarrow\mathcal G_L\cup\{g\},\qquad
       \operatorname{pa}(g)\leftarrow r,\qquad h\leftarrow g.
       \label{eq:app-atomic-commit}
    \end{equation}
\end{enumerate}

If construction or validation fails, Eq.~\eqref{eq:app-atomic-commit} is not
performed: $g$ is not a legal candidate and $U$ remains an unclosed
\textsc{Open} event.  This ordering prevents a partially written successor
from becoming the live issue.

Retirement is a separate authenticated transaction: setting
$h\leftarrow\dagger$ leaves the issue set, images, and embedded payloads
unchanged while making every issue non-current.  A later successful
issuance transaction restores an active issue through $h\leftarrow g$.

\subsection{Proof of the Live-Reference Necessity}

Consider two worlds $W_0$ and $W_1$.  In $W_0$, issued bitmap
$\widetilde X_a$ is the live issue; in $W_1$, either an authorized successor
has been committed or the lineage has been retired, so the correct status of
the same $\widetilde X_a$ is \textsc{Superseded}.  Hold fixed the bitmap, key,
verifier program, immutable
registry material, and every other trusted verifier input.  A deterministic
verifier receives the same input tuple in both worlds and must return the same
answer.  A randomized verifier likewise induces the same output distribution
because its conditional input distribution is unchanged.  Neither can return
\textsc{Current} in $W_0$ and \textsc{Superseded} in $W_1$ with certainty.
Therefore at least one trusted input must vary when the authorization relation
changes.

For a fixed issue set $\mathcal G_L$, the head must distinguish
$|\mathcal G_L|$ possible active issues and one inactive value.  A mutable
coordinate with $b$ bits has at most $2^b$ distinguishable states.  Hence
$2^b\geq|\mathcal G_L|+1$ and
$b\geq\lceil\log_2(|\mathcal G_L|+1)\rceil$.  This proves the lower bound
stated in Section~\ref{sec:problem}.  A minimal mutable coordinate is therefore
sufficient to give otherwise static image evidence a live authorization
meaning.

\subsection{Security Assumptions}
\label{app:security-assumptions}

The issuer controls the watermark key and an authenticated registry that
provides integrity and freshness for $h$ and the finite candidate set
$\mathcal G_L$.  Historical bitmaps may be copied, replayed, post-processed,
or edited, but they cannot become a new issue unless the issuer
authenticates the issuance transaction.  A replayed predecessor is therefore
\textsc{Superseded}, an unissued derivative remains \textsc{Open}, and
evidence with no supported lineage association is \textsc{Unknown}
(Appendix~\ref{app:negative-queries}).  The
image experiments evaluate these lifecycle events and the delivery
transformations specified in Appendices~\ref{app:evaluation-instantiation}
and~\ref{app:rendition-robustness}.  \rev{Appendix~\ref{app:active-attacks} evaluates removal,
disruption, and forgery by a party without the key.}

\subsection{Frozen Components and Traceability}

Table~\ref{tab:app-artifacts} identifies the frozen components that determine
the reported results.  The anonymous artifact manifest records versions,
producing scripts, and table/figure dependencies.

\begin{table}[!htbp]
\caption{Frozen components of the evaluation.}
\label{tab:app-artifacts}
\centering
\small
\begin{tabular}{ll}
\toprule
Artifact & Role \\
\midrule
Writer--decoder checkpoint & physical two-bit channel \\
Carrier/verification core & layout, coding, synchronization \\
Protocol implementation & issue evidence and trajectory execution \\
Spatial-reader checkpoint & candidate-equivariant recovery \\
Trajectory assignment & roots, prompts, parents, seeds \\
Evaluation image set & roots and derived lifecycle images \\
Evaluation rows & status, robustness, reader, and maps \\
Localization controls & matched endpoint comparison \\
Editor repositories & complete-output generators \\
\bottomrule
\end{tabular}
\end{table}

Together, the frozen components, event assignments, query rows, and image
manifest connect every reported table to one fixed issuance graph.

\section{GSF Instantiation and Training Details}
\label{app:frozen-instantiation}

This section specifies the physical instantiation used throughout the evaluation.
One shared writer--decoder carries lineage evidence, issue evidence, and a
spatial generation field at a fixed operating point.

\subsection{Canonical Grid and Payload Allocation}

The verifier first represents an image on a canonical $512\times512$ canvas.
The decoder produces a $16\times16$ grid, so one cell corresponds to a
$32\times32$ pixel block and carries two binary logits plus a reliability
logit.  The 256 cells are partitioned into 32 lineage cells, 64 issue-evidence
cells, and 160 generation-field cells (Table~\ref{tab:app-coding}).  The
lineage and issue cells are
sampled deterministically from the interior; the generation field contains
the remaining 100 interior cells and 60 outer-ring cells.  Thus, after
reserving global evidence for status verification, the same carrier still
provides 160 history-bearing regions.

\begin{table}[!htbp]
\caption{Coding and layout of the frozen carrier.  A two-bit generation entry
is a keyed candidate fingerprint, not a binary generation number.}
\label{tab:app-coding}
\centering
\small
\begin{tabular}{lrrl}
\toprule
Component & Cells & Physical slots & Coding \\
\midrule
Lineage evidence & 32 & 64 & 36 bits $\rightarrow$ BCH$(63,36,t{=}5)$ \\
Issue evidence & 64 & 128 & 64 bits $\rightarrow$ BCH$(127,64,t{=}10)$ \\
Generation field & 160 & 320 & two keyed bits per cell \\
\midrule
Total & 256 & 512 & shared writer/decoder \\
\bottomrule
\end{tabular}
\end{table}

The 36-bit lineage identifier is BCH encoded \citep{hocquenghem1959codes,bose1960bch}, XORed with an HMAC-SHA256 stream \citep{krawczyk1997hmac,nist2015sha}
in the lineage-key domain, and placed in 63 of the 64 available slots.  Issue
evidence begins with a low-frequency content descriptor: the issued image is
converted to grayscale, resized to $128\times128$, partitioned into a
$4\times4$ layout, and quantized to four bits per block.  Its 64 bits are BCH
encoded and keyed by the candidate issue label.  At verification, every
issued candidate is decoded and the correctable candidate with the
fewest corrections is selected.

For a generation candidate $a$ and field cell $i$, the implementation computes
\begin{equation}
 c_{L,a,i}=\operatorname{LSB}_2
 \operatorname{HMAC\mbox{-}SHA256}(K,L\Vert\texttt{g}a\Vert i).
 \label{eq:app-prf-implementation}
\end{equation}
The operation is deterministic for fixed $(K,L,a,i)$.  Reissuance is fresh at
the embedding level, not at the fingerprint-value level.

\subsection{Logical Inheritance and Physical Full Rewrite}
\label{app:logical-physical-rewrite}

The logical update is local, but the physical write is complete.  Given the
committed target $\tau_g$, the payload supplied to the writer is
\begin{equation}
 C_g(i)=c_{L,\tau_g(i),i}\quad\text{for every field cell }i.
\end{equation}
Cells inside $A_g$ therefore carry fingerprints of $g$, while cells outside
$A_g$ freshly embed the deterministic fingerprints of their inherited labels.
Issuance never relies on residual pixels from the predecessor and does not
use masked physical writing.  The optional synchronization template of the
training code is disabled at issuance; alignment is described below.

\subsection{Writer and Decoder Networks}

Both networks are content conditioned and fully convolutional
(Tables~\ref{tab:app-writer-architecture}
and~\ref{tab:app-decoder-architecture}).  A
convolution--GroupNorm(8)--GELU block \citep{wu2018groupnorm,hendrycks2016gelu} is used throughout unless stated
otherwise.  The writer receives RGB concatenated with a nearest-neighbor
expansion of the two-bit field and predicts a bounded RGB residual.

\begin{table}[!htbp]
\caption{Frozen writer architecture.  All convolutions use padding that
preserves size unless stride is two.}
\label{tab:app-writer-architecture}
\centering
\small
\begin{tabular}{llll}
\toprule
Stage & Operation & Channels & Spatial size \\
\midrule
Input & RGB $+$ two-bit map & 5 & $512^2$ \\
Stem & $3\times3$, stride 1 & $5\rightarrow64$ & $512^2$ \\
Down 1 & $3\times3$, stride 2 & $64\rightarrow64$ & $256^2$ \\
Down 2 & $3\times3$, stride 2 & $64\rightarrow128$ & $128^2$ \\
Middle & two $3\times3$ blocks & $128\rightarrow128$ & $128^2$ \\
Up 2 & nearest upsample, skip, $3\times3$ & $192\rightarrow64$ & $256^2$ \\
Up 1 & nearest upsample, skip, $3\times3$ & $128\rightarrow64$ & $512^2$ \\
Output & $3\times3$, zero initialized, $\tanh$ & $64\rightarrow3$ & $512^2$ \\
\bottomrule
\end{tabular}
\end{table}

The predicted residual is multiplied by 0.1 and added to the normalized image
before clipping.  The writer contains 596,355 trainable parameters.

\begin{table}[!htbp]
\caption{Frozen decoder architecture.  The final three channels are two bit
logits and one confidence logit.}
\label{tab:app-decoder-architecture}
\centering
\small
\begin{tabular}{llll}
\toprule
Stage & Operation & Channels & Spatial size \\
\midrule
Stem & $3\times3$, stride 1 & $3\rightarrow64$ & $512^2$ \\
Down 1--2 & two $3\times3$, stride 2 & $64\rightarrow64\rightarrow128$ & $128^2$ \\
Down 3--5 & three $3\times3$, stride 2 & $128\rightarrow128\rightarrow256\rightarrow256$ & $16^2$ \\
Context & two $3\times3$ blocks & $256\rightarrow256$ & $16^2$ \\
Readout & $1\times1$ convolution & $256\rightarrow3$ & $16^2$ \\
\bottomrule
\end{tabular}
\end{table}

The decoder contains 2,329,155 trainable parameters.  Candidate likelihoods
use the two bit logits; the confidence logit and hard-bit margin enter the
reader evidence tensor in Appendix~\ref{app:spatial-reader-details}.

\subsection{Writer--Decoder Training}

We train for 100,000 optimizer steps with batch size 16 and
AdamW \citep{loshchilov2019adamw} with learning rate $10^{-4}$, default betas $(0.9,0.999)$, default weight
decay $0.01$, and gradient-norm clipping at 1.0.  Images were center cropped,
resized to $512\times512$, and sampled from a 25,000-image COCO training
directory disjoint from the 480 evaluation roots.  The experiment assigns
each image set a single role: these 25,000 images fit the writer and decoder; a
separate 24-image COCO validation set checks the frozen physical channel
(Table~\ref{tab:app-carrier-gate});
development and calibration roots determine the reader and routing rules;
and the 480 evaluation roots contribute only end-to-end scores.  The
24-image set has zero image-ID overlap with the evaluation roots.

Let $b$ denote the sampled payload and $q$ the confidence target indicating
whether both hard bits were correct at a cell.  The training objective is
\begin{equation}
\begin{aligned}
 \mathcal L={}&\operatorname{BCE}(D_{1:2}(T(\widetilde X)),b)\\
 &+0.1\operatorname{BCE}(D_3(T(\widetilde X)),q)\\
 &+\lambda_q\bigl(100\operatorname{MSE}(\widetilde X,X)
 +\operatorname{LPIPS}(\widetilde X,X)\bigr).
\end{aligned}
 \label{eq:app-writer-loss}
\end{equation}
For the first 2,000 steps the channel is identity; over the next 10,000 steps
the frozen distortion mixture is introduced, after which its six members are
drawn uniformly.  They are identity, real JPEG with quality uniformly sampled
from 50--90 using a straight-through gradient, down/up resize with scale
0.5--1.0, Gaussian noise up to $2/255$, circular shifts of at most four pixels
per axis, and a screenshot simulation (resize to 384, JPEG 70, resize to 512).

The quality weight starts at $\lambda_q=0.05$.  Every 100 steps, it is multiplied
by 1.05 (capped at 200) if PSNR or LPIPS \citep{zhang2018lpips} violates the quality target;
otherwise, if BER exceeds 0.02, it is multiplied by 0.97 (floored at 0.05).
The controller derives its quality targets from an independently measured
reference-carrier reproduction under the same image normalization: the PSNR
floor is the reference median minus 1 dB, and the LPIPS ceiling is twice the
reference median.  These rules yield PSNR $\geq35.2439$ dB and LPIPS
$\leq0.007310$.  The BER target of 0.02 governs the adaptive quality controller
during training; status decisions use the frozen evidence-routing rules
below.  We retain the rule-derived precision so the training run can be
reproduced exactly.

\begin{table}[!htbp]
\caption{Physical-channel characterization of the frozen checkpoint on 24
held-out COCO validation images, disjoint from both writer training and the
480 evaluation roots.  BER is measured over the full grid before geometric
support accounting or status routing.}
\label{tab:app-carrier-gate}
\centering
\scriptsize
\setlength{\tabcolsep}{3pt}
\begin{tabular}{@{}p{2.9cm}p{3.3cm}p{6.9cm}@{}}
\toprule
Check & Observation & Protocol significance \\
\midrule
Clean output & PSNR $\uparrow$: 45.646 dB; LPIPS $\downarrow$: 0.003431;
median maximum error $\downarrow$: 13
& Establishes the physical fidelity of the frozen carrier; the fidelity
distribution of the 480 evaluation roots appears in Appendix~\ref{app:fidelity-statistics}. \\
JPEG 75, resize to 384 and back, rotate $2^\circ$
& BER $\downarrow$: 0.00098, 0.00000, and 0.00081
& Confirms payload readout under representative delivery operators before
status routing. \\
Screenshot simulation & BER $\downarrow$: 0.01571
& Measures the harder composite delivery channel used in the training
curriculum. \\
Crop 90\% and resize & all-grid BER $\downarrow$: 0.07723
& Exposes missing border support; the verifier accounts for that support
before interpreting the payload. \\
\bottomrule
\end{tabular}
\end{table}

Crop makes the distinction between raw bit error and status evidence
explicit.  A separate controlled support-accounting probe on development
roots localized every crop error to the 60 border cells removed from view and
measured zero BER on the common-view cells.  The verifier records those border
locations as predictable erasures before BCH decoding.  The 0.07723 all-grid
value therefore quantifies geometric support loss, while the status
router consumes the erasure-aware evidence tensor.
Appendices~\ref{app:lifecycle-results} and~\ref{app:rendition-robustness}
measure the corresponding status behavior.

\subsection{Blind Alignment and Frozen Evidence Routing}

Alignment uses a deterministic search lattice and no reference image.
Rotation and scale maximize the mean absolute decoder margin, a key-free
score: the coarse search evaluates 13 rotations in $[-3^\circ,3^\circ]$ and
11 isotropic scales in $[0.85,1.15]$, and a second stage evaluates a
$9\times9$ grid within $\pm0.25^\circ$ and $\pm0.015$ scale of the coarse
optimum.  The verifier then searches integer translations with
$(t_x,t_y)\in\{-16,-12,\ldots,16\}^2$, scoring each candidate by agreement
with the keyed expected signs of Eq.~\eqref{eq:app-prf-implementation}, and
refines each coordinate within $\pm3$ pixels at unit steps.  Cells brought
in from outside the canvas are recorded as predictable erasures for BCH
decoding.  Evidence routing begins after canonicalization.

Threshold placement was completed on development data before the 480-root
evaluation.  Table~\ref{tab:app-verifier-parameters} groups the resulting rules
by evidence function and connects each placement rule to its end-to-end role.
Together they form one frozen evidence router shared across transforms.

\begin{table}[!htbp]
\caption{Frozen evidence-routing rules.  Development and calibration roots
place every value; the resulting operating point remains fixed across all 480
evaluation roots, the clean images, and 31 delivery transformations.}
\label{tab:app-verifier-parameters}
\centering
\scriptsize
\setlength{\tabcolsep}{3pt}
\begin{tabular}{@{}p{2.1cm}p{2.5cm}p{5.3cm}p{3.1cm}@{}}
\toprule
Evidence family & Frozen rule & Placement rule (development data) & Protocol role \\
\midrule
Cell readability
& readable iff margin $>c_\star$, $c_\star=4.2772$
& Maximum of the per-root first-percentile margins over 12 development roots
and five delivery transforms after predictable erasures (14,304 cell
observations).
& Produces the erasure mask consumed by downstream status inference. \\
Damage routing on clean queries
& $G_{\mathrm{dead}}=1$, $C_{\mathrm{cc}}=1$,
$B_{\mathrm{bg}}=1/4$
& The three coordinates were fixed jointly at one operating point on 144
clean queries from 12 development roots and 20 additional roots selected by
a fixed hash rule.  The status objective maximized event-defined
\textsc{Open} while preserving status correctness on issued images and
bounding false \textsc{Current} and issued-image \textsc{Open} outcomes.
& Separates localized disruption supporting \textsc{Open} from complete
loss routed to \textsc{Unknown}. \\
Unmatched field candidates
& $C_{\mathrm{nokey}}=1$
& Smallest integer covering the benign calibration maximum while keeping the
row-level false-trigger bound below $10^{-3}$; confirmed on a separate
12-root calibration split.
& More than one live cell matching no issued candidate supplies positive
\textsc{Open} evidence. \\
Issue consistency
& $\tau_{\mathrm{issue}}=17/64$
& 99th percentile over 359 benign queries from 60 development roots, rounded
up to the native $1/64$ distance grid.
& Excess mismatch supplies positive \textsc{Open} evidence. \\
Current-issue safeguard
& $\tau_{\mathrm{id}}=1/64$ within
$|\theta|<0.5^\circ$, $|s-1|\leq0.001$
& The pose predicate was fixed before the evaluation; the mismatch
threshold is the rounded-up 99th percentile of 240 benign identity-pose
queries on the same 60 roots.
& Withholds \textsc{Current} when current-issue evidence is insufficient,
routing the internal rejection to \textsc{Unknown}. \\
\bottomrule
\end{tabular}
\end{table}

The router composes these signals in a fixed order.  Low-margin cells first
become erasures.  Unmatched live candidates, undecodable issue evidence, a
localized disrupted cluster, and issue inconsistency then provide positive
\textsc{Open} evidence.  Broad degradation without an issue match supports
\textsc{Unknown}.  Once an issue is matched, equality with $h$ yields
\textsc{Current}, whereas a different issued label yields
\textsc{Superseded}.  Appendix~\ref{app:lifecycle-results} evaluates this one
operating point by status accuracy on issued images, \textsc{Open} recall, and false
issue acceptance.  Appendix~\ref{app:rendition-robustness} then applies it to
the clean images and 31 delivery transformations without per-condition
adjustment.

\paragraph{Static cost.}
The image-resident writer and decoder together contain 2.93M trainable
parameters; the default spatial reader adds only 4,633.  Registry and storage
growth are reported in Appendix~\ref{app:fidelity-statistics}.

\section{Complete-Output Reissuance Evaluation}
\label{app:evaluation-instantiation}

We instantiate the reissuance protocol on COCO-derived sources and retain the
complete RGB image returned by each editor.  Every trajectory therefore
passes through editing, review, authorization, reissuance, branching, and
later head-relative verification.  This design evaluates reissuable provenance across a
realistic complete-output workflow rather than restoring generated regions to
an unchanged watermarked background.

\subsection{Roots, Trajectories, and Edit Events}

We randomly sampled 480 images from the COCO dataset \citep{lin2014coco} as lineage roots.  Each
image was center cropped and resized to $512\times512$; every prompt, editor,
seed, parent, and branch assignment was fixed before generation.

For each root we store
\begin{equation}
 \text{source}\rightarrow I_0\rightarrow U_0\rightarrow I_1
 \rightarrow U_1\rightarrow I_2\rightarrow U_2\rightarrow I_3
 \rightarrow U_3,
 \label{eq:app-main-chain}
\end{equation}
together with one unissued derivative $U_p$ from either earlier issue $I_1$ or $I_2$
(Table~\ref{tab:app-trajectory-composition}).
The issues are $I_0,\ldots,I_3$; $U_0,\ldots,U_3,U_p$ are complete
editor returns prior to any commit.  The first three main-chain outputs are
later closed by successors.  The final output and earlier-issue derivative remain
unclosed, demonstrating that \textsc{Open} does not presuppose a future issue.

\begin{table}[!htbp]
\caption{Composition of the complete-output evaluation.}
\label{tab:app-trajectory-composition}
\centering
\small
\begin{tabular}{lr}
\toprule
Object & Count \\
\midrule
COCO-derived roots & 480 \\
Issues per root & 4 \\
Complete-output edits per root & 5 \\
Issue operations & $4\times480$ \\
Complete-output editor calls & 2,400 \\
Stored images per root & 10 \\
Homogeneous-editor chains & 48 (12 per editor) \\
Mixed-editor chains & 432 \\
Branches from issue 1 / issue 2 & 249 / 231 \\
\bottomrule
\end{tabular}
\end{table}

\begin{figure}[!htbp]
\centering
\includegraphics[width=\linewidth]{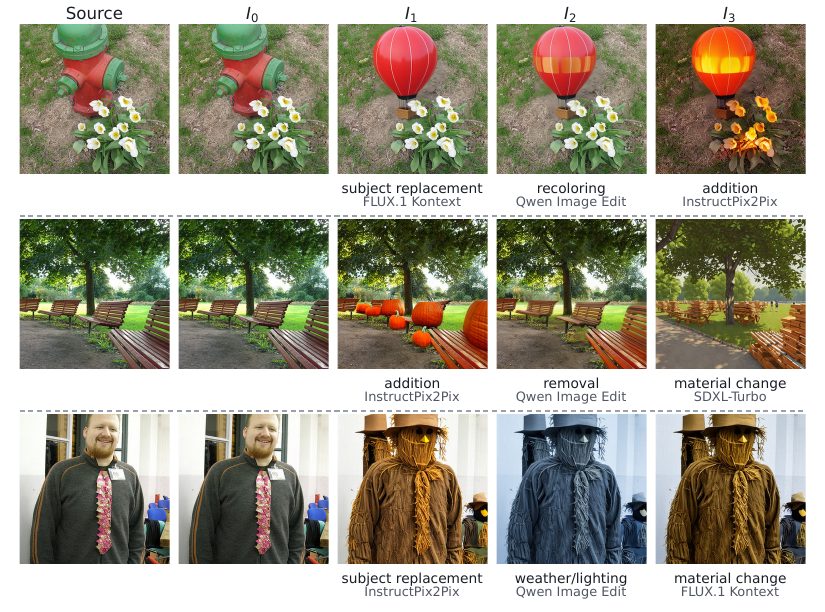}
\caption{Three issued lineages from the evaluation: the source image, the
initial issue $I_0$, and the reissues $I_1$--$I_3$, with the operation class
and the editor of each reissue.  Every reissue is a complete editor return,
not a pasted patch, so the writer rewrites the whole field at each issuance.}
\label{fig:app-lineages}
\end{figure}

The operation schedule contains 480 subject replacements, 168 recolorings,
120 material changes, 480 additions, 192 removals, 384 background
replacements, 384 weather/lighting changes, and 192 global-style changes.
Every root contains at least three operation classes and no immediate repeated
class; removals are scheduled only when an added object exists.  The frozen
schedule was executed with the editor configurations of
Table~\ref{tab:app-editor-config}, without outcome-dependent regeneration,
seed replacement, or sample deletion.  Figure~\ref{fig:app-lineages} shows
three of the resulting lineages with the operation and editor of each
reissue.

\begin{table}[!htbp]
\caption{Frozen complete-output editor configurations.  All returned images
were converted to RGB, resized to $512\times512$ with Lanczos when necessary,
and stored losslessly as PNG.}
\label{tab:app-editor-config}
\centering
\scriptsize
\begin{tabular}{lllr}
\toprule
Editor & Precision & Inference configuration & Edges \\
\midrule
FLUX.1-Kontext-dev & bfloat16 & 28 steps, guidance 2.5 & 592 \\
InstructPix2Pix & float16 & 30 steps, text guidance 8.0, image guidance 1.5 & 593 \\
Qwen-Image-Edit-2509 & bfloat16 & fused Lightning LoRA, 8 steps, CFG 1.0 & 591 \\
SDXL-Turbo img2img & float16 & 8 steps, guidance 0.0, strength 0.50 & 624 \\
\bottomrule
\end{tabular}
\end{table}

FLUX.1 Kontext, InstructPix2Pix, and Qwen-Image-Edit are instruction editors
\citep{blackforestlabs2025kontext,brooks2023instructpix2pix,wu2025qwenimage};
SDXL-Turbo \citep{sauer2024adversarial}, distilled from SDXL
\citep{podell2024sdxl}, is used as an image-to-image editor in the SDEdit
manner \citep{meng2022sdedit}.  All 591 Qwen rows use the fused Lightning
LoRA, eight steps, and CFG 1.0.

\subsection{Complete Returns Change the Whole Carrier}

For every edit we measured the mean absolute RGB change (8-bit units) and the
fraction of $32\times32$ cells containing physical pixel change.  Table
\ref{tab:app-complete-change} shows that the median changed-cell fraction is
one for all four editors.  This does not mean every change is semantically
intended; it establishes that an unchanged-background assumption is not
available when the complete return is retained.

\begin{table}[!htbp]
\caption{Physical change in retained complete outputs.  Entries are computed
from all 2,400 editor calls.}
\label{tab:app-complete-change}
\centering
\small
\begin{tabular}{lrrrrrr}
\toprule
& \multicolumn{3}{c}{Mean absolute RGB change} &
\multicolumn{3}{c}{Changed-cell fraction} \\
\cmidrule(lr){2-4}\cmidrule(lr){5-7}
Editor & p10 & median & p90 & minimum & p10 & median \\
\midrule
FLUX.1 Kontext & 4.95 & 16.18 & 81.06 & 0.09 & 0.80 & 1.00 \\
InstructPix2Pix & 13.82 & 34.77 & 70.59 & 0.60 & 1.00 & 1.00 \\
Qwen Image Edit & 5.28 & 17.26 & 51.23 & 0.18 & 0.74 & 1.00 \\
SDXL-Turbo & 24.75 & 32.55 & 41.64 & 1.00 & 1.00 & 1.00 \\
\bottomrule
\end{tabular}
\end{table}

\subsection{From a Physical Proposal to an Authorized Scope}
\label{app:scope-construction}

At reissuance, the issuer supplies a proposed revision scope and commits the
region it approves.  GSF accepts proposals derived from an edit mask, layer
history, an authenticated operation record, pixel difference, or watermark
disruption.  For the complete-return experiment, we instantiate this input
with deterministic \emph{blockwise pixel difference} between the issued
parent and editor return, computed before verification.

For an issued parent image $I_r$ and complete editor return $U$, define
\begin{equation}
 d(p)=\frac{1}{3}\sum_c |U_c(p)-I_{r,c}(p)|,\qquad
 \bar d_i=\frac{1}{|B_i|}\sum_{p\in B_i}d(p).
\end{equation}
The initial proposal selects cells whose mean change exceeds a fixed relative
and absolute threshold:
\begin{equation}
 R_i^{(0)}=\mathbf 1\!\left[\bar d_i>
 \max\{2,Q_{0.75}(\{\bar d_j\})\}\right].
 \label{eq:app-scope-proposal}
\end{equation}
If fewer than 32 eligible interior generation cells are covered, the binary
grid is dilated with a $3\times3$ kernel for at most eight iterations.  This
fixed construction supplies one reproducible proposed region for every
complete return.

The issuer commits the approved region as $A_g\leftarrow R_g$, thereby
updating $\tau_g$ and recording the successor.  GSF then writes the
committed scope into the successor and carries it forward as regional
authorization history.  A human-provided mask, layer record, or authenticated
edit record enters the same commit-and-reissue path, allowing each deployment
to choose proposal evidence suited to its authoring workflow.

The 256-cell logical grid defines the complete authorized revision field.  In
the evaluated carrier, 160 cells are configured as history-bearing
observations, and generation-map metrics score all 160 of them; the remaining
96 cells provide lineage and issue evidence.

\subsection{Event-Defined Query Schedule}

Table~\ref{tab:app-query-schedule} enumerates all clean lifecycle queries.  An
issued bitmap can be queried at multiple later heads without being
regenerated; this is precisely the test of head-relative semantics.

\begin{table}[!htbp]
\caption{Complete clean query schedule.  Earlier-issue derivatives are evaluated at
the heads for which their issued parent has the stated relation to the
live issue.}
\label{tab:app-query-schedule}
\centering
\small
\begin{tabular}{llr}
\toprule
Ground-truth class & Query condition & Query set \\
\midrule
\textsc{Current} & each issue under its own live head & $4\times480$ \\
\textsc{Superseded} & each predecessor under every later head & $6\times480$ \\
\textsc{Open} & pre-commit outputs later closed on the main chain & $3\times480$ \\
\textsc{Open} & unissued output from the terminal issue & $1\times480$ \\
\textsc{Open} & derivative from issue 1, queried at head 1 & 249 \\
\textsc{Open} & derivative from issue 1, queried at head 2 & 249 \\
\textsc{Open} & derivative from issue 2, queried at head 2 & 231 \\
\textsc{Open} & all earlier-issue derivatives at final head 3 & 480 \\
\bottomrule
\end{tabular}
\end{table}

For an issued successor $I_g$, regional truth is $\tau_g$.  For an unissued
output from $r$, global truth is \textsc{Open} regardless of decoder success,
and the regional target remains $\tau_r$.  A regional output is scored only
where field evidence is declared observable; elsewhere the verifier returns
$\bot$.

\subsection{Metrics and Statistical Unit}

Global results report per-query status accuracy or recall, false acceptance of
an unissued output as an issue, and \textsc{Unknown} abstention.
Regional results report ambiguous-cell accuracy, all-field accuracy, mean IoU
over generation labels present in the target, generation-specific IoU, and
IoU/F1 of the newest-generation slice.  The global and regional metrics in the
joint table share the same successor and live head, directly pairing
issue-level and regional-history evidence.

The root is the independent sampling unit.  Heads, query conditions, cells,
and post-processing settings are repeated measurements within a root.  Tables state when
an entry is a pooled count, a median over images, or a presence-conditioned
IoU.  Where bootstrap intervals \citep{efron1979bootstrap} are used, roots are sampled with replacement
and every within-root observation is kept together; cells are never treated
as independent samples for uncertainty estimation.

\section{How Reissuance Closes the Transition}
\label{app:lifecycle-results}

The complete result reveals a semantic transition that static watermark
accuracy cannot express: an unchanged issue remains decodable while changing
from \textsc{Current} to \textsc{Superseded}, and a destroyed predecessor
field becomes an unresolved \textsc{Open} transition until reissuance creates
its successor.  The tables below establish both halves on the complete query
schedule.  Their labels are assigned from the event graph before
verification, so weak image evidence can cause an \textsc{Open} event to be
missed but can never remove it from the denominator.

\subsection{A Live Head Makes Static Issues Expire}

Table~\ref{tab:app-registered-legs} follows each unchanged issued bitmap
as the live head advances.  The same issue is \textsc{Current} at its own
head and \textsc{Superseded} at every later head; no image is regenerated for
the later queries.  This makes the experiment a direct test of head-relative
semantics rather than a new decoding trial on a modified predecessor.

\begin{table}[!htbp]
\caption{Issued images under their own and later live heads.  Entries are
counts over 480 roots.  ``Open'' and ``Unk.'' are the two observed error
destinations; there is no \textsc{Current}$\leftrightarrow$\textsc{Superseded}
confusion.}
\label{tab:app-registered-legs}
\centering
\small
\setlength{\tabcolsep}{3pt}
\begin{tabular}{llrrrr}
\toprule
Ground truth & Query condition & Correct & Open & Unk. & Accuracy (\%) $\uparrow$ \\
\midrule
\textsc{Current} & issue 0 at head 0 & 478 & 2 & 0 & 99.58 \\
                  & issue 1 at head 1 & 479 & 1 & 0 & 99.79 \\
                  & issue 2 at head 2 & 479 & 1 & 0 & 99.79 \\
                  & issue 3 at head 3 & 475 & 2 & 3 & 98.96 \\
\cmidrule(lr){2-6}
                  & all conditions under the issue's own head ($4\times480$) & 1911 & 6 & 3 & 99.53 \\
\midrule
\textsc{Superseded} & issue 0 at head 1 & 478 & 2 & 0 & 99.58 \\
                    & issue 0 at head 2 & 479 & 1 & 0 & 99.79 \\
                    & issue 1 at head 2 & 479 & 1 & 0 & 99.79 \\
                    & issue 0 at head 3 & 479 & 1 & 0 & 99.79 \\
                    & issue 1 at head 3 & 479 & 1 & 0 & 99.79 \\
                    & issue 2 at head 3 & 479 & 1 & 0 & 99.79 \\
\cmidrule(lr){2-6}
                    & all conditions under a later head ($6\times480$) & 2873 & 7 & 0 & 99.76 \\
\bottomrule
\end{tabular}
\end{table}

Across the ten issued-image query conditions, the combined status accuracy is
99.67\%.  More importantly for the supersession claim, the verifier never calls
an issued predecessor \textsc{Current} after the live head has advanced,
and never calls the live issue \textsc{Superseded}.  Its residual errors on
issued images all withhold a status decision: 13 return \textsc{Open} and
three return \textsc{Unknown}.  None reverses the current--superseded relation
defined by the live head.

\subsection{Reissuance Separates Unissued Edits from Issues}

The \textsc{Open} set includes both transitions that are later closed by a
successor and outputs that remain unclosed.  Table~\ref{tab:app-open-legs}
reports every condition separately.  The first three rows are the retained
editor outputs immediately before issues 1--3 are committed.  The final
unissued output and derivatives from earlier issues have no assumed successor; their truth
therefore cannot depend on a future reissue.

\begin{table}[!htbp]
\caption{Event-defined \textsc{Open} queries.  ``False issue'' combines false
\textsc{Current} and false \textsc{Superseded} acceptance.  Counts in each
row sum to the stated query set.}
\label{tab:app-open-legs}
\centering
\small
\setlength{\tabcolsep}{3pt}
\begin{tabular}{lrrrrr}
\toprule
Query condition & Query set & Open & False issue & Unk. & Open recall (\%) $\uparrow$ \\
\midrule
pre-issue-1 output at head 0 & 480 & 461 & 9  & 10 & 96.04 \\
pre-issue-2 output at head 1 & 480 & 452 & 11 & 17 & 94.17 \\
pre-issue-3 output at head 2 & 480 & 446 & 10 & 24 & 92.92 \\
\cmidrule(lr){1-6}
pre-reissue editor outputs & $3\times480$ & 1359 & 30 & 51 & 94.38 \\
final unissued output at head 3 & $1\times480$ & 440 & 12 & 28 & 91.67 \\
\midrule
derivative from issue 1 at head 1 & 249 & 234 & 6  & 9  & 93.98 \\
derivative from issue 1 at head 2 & 249 & 238 & 9  & 2  & 95.58 \\
derivative from issue 2 at head 2 & 231 & 219 & 1  & 11 & 94.81 \\
all earlier-issue derivatives at head 3 & 480 & 454 & 12 & 14 & 94.58 \\
\cmidrule(lr){1-6}
earlier-issue derivatives & 1209 & 1145 & 28 & 36 & 94.71 \\
\midrule
all unissued outputs & 3129 & 2944 & 70 & 115 & \textbf{94.09} \\
\bottomrule
\end{tabular}
\end{table}

Only 2.24\% of unissued outputs are falsely accepted as an issue;
3.68\% abstain as \textsc{Unknown}.  The earlier-issue results also show why an
\textsc{Open} status cannot be defined as ``the parent of an observed
successor'': the same semantics remain recoverable on final and earlier-issue
outputs that are never committed.

The edit generators are not concentrated in one favorable chain type.  The
\textsc{Open} recall on homogeneous FLUX.1 Kontext, InstructPix2Pix, Qwen
Image Edit, and SDXL-Turbo chains is 93.42\% ($n=76$), 97.47\% ($n=79$),
96.15\% ($n=78$), and 97.40\% ($n=77$), respectively.  The much larger set of
mixed-editor chains obtains 93.86\% ($n=2819$).

\subsection{Issue Evidence Separates Unissued Outputs from Issues}
\label{app:lifecycle-evidence-ablation}

Issue evidence decodes on issued images and fails on edited ones: it
matches an issue on 99.67\% of issued queries, whereas on 74.94\% of
unissued outputs it does not decode
(Table~\ref{tab:app-open-evidence-routes}), and it keeps decoding under
ordinary delivery (Appendix~\ref{app:rendition-robustness}).  How the
router reads that failure sets the operating point
(Table~\ref{tab:lifecycle-evidence}).  Abstaining on broad field loss
forfeits most of it, at 14.29\% recall; abstaining only on complete loss
reads it as \textsc{Open}, 88.65\%, at unchanged issued-image accuracy, so
the recall gained comes from unissued outputs alone; counting any disrupted
cluster as an unresolved change, the operating point of
Table~\ref{tab:app-verifier-parameters}, catches most outputs whose issue
evidence survives an edit, reaching 94.09\% and halving false issue
acceptance for 0.21 points of issued-image accuracy.

\begin{table}[!htbp]
\caption{One routing chain at three operating points (\%); the last is the
point used throughout.  The oracle is a diagnostic upper bound on the final
unissued and earlier-issue subset ($2\times480$), not a fourth operating
point.}
\label{tab:lifecycle-evidence}
\centering
\small
\begin{tabular}{@{}lrrr@{}}
\toprule
Routing rule & Issued acc. $\uparrow$ & \textsc{Open} recall $\uparrow$ &
Issue FAR $\downarrow$ \\
\midrule
Abstain on broad field loss                              & \textbf{99.88} & 14.29 & 4.92 \\
Abstain only on complete field loss                      & \textbf{99.88} & 88.65 & 4.92 \\
$+$ read any disrupted cluster as \textsc{Open}          & 99.67 & \textbf{94.09} & \textbf{2.24} \\
\midrule
Oracle upper bound (subset)       & -- & 98.23 & -- \\
\bottomrule
\end{tabular}
\end{table}

Undecodable issue evidence or a disrupted cluster opens the transition; a
decoded issue match is read against the live head as \textsc{Current} or
\textsc{Superseded}; reissuance then closes the transition with a new
issue.  The 98.23\% oracle is a diagnostic upper bound on the
stated $2\times480$ subset.

\subsection{Abstention Separates Missing Evidence from Authorization Claims}

Table~\ref{tab:app-open-evidence-routes} partitions all 3,129 unissued
queries by their final decision and the dominant evidence route.  It is a
diagnostic decomposition, not a redefinition of truth.  In particular, the
115 abstentions remain false negatives for \textsc{Open} recall.

\begin{table}[!htbp]
\caption{Dominant evidence routes for every event-defined unissued output.}
\label{tab:app-open-evidence-routes}
\centering
\small
\begin{tabular}{llrr}
\toprule
Final prediction & Dominant evidence route & Count & Fraction (\%) \\
\midrule
\textsc{Open} & issue evidence does not decode & 2345 & 74.94 \\
               & localized carrier-disruption support & 543 & 17.35 \\
               & additional inconsistency evidence & 56 & 1.79 \\
\midrule
\textsc{Unknown} & insufficient or broadly degraded evidence & 115 & 3.68 \\
\midrule
false issue & false \textsc{Current} match & 49 & 1.57 \\
                       & false \textsc{Superseded} match & 21 & 0.67 \\
\bottomrule
\end{tabular}
\end{table}

The dominant successful route is not literal pixelwise survival of the
predecessor payload.  It is the absence of a committed issue match and,
where needed, localized carrier destruction.  Fresh issuance reverses that
condition by writing a complete
successor payload and atomically advancing the live head.  The evidence
analysis therefore supports the central lifecycle mechanism: editing opens a
transition, authorization closes it, and supersession changes the meaning of
the otherwise unchanged predecessor.

\section{Delivery Robustness: Status Decisions, Regional History, and the Brightening Boundary}
\label{app:rendition-robustness}

The verifier carries three responsibilities through delivery.  It
must keep reading an issued image as \textsc{Current} under its own head and
as \textsc{Superseded} under a later head, it must keep reading an unissued
output as \textsc{Open}, and it must keep the regional history of an issued
image readable.  Either of the first two alone is incomplete: a verifier
that calls every damaged image \textsc{Open} obtains high \textsc{Open}
recall while failing every issue it was meant to certify.  This appendix
therefore reports the complete matrices behind
Figure~\ref{fig:delivery-robustness} and
Section~\ref{sec:results-robustness}: four endpoints and the generation map
at depths 1--3, the clean images and 31 transformations, one frozen verifier
and one frozen reader.  It then documents the one boundary that matters for
ordinary delivery, brightening, and the image property that explains why
earlier issues cross it first.

\subsection{Post-Processing Operators and Paired Endpoints}
\label{app:robustness-setup}

All inputs are stored and decoded as 8-bit RGB.  Brightness multiplies RGB
values followed by clipping; contrast scales values about the whole-image
mean; gamma applies $255(x/255)^\gamma$; resizing is bicubic downsampling to
the stated side length followed by bicubic restoration to $512\times512$;
blur is Gaussian with the stated $\sigma$; JPEG uses the stated quality
factor.  The 31 transformations are JPEG at quality 95, 90, 85, 75, 50, 40,
and 30; double JPEG 75, which applies the quality-75 encode/decode twice;
resizing to 448, 384, 320, 256, and 192 pixels; blur with $\sigma$ of 0.25,
0.5, 0.75, and 1.0; brightness $\times0.85$, $0.90$, $0.95$, $1.05$,
$1.10$, $1.15$, and $1.30$; contrast $\times0.80$, $1.25$, and $1.50$; gamma
$0.80$ and $1.25$; and two composite platform proxies,
\begin{align*}
\text{Platform A}:&\quad \operatorname{JPEG}_{60}
  (\operatorname{Resize}_{384}(x)),\\
\text{Platform B}:&\quad \operatorname{JPEG}_{70}
  (\operatorname{Resize}_{256}(\operatorname{JPEG}_{80}(x))).
\end{align*}
These are deterministic image pipelines, not measurements from named
commercial services.  Figure~\ref{fig:delivery-robustness}(a--d) shows the
JPEG, resize, blur, and brightness families up to JPEG 75, 256 pixels,
$\sigma=1.0$, and $\times1.15$, with 95\% Clopper--Pearson intervals \citep{clopper1934confidence} for
the status lines and a bootstrap interval over roots for the map line; the
tables below add the remaining levels and families.

Every condition is applied to four endpoints, all queried at head 3 and each
with 480 roots: the current issue $I_3$ (truth \textsc{Current}), the
initial issue $I_0$ (truth \textsc{Superseded}), the never-issued output
after $I_3$, written $U_3$, and the branch from $I_1$ or $I_2$, written
$U_p$ (both truth \textsc{Open}).  These are the head-3 queries of
Tables~\ref{tab:app-registered-legs} and~\ref{tab:app-open-legs} under
transformation, $4\times32\times480=61{,}440$ verifier queries in all.
Every query uses the same verifier and the frozen rules of
Table~\ref{tab:app-verifier-parameters}; there is no condition-specific
calibration.  The decisions were collected in two runs without new editor
calls: one decoded $I_3$, $U_3$, and $U_p$ under the clean images and the
21 coarser transformations; a later run added $I_0$ under those conditions
and all four endpoints under the ten finer levels (JPEG 95, 90, and 85;
brightness $\times0.90$, $0.95$, $1.05$, and $1.10$; blur 0.25 and 0.75;
resize 448).  The later run reproduced the earlier run's 1,920 clean
decisions exactly, so the two runs are reported as one matrix.

\paragraph{Regional history under the same transformations.}
The generation map is read on the current issue at each depth, $I_d$ at
head $d$, as in Table~\ref{tab:generation-map-results}, on the clean
images and all 31 transformations: $480\times3\times32=46{,}080$ readings
with the spatial reader (Appendix~\ref{app:spatial-reader-details}).
Post-processing does not change which generation wrote which cell, so the
ground truth of every reading is the clean field of its root.  The clean
readings reproduce Table~\ref{tab:generation-map-results}, and the head-3
status decisions of the same run reproduce
Table~\ref{tab:app-robust-matrix} on all 15,360 current-issue queries.
The same run also decided $I_1$ at head 1 and $I_2$ at head 2 under every
condition, the own-head queries Table~\ref{tab:app-robust-matrix} lacks;
Table~\ref{tab:app-map-delivery} reports them beside the map.  Reading the
46,080 maps took about 20 GPU-hours and no new editor call.

\subsection{The Complete Matrix}
\label{app:robustness-matrix}

Table~\ref{tab:app-robust-matrix} reports all four endpoints.  For the
issued endpoints the truth is one issued status, so every \textsc{Open} or
\textsc{Unknown} entry is an error, not a conservative success; the two
columns that would record a \textsc{Current}--\textsc{Superseded} reversal
are zero in every row and are omitted.  For the unissued endpoints the
pooled \textsc{Open} column is the \textsc{Open} line of
Figure~\ref{fig:delivery-robustness}.

\begin{table}[!htbp]
\caption{The complete delivery matrix at head 3 (\% of 480 roots per
endpoint).  $I_3$ is \textsc{Current} in truth and $I_0$
\textsc{Superseded}; the omitted columns, $I_3$ predicted
\textsc{Superseded} and $I_0$ predicted \textsc{Current}, are zero in every
row.  $U_3$ and $U_p$ are \textsc{Open} in truth; the pooled columns
combine them (960 queries), with false issue a prediction of
\textsc{Current} or \textsc{Superseded}.  Unk.\ is \textsc{Unknown},
including decisions withheld by the current-issue safeguard.}
\label{tab:app-robust-matrix}
\centering
\scriptsize
\setlength{\tabcolsep}{3pt}
\resizebox{\linewidth}{!}{%
\begin{tabular}{lrrr|rrr|rr|rrr}
\toprule
& \multicolumn{3}{c|}{Current issue $I_3$} &
  \multicolumn{3}{c|}{Initial issue $I_0$} &
  \multicolumn{2}{c|}{Open $\uparrow$} &
  \multicolumn{3}{c}{Pooled $U_3$ and $U_p$} \\
\cmidrule(lr){2-4}\cmidrule(lr){5-7}\cmidrule(lr){8-9}\cmidrule(lr){10-12}
Condition & Current $\uparrow$ & Open $\downarrow$ & Unk.\ $\downarrow$ &
Superseded $\uparrow$ & Open $\downarrow$ & Unk.\ $\downarrow$ &
$U_3$ & $U_p$ & Open $\uparrow$ & False issue $\downarrow$ & Unk.\ $\downarrow$ \\
\midrule
Clean & 98.96 & 0.42 & 0.63 & 99.79 & 0.21 & 0.00 & 91.67 & 94.58 & 93.13 & 2.50 & 4.38 \\
\addlinespace[2pt]
JPEG 95 & 99.58 & 0.00 & 0.42 & 99.79 & 0.21 & 0.00 & 92.50 & 96.46 & 94.48 & 2.40 & 3.13 \\
JPEG 90 & 98.96 & 0.42 & 0.63 & 99.38 & 0.63 & 0.00 & 93.13 & 96.25 & 94.69 & 2.08 & 3.23 \\
JPEG 85 & 97.92 & 1.25 & 0.83 & 99.17 & 0.83 & 0.00 & 94.38 & 96.46 & 95.42 & 1.88 & 2.71 \\
JPEG 75 & 95.63 & 3.96 & 0.42 & 96.46 & 3.54 & 0.00 & 95.00 & 97.92 & 96.46 & 1.77 & 1.77 \\
JPEG 50 & 35.00 & 64.79 & 0.21 & 38.96 & 61.04 & 0.00 & 98.54 & 98.96 & 98.75 & 0.52 & 0.73 \\
JPEG 40 & 1.67 & 94.79 & 3.54 & 4.17 & 94.79 & 1.04 & 98.96 & 99.79 & 99.38 & 0.00 & 0.63 \\
JPEG 30 & 0.00 & 91.67 & 8.33 & 0.00 & 90.42 & 9.58 & 99.58 & 99.58 & 99.58 & 0.00 & 0.42 \\
Double JPEG 75 & 95.21 & 3.96 & 0.83 & 94.17 & 5.83 & 0.00 & 95.42 & 98.13 & 96.77 & 1.77 & 1.46 \\
\addlinespace[2pt]
Resize 448 & 98.75 & 0.63 & 0.63 & 99.58 & 0.42 & 0.00 & 92.29 & 95.83 & 94.06 & 2.60 & 3.33 \\
Resize 384 & 99.38 & 0.21 & 0.42 & 99.79 & 0.21 & 0.00 & 91.88 & 96.46 & 94.17 & 1.98 & 3.85 \\
Resize 320 & 98.96 & 0.63 & 0.42 & 99.58 & 0.42 & 0.00 & 93.96 & 97.92 & 95.94 & 1.56 & 2.50 \\
Resize 256 & 99.38 & 0.21 & 0.42 & 99.79 & 0.21 & 0.00 & 92.71 & 96.46 & 94.58 & 1.35 & 4.06 \\
Resize 192 & 1.04 & 94.17 & 4.79 & 0.42 & 92.29 & 7.29 & 96.88 & 98.33 & 97.60 & 0.00 & 2.40 \\
\addlinespace[2pt]
Blur $0.25$ & 99.17 & 0.21 & 0.63 & 99.79 & 0.21 & 0.00 & 92.08 & 94.79 & 93.44 & 2.50 & 4.06 \\
Blur $0.5$ & 98.96 & 0.42 & 0.63 & 99.79 & 0.21 & 0.00 & 91.67 & 96.04 & 93.85 & 2.29 & 3.85 \\
Blur $0.75$ & 99.17 & 0.42 & 0.42 & 99.79 & 0.21 & 0.00 & 92.71 & 95.83 & 94.27 & 2.29 & 3.44 \\
Blur $1.0$ & 98.13 & 0.83 & 1.04 & 100.00 & 0.00 & 0.00 & 95.63 & 97.50 & 96.56 & 1.25 & 2.19 \\
\addlinespace[2pt]
Brightness $\times0.85$ & 99.38 & 0.42 & 0.21 & 99.79 & 0.21 & 0.00 & 92.92 & 95.21 & 94.06 & 2.29 & 3.65 \\
Brightness $\times0.90$ & 98.96 & 0.42 & 0.63 & 99.79 & 0.21 & 0.00 & 91.25 & 95.21 & 93.23 & 2.29 & 4.48 \\
Brightness $\times0.95$ & 98.96 & 0.42 & 0.63 & 99.79 & 0.21 & 0.00 & 91.88 & 95.83 & 93.85 & 2.40 & 3.75 \\
Brightness $\times1.05$ & 92.92 & 6.04 & 1.04 & 73.54 & 26.46 & 0.00 & 90.83 & 95.00 & 92.92 & 2.40 & 4.69 \\
Brightness $\times1.10$ & 84.38 & 13.33 & 2.29 & 61.46 & 38.54 & 0.00 & 91.25 & 95.21 & 93.23 & 2.19 & 4.58 \\
Brightness $\times1.15$ & 72.50 & 21.88 & 5.63 & 50.42 & 49.58 & 0.00 & 91.88 & 95.83 & 93.85 & 1.77 & 4.38 \\
Brightness $\times1.30$ & 39.38 & 50.42 & 10.21 & 24.58 & 75.21 & 0.21 & 94.17 & 97.29 & 95.73 & 0.94 & 3.33 \\
\addlinespace[2pt]
Contrast $\times0.80$ & 99.17 & 0.21 & 0.63 & 99.79 & 0.21 & 0.00 & 91.46 & 95.42 & 93.44 & 2.40 & 4.17 \\
Contrast $\times1.25$ & 64.58 & 29.79 & 5.63 & 31.04 & 68.96 & 0.00 & 91.04 & 95.63 & 93.33 & 1.46 & 5.21 \\
Contrast $\times1.50$ & 31.25 & 61.04 & 7.71 & 10.42 & 89.38 & 0.21 & 94.58 & 95.83 & 95.21 & 0.94 & 3.85 \\
\addlinespace[2pt]
Gamma $0.80$ & 75.00 & 0.21 & 24.79 & 99.79 & 0.21 & 0.00 & 91.04 & 94.79 & 92.92 & 2.50 & 4.58 \\
Gamma $1.25$ & 66.67 & 3.33 & 30.00 & 93.96 & 6.04 & 0.00 & 92.08 & 95.83 & 93.96 & 1.77 & 4.27 \\
\addlinespace[2pt]
Platform A & 58.13 & 41.88 & 0.00 & 44.38 & 55.63 & 0.00 & 98.75 & 97.92 & 98.33 & 0.52 & 1.15 \\
Platform B & 2.50 & 95.63 & 1.88 & 0.63 & 96.46 & 2.92 & 98.96 & 99.58 & 99.27 & 0.00 & 0.73 \\
\bottomrule
\end{tabular}}
\end{table}

\subsection{Envelope and Boundary}
\label{app:robustness-envelope}

\paragraph{Eighteen conditions keep all three statuses above 93\%.}
The clean images; JPEG at quality 95, 90, 85, and 75; double JPEG 75; resize
448, 384, 320, and 256; blur at every tested $\sigma$; darkening to
$\times0.95$, $0.90$, and $0.85$; and contrast $\times0.80$ all keep
\textsc{Current} retention at 95.21--99.58\%,
\textsc{Superseded} retention at 94.17--100.00\%, and
pooled \textsc{Open} recall at 93.13--96.77\%.
Moderate compression, resizing, blur, repeated JPEG encoding, darkening, and
contrast reduction therefore do not force a choice between preserving an
issue and recognizing an unissued output; the three-way distinction
survives a non-pristine delivery path.  The depth-3 map accuracy is at least its clean value of 93.13\% in
every one of these conditions (Table~\ref{tab:app-map-delivery}).

\paragraph{Fourteen conditions cross the boundary, and issued images fail
toward \textsc{Open} or \textsc{Unknown}.}
The boundary is sharpest under JPEG 30, resize 192, and Platform B, where
\textsc{Current} retention falls to 0.00--2.50\% and
\textsc{Superseded} retention to 0.00--0.63\%,
with 91.67--95.63\% of current issues read as \textsc{Open}.
JPEG crosses between quality 75 and 50 (95.63\% to 35.00\%
\textsc{Current}), resizing between 256 and 192 pixels
(99.38\% to 1.04\%), and blur does not cross within
$\sigma\leq1.0$.  Brightening and contrast gain cross more gradually:
\textsc{Current} retention under brightness $\times1.05$, $1.10$, $1.15$,
and $1.30$ is 92.92\%, 84.38\%, 72.50\%, and 39.38\%, and
\textsc{Superseded} retention of the initial issue 73.54\%, 61.46\%, 50.42\%, and 24.58\%.
Gamma is the one family whose errors go to \textsc{Unknown} rather than
\textsc{Open}: at gamma $0.80$ and $1.25$ the current issue is withheld in
24.79\% and 30.00\% of cases while the initial issue stays
\textsc{Superseded} in 99.79\% and 93.96\% (Section~\ref{app:robustness-mechanism}).
In no condition does either issued endpoint acquire the other issued
status: over the 30,720 issued queries of Table~\ref{tab:app-robust-matrix}
there is no \textsc{Current}--\textsc{Superseded} reversal.  Broad
degradation removes the positive issue match, so the verifier withholds
the status it can no longer support; it never assigns the wrong one.  Where the status crosses, the map
does not: JPEG 50 leaves 35.00\% of current issues \textsc{Current} but
93.13\% of their cells correctly labeled.

\paragraph{The map outlasts the status.}
Across the 31 transformations, the depth-3 map accuracy of the current
issue falls below its clean value only where cells become unreadable: to
92.50\% under Platform B, 91.88\% under brightness $\times1.30$ and
contrast $\times1.50$, 91.25\% under JPEG 40, 87.50\% under resize 192, and
85.00\% under JPEG 30 (means 92.12\%, 90.58\%, 90.47\%, 90.08\%, 85.53\%,
and 82.89\% against 92.89\% clean).  The largest losses at depths 1 and 2
fall under the same two conditions, JPEG 30 (93.13\% and 88.75\%) and
resize 192 (92.50\% and 90.00\%), and the depth order of
Table~\ref{tab:generation-map-results}, mIoU falling from depth 1 to
depth 3, holds in every condition.  The status decision fails earlier
because it reacts to evidence the map does not need.  A localized cluster
of unreadable cells is positive \textsc{Open} evidence
(Appendix~\ref{app:frozen-instantiation}): under JPEG 50 such a cluster
appears in 65.0\% of current issues while the median share of readable
field cells is 96.9\%, so 35.00\% remain \textsc{Current} and the map is
unchanged; under gamma at least nine current issues in ten keep every cell
readable, the current-issue safeguard withholds 24.79\% and 30.00\%
(Section~\ref{app:robustness-mechanism}), and the map is again unchanged.
Where cells are lost, the map loses fewer labels than the field loses
readable cells: at JPEG 30 the median readable share is 61.3\% and 85.00\%
of cells are still labeled correctly, and the spatial reader's gain over
independent decoding on ambiguous cells stays between 33.4 and 63.1 points
in every condition and depth.

\paragraph{A surviving status carries the clean map.}
For images whose status survives a transformation, the map is the clean
map: on the roots read as \textsc{Current}, the paired change in map
accuracy relative to the clean image has a median of 0.00 points at depths
1--3 in every condition with at least 30 such roots, and a mean within
$\pm0.25$ points.  A \textsc{Current} decision under delivery is therefore
accompanied by regional history as accurate as on the clean image.

\begin{table}[!htbp]
\caption{Regional history under delivery (medians over 480 roots, \%).  The
current issue at each depth, $I_d$ at head $d$, is transformed and read with
the spatial reader: map
accuracy and mIoU over the 160 field cells at depths 1--3 and the
newest-slice F1 against $A_3$ at depth 3.  The last three columns give,
for the same queries, \textsc{Current} retention of $I_1$, $I_2$, and $I_3$
at their own heads; the $I_3$ column repeats the delivery matrix.  The
clean row is the untransformed result.}
\label{tab:app-map-delivery}
\centering
\scriptsize
\setlength{\tabcolsep}{3pt}
\begin{tabular}{lrrr|rrr|r|rrr}
\toprule
& \multicolumn{3}{c|}{Map acc.\ $\uparrow$} &
  \multicolumn{3}{c|}{Map mIoU $\uparrow$} &
  \multicolumn{1}{c|}{Newest F1 $\uparrow$} &
  \multicolumn{3}{c}{\textsc{Current} at own head $\uparrow$} \\
\cmidrule(lr){2-4}\cmidrule(lr){5-7}\cmidrule(lr){8-8}\cmidrule(lr){9-11}
Condition & $d{=}1$ & $d{=}2$ & $d{=}3$ & $d{=}1$ & $d{=}2$ & $d{=}3$ & $d{=}3$ & $I_1$ & $I_2$ & $I_3$ \\
\midrule
Clean & 97.19 & 95.00 & 93.13 & 93.87 & 87.27 & 78.96 & 96.30 & 99.79 & 99.79 & 98.96 \\
\addlinespace[2pt]
JPEG 95 & 97.50 & 95.00 & 93.13 & 93.84 & 87.11 & 79.18 & 96.30 & 99.58 & 99.79 & 99.58 \\
JPEG 90 & 96.88 & 95.00 & 93.13 & 93.87 & 87.11 & 79.12 & 96.37 & 99.58 & 99.38 & 98.96 \\
JPEG 85 & 96.88 & 95.00 & 93.13 & 93.85 & 87.00 & 79.03 & 96.34 & 98.75 & 97.92 & 97.92 \\
JPEG 75 & 96.88 & 95.00 & 93.13 & 93.81 & 86.96 & 79.05 & 96.32 & 96.25 & 96.67 & 95.63 \\
JPEG 50 & 96.88 & 95.00 & 93.13 & 93.62 & 86.58 & 78.32 & 96.06 & 41.88 & 36.25 & 35.00 \\
JPEG 40 & 96.25 & 93.13 & 91.25 & 92.33 & 82.48 & 72.76 & 94.74 & 4.58 & 3.75 & 1.67 \\
JPEG 30 & 93.13 & 88.75 & 85.00 & 86.40 & 73.78 & 61.45 & 90.25 & 0.00 & 0.21 & 0.00 \\
Double JPEG 75 & 96.88 & 95.00 & 93.13 & 93.83 & 86.97 & 79.08 & 96.32 & 95.21 & 95.21 & 95.21 \\
\addlinespace[2pt]
Resize 448 & 96.88 & 95.00 & 93.13 & 93.87 & 87.34 & 79.13 & 96.30 & 99.79 & 99.58 & 98.75 \\
Resize 384 & 96.88 & 95.00 & 93.13 & 93.82 & 87.31 & 79.13 & 96.31 & 99.38 & 99.58 & 99.38 \\
Resize 320 & 96.88 & 95.00 & 93.13 & 93.87 & 87.13 & 79.23 & 96.32 & 99.58 & 99.58 & 98.96 \\
Resize 256 & 97.19 & 95.00 & 93.75 & 93.85 & 87.13 & 79.59 & 96.37 & 99.58 & 99.58 & 99.38 \\
Resize 192 & 92.50 & 90.00 & 87.50 & 85.25 & 75.05 & 67.31 & 91.46 & 0.21 & 0.21 & 1.04 \\
\addlinespace[2pt]
Blur $0.25$ & 97.19 & 95.00 & 93.13 & 93.87 & 87.27 & 79.03 & 96.30 & 99.79 & 99.79 & 99.17 \\
Blur $0.5$ & 96.88 & 95.00 & 93.13 & 93.81 & 87.37 & 79.06 & 96.30 & 100.00 & 99.79 & 98.96 \\
Blur $0.75$ & 97.19 & 95.00 & 93.13 & 93.87 & 87.19 & 79.11 & 96.30 & 99.79 & 99.58 & 99.17 \\
Blur $1.0$ & 96.88 & 95.00 & 93.13 & 93.79 & 87.00 & 79.29 & 96.32 & 99.17 & 99.38 & 98.13 \\
\addlinespace[2pt]
Brightness $\times0.85$ & 97.19 & 95.00 & 93.13 & 93.87 & 87.27 & 79.09 & 96.30 & 99.79 & 99.58 & 99.38 \\
Brightness $\times0.90$ & 97.50 & 95.00 & 93.13 & 93.87 & 87.27 & 78.98 & 96.33 & 99.58 & 99.58 & 98.96 \\
Brightness $\times0.95$ & 97.50 & 95.00 & 93.13 & 93.87 & 87.27 & 79.03 & 96.30 & 99.79 & 99.79 & 98.96 \\
Brightness $\times1.05$ & 96.88 & 95.00 & 93.13 & 93.69 & 86.71 & 78.79 & 96.24 & 81.88 & 89.79 & 92.92 \\
Brightness $\times1.10$ & 96.88 & 95.00 & 93.13 & 93.56 & 86.40 & 78.28 & 96.08 & 69.38 & 78.13 & 84.38 \\
Brightness $\times1.15$ & 96.88 & 95.00 & 93.13 & 93.20 & 86.08 & 78.13 & 96.00 & 54.17 & 66.25 & 72.50 \\
Brightness $\times1.30$ & 96.25 & 93.75 & 91.88 & 92.14 & 83.84 & 75.86 & 95.51 & 25.42 & 32.50 & 39.38 \\
\addlinespace[2pt]
Contrast $\times0.80$ & 97.19 & 95.00 & 93.13 & 93.87 & 87.35 & 78.98 & 96.31 & 99.58 & 99.79 & 99.17 \\
Contrast $\times1.25$ & 96.88 & 94.38 & 93.13 & 93.15 & 85.59 & 77.72 & 95.96 & 42.71 & 56.46 & 64.58 \\
Contrast $\times1.50$ & 95.63 & 93.75 & 91.88 & 91.42 & 83.30 & 74.87 & 95.34 & 16.46 & 24.58 & 31.25 \\
\addlinespace[2pt]
Gamma $0.80$ & 97.50 & 95.00 & 93.13 & 93.83 & 87.18 & 78.96 & 96.30 & 74.17 & 71.25 & 75.00 \\
Gamma $1.25$ & 97.50 & 95.00 & 93.13 & 93.87 & 87.18 & 79.09 & 96.30 & 63.13 & 66.25 & 66.67 \\
\addlinespace[2pt]
Platform A & 96.88 & 95.00 & 93.13 & 93.72 & 86.67 & 79.23 & 96.30 & 61.88 & 56.25 & 58.13 \\
Platform B & 96.25 & 94.38 & 92.50 & 92.56 & 84.63 & 77.85 & 95.65 & 1.25 & 2.50 & 2.50 \\
\bottomrule
\end{tabular}
\end{table}

\paragraph{\rev{Unissued outputs are rarely accepted as issues.}}
Pooled \textsc{Open} recall never falls below 92.92\% (the single-endpoint
minimum is 90.83\%, $U_3$ under brightness $\times1.05$) and rises under
heavy damage to 99.58\% at JPEG 30, because damage adds disruption
evidence.  Pooled false acceptance of an
unissued output as an issue stays at or below 2.60\% (resize 448)\rev{, against
2.50\% on the clean images,} and
pooled \textsc{Unknown} at or below 5.21\% (contrast $\times1.25$).

\paragraph{The head changes the label, not the evidence.}
The initial issue was also queried at head 0, where its truth is
\textsc{Current}, under the clean images and the 21 coarser transformations
in the earlier run.  Pairing those decisions with the head-3 decisions of the
same 10,560 images, 99.19\% agree up to relabeling (\textsc{Current} or
withheld at head 0 against \textsc{Superseded} at head 3, \textsc{Open}
against \textsc{Open}, \textsc{Unknown} against \textsc{Unknown}).  The
transformed-image results therefore hold for the head-relative decision as
such, not for one head.

\subsection{Why Earlier Issues Fail First under Brightening}
\label{app:robustness-mechanism}

Under every brightening and contrast-gain level, retention falls with the
age of the issue: at brightness $\times1.15$ the initial issue keeps
\textsc{Superseded} in 50.42\% of cases, and $I_1$, $I_2$, and $I_3$ keep
\textsc{Current} at their own heads in 54.17\%, 66.25\%, and 72.50\%
(Table~\ref{tab:app-map-delivery}); at contrast $\times1.25$ the four
figures are 31.04\%, 42.71\%, 56.46\%, and 64.58\%.  This runs against the
natural expectation that the least edited image is the most robust
carrier.  A post-hoc measurement of the untransformed images,
Table~\ref{tab:app-saturation}, gives the reason.  Brightening multiplies
8-bit values and clips at 255, so it destroys carrier evidence in exactly
the pixels that were already bright, and every pass through a diffusion
editor compresses the highlight tail: at nearly the same mean intensity,
the 99th-percentile intensity is 242 for $I_0$ and 232, 228, and 222 for
$I_1$, $I_2$, and $I_3$, with the unissued outputs $U_3$ and $U_p$ in the
same range.  At $\times1.05$ the initial issue therefore clips 2.34 times
as many pixels as the current issue, and $I_1$ clips more than $I_3$ in
63.8\% of the roots.  What separates the images is the number of editor
passes they have been through, not their generation label as such.
Darkening produces no clipping, and none of the four endpoints loses status
under it.

\begin{table}[!htbp]
\caption{Highlight statistics of the untransformed issues and unissued
outputs over 480 roots (8-bit intensities; clipped-pixel shares in \%).
Brightening saturates every pixel above $255/c$.}
\label{tab:app-saturation}
\centering
\small
\setlength{\tabcolsep}{5pt}
\begin{tabular}{lrrrrrr}
\toprule
& & & \multicolumn{4}{c}{Clipped pixels (\%) at brightness $\times c$} \\
\cmidrule(lr){4-7}
Image & Mean & 99th pct.\ & $1.05$ & $1.10$ & $1.15$ & $1.30$ \\
\midrule
$I_0$ (initial issue) & 109.06 & 241.53 & 3.83 & 5.79 & 7.98 & 14.62 \\
$I_1$ (issue 1) & 107.56 & 231.72 & 2.65 & 4.53 & 6.64 & 13.22 \\
$I_2$ (issue 2) & 108.99 & 227.60 & 1.99 & 3.92 & 5.95 & 12.58 \\
$I_3$ (current issue) & 106.94 & 221.84 & 1.64 & 3.12 & 4.86 & 11.26 \\
$U_3$ (output after $I_3$) & 108.50 & 221.92 & 1.68 & 3.36 & 5.24 & 12.03 \\
$U_p$ (branch) & 110.38 & 227.65 & 1.89 & 3.55 & 5.77 & 13.14 \\
\bottomrule
\end{tabular}
\end{table}

Gamma reverses the order because it clips nothing: it is a monotone map
within the 8-bit range, and what fails under it is the current-issue
safeguard of Table~\ref{tab:app-verifier-parameters}, which withholds
\textsc{Current} when the current-issue evidence is insufficient.  A
\textsc{Superseded} decision passes through no such safeguard, so under
gamma the initial issue at head 3 keeps 99.79\% and 93.96\% while
the current issue keeps 75.00\% and 66.67\%, and the map is unchanged
(93.13\% at depth 3).  Three limits apply.
$I_0$--$I_3$ are different images, so the brightening gap is a property
of the carrier images---a source set with fewer highlights would narrow
it---rather than of the protocol.  Clipping accounts for most but not all
of the ordering: a model of status loss on the clipped share alone leaves
a smaller but nonzero depth effect.  And only the input highlight
distribution was measured, not where the writer places its energy; a
preference for high-contrast regions would add a second cause that the
present measurement cannot separate.

\subsection{Beyond the Boundary, Destruction Is Not an Accusation}
\label{app:robustness-abstention}

Outside the envelope, a severely degraded issue and a complete-output edit
produce the same observable pattern: issue evidence no longer decodes and
the field is disrupted.  The frozen router does not always separate the
two, which is why heavy damage sends issued images to \textsc{Open} rather
than to \textsc{Unknown}.  The map quantifies what survives such damage:
at JPEG 30 the field still labels 85.00\% of cells correctly and the
newest slice keeps 90.25\% F1.  Whether that surviving history can also
separate benign degradation from an unissued edit is untested here: an
unissued output keeps its parent's regional history by construction, so a
well-recovered map is compatible with either case.  Reporting only payload survival would miss this
collision, and reporting only \textsc{Open} recall would count it as a
success; the paired matrix makes it explicit.  \rev{Within the envelope,
destruction does not read as an accusation: all three statuses stay above
93\%, at most 2.60\% of unissued outputs are accepted as issues, and the
failure direction for issued images is toward withholding.}

\begin{revblock}
\subsection{Retention and False Acceptance across Operating Points}
\label{app:robustness-sweep}

The operating point of Table~\ref{tab:app-verifier-parameters} keeps all
three statuses above 93\% in the eighteen conditions of
Appendix~\ref{app:robustness-envelope}.  JPEG 50 and Platform~B lie outside
them: there it keeps 35.0 and 2.5\% of current issues \textsc{Current}
(Table~\ref{tab:app-robust-matrix}), whereas TrustMark-Q, carrying our
reissuance protocol, keeps 98.5 and 99.4\%
(Table~\ref{tab:app-cross-method-delivery}).  Retention is one coordinate of
an operating point; the other is what the system accepts as an issue under
the same conditions, and Table~\ref{tab:rev-sweep} reports both.  Under
JPEG 50 the generation map is unchanged, and the status is lost because a
disrupted cluster, which appears in 65.0\% of the current issues, is read
as \textsc{Open} evidence (Appendix~\ref{app:robustness-envelope}).  The
table therefore varies $C_{\mathrm{cc}}$, the number of cells a disrupted
cluster must exceed to supply \textsc{Open} evidence, over ten values with
the other rules fixed: after the evaluation, all 61,440 queries of
Appendix~\ref{app:robustness-setup} were re-decided at each value from their
stored evidence, without new images.  TrustMark-Q's images were decoded
under the three conditions of the table: for 398 roots the images of
Appendix~\ref{app:cross-method-comparison}, and for the other 82
regenerations with the same instructions, so its retention under JPEG 50
and Platform~B differs from Table~\ref{tab:app-cross-method-delivery} by one
root each.

\begin{table}[!htbp]
\revtable{R7}
\caption{\textsc{Current} retention $\uparrow$ / false issue acceptance $\downarrow$ across values of $C_{\mathrm{cc}}$ (\%).  Other rules as in Table~\ref{tab:app-verifier-parameters}; unissued outputs: $U_3$ and $U_p$ at head 3 ($2\times480$ per condition).}
\label{tab:rev-sweep}
\centering
\small
\revfit{%
\begin{tabular}{@{}lccc@{}}
\toprule
$C_{\mathrm{cc}}$ & JPEG 50 & Platform~B & Clean \\
\midrule
0 & 1.9 / 0.0 & 0.0 / 0.0 & 98.1 / 1.0 \\
1 (Table~\ref{tab:app-verifier-parameters}) & 35.0 / 0.5 & 2.5 / 0.0 & 99.0 / 2.5 \\
2 & 54.0 / 0.8 & 11.7 / 0.1 & 99.4 / 3.1 \\
3 & 68.5 / 1.0 & 29.8 / 0.4 & 99.4 / 3.8 \\
4 & 75.2 / 1.6 & 46.9 / 0.5 & 99.4 / 4.6 \\
6 & 84.8 / 2.2 & 68.3 / 0.8 & 99.4 / 6.1 \\
8 & 89.8 / 2.6 & 82.1 / 1.3 & 99.4 / 7.0 \\
12 & 93.8 / 3.4 & 90.4 / 2.1 & 99.4 / 8.4 \\
16 & 94.6 / 4.6 & 93.8 / 3.0 & 99.4 / 9.7 \\
24 & 95.2 / 5.8 & 95.6 / 4.3 & 99.4 / 10.6 \\
\midrule
$+$ TrustMark-Q & 98.3 / \textbf{24.0} & 99.2 / \textbf{26.1} & 99.8 / \textbf{29.6} \\
\bottomrule
\end{tabular}}
\end{table}

TrustMark-Q keeps 98.3 and 99.2\% of its current issues under JPEG 50 and
Platform~B because its global message survives them, and for the same reason
it accepts 24.0 and 26.1\% of unissued outputs as issues there, where GSF at
the operating point of Table~\ref{tab:app-verifier-parameters} accepts
0.5 and 0.0\%.  For GSF, $C_{\mathrm{cc}}=12$ is the smallest swept
value with retention above 93\% under JPEG 50; on the clean images it raises
false issue acceptance from 2.5 to 8.4\% and lowers \textsc{Open} recall
from 93.1 to 82.7\%.  No swept value keeps all three statuses above 93\%
both on the clean images and under either condition, and at every swept
value none of the 30,720 issued-image queries is given the other issued
status.  Outside the envelope a degraded issue and an unissued edit produce
the same evidence (Appendix~\ref{app:robustness-abstention}), here a
disrupted cluster, so the threshold moves errors from issued images to
unissued outputs without separating the two.
\end{revblock}

\section{Why Spatial Context Recovers Constant-Rate History}
\label{app:spatial-reader-details}

Representing an expanding authorized history at a fixed local rate creates a
predictable ambiguity: several legal generations can share the same local
fingerprint.  The spatial reader resolves that ambiguity by recalibrating
candidate evidence and propagating reliable temporal anchors across coherent
revision regions.  We establish the full chain from the collision law, through
independent-cell degradation, to spatial recovery.

\subsection{Collision Probability and the Fixed-Slot Choice}

For a $b$-bit keyed fingerprint and $d$ other issued candidates, the ideal
PRF model gives
\begin{equation}
p^{(b)}_{\mathrm{amb}}(d)
=1-\left(1-2^{-b}\right)^d.
\label{eq:app-rate-collision}
\end{equation}
This per-region probability alone does not select a width: under a fixed
carrier budget, wider fingerprints also produce fewer and coarser logical
regions.  Appendix~\ref{app:bitwidth-ladder} proves the resulting fixed-slot
anchor law and evaluates the alternative rate--granularity allocations at
$b\in\{1,2,3,4,8\}$.  This appendix instead asks the downstream question:
given the selected two-bit field, how does spatial inference recover the
history compressed by its measured collisions?

\subsection{Observed Ambiguity Follows the Collision Law}

\begin{figure}[!htbp]
\centering
\includegraphics[width=0.55\linewidth]{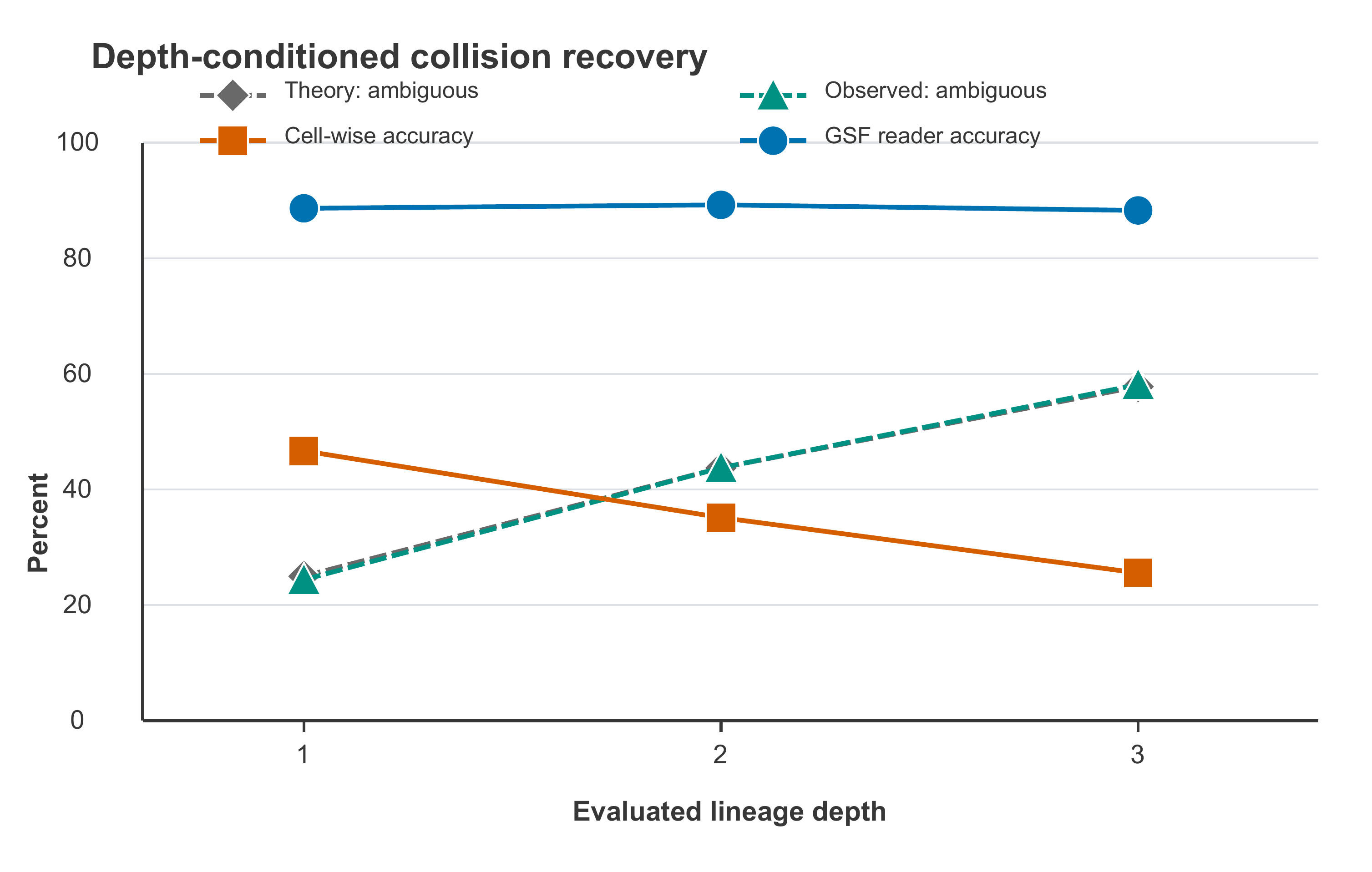}
\caption{Depth-conditioned collision formation and recovery.  Predicted and
observed ambiguity increase together, independent cell-wise accuracy falls,
and spatial recovery remains stable.  Each point summarizes 480 issued
successors.}
\label{fig:app-reader-depth}
\end{figure}

Table~\ref{tab:app-collision-depth} compares the prediction with the frozen
two-bit field on the three issued successor depths.  Every row contains one
own-head issue query over 480 independent roots
(Figure~\ref{fig:app-reader-depth}).  The empirical median differs
from theory by at most 0.63 percentage points.  The decline of cell-wise
inference is therefore a direct representational effect: the hard local
observation genuinely admits multiple legal generations.

\begin{table}[!htbp]
\caption{Collision formation and recovery by lineage depth.  ``Cells'' is the
median number of ambiguous cells among the 160 generation-field cells.
Accuracies are medians over images and are restricted to ambiguous cells.}
\label{tab:app-collision-depth}
\centering
\small
\begin{tabular}{crrrrrr}
\toprule
& \multicolumn{3}{c}{Ambiguity} &
\multicolumn{3}{c}{Ambiguous-cell accuracy $\uparrow$} \\
\cmidrule(lr){2-4}\cmidrule(lr){5-7}
Depth & Theory (\%) & Observed (\%) & Cells & Local & Spatial & Gain $\uparrow$ \\
\midrule
1 & 25.00 & 24.38 & 39 & 46.67 & 88.64 & +41.97 \\
2 & 43.75 & 43.75 & 70 & 35.14 & 89.23 & +54.09 \\
3 & 57.81 & 58.13 & 93 & 25.56 & 88.27 & +62.71 \\
\bottomrule
\end{tabular}
\end{table}

As unique anchors become scarcer, independent-cell accuracy approaches chance
among the colliding candidates.  Spatial-reader accuracy remains near 89\%
through all three measured depths, so the absolute gain grows rather than
shrinks.  The candidate interface can score any legal fingerprint supplied to
it; the concrete batched implementation pads candidates into eight planes and
masks unused planes.  The experiments use at most four legal generations.
Because all candidate-plane weights are shared, the learned parameterization
contains no generation-index-specific parameter.  The evaluation uses
four legal generations; Appendix~\ref{app:bitwidth-ladder} characterizes how
the coding trade-off evolves at greater candidate depth.

\subsection{Reader Instantiation}

The frozen image decoder produces two bit logits and one confidence logit per
cell.  For legal candidate $a$, the exact summed local likelihood referenced in
the main text is
\begin{equation}
u_{i,a}=\sum_{b=1}^{2}
\log\sigma\!\left((2c_{L,a,i,b}-1)\ell_{i,b}\right).
\label{eq:local-likelihood}
\end{equation}
The implementation uses five channels per legal candidate: mean two-bit log
likelihood, the two individual bit log likelihoods, decoder confidence, and a
logit-margin statistic.  The batched tensor has shape
$B\times8\times5\times16\times16$, with a Boolean mask selecting legal
candidates.

The deployed reader has 4,633 trainable parameters.  A shared $1\times1$
stem maps five channels to 24 channels and applies GroupNorm and GELU.  Three
shared depthwise-separable residual blocks use $3\times3$ kernels with
dilations 1, 2, and 4.  Candidate-set context is formed only by masked mean
and log-sum-exp pooling; a shared pointwise head returns one score per candidate
and cell.  There is no candidate embedding or candidate-order parameter.

These shared candidate operations and symmetric pooling give
\begin{equation}
\mathbf p_\psi(\pi Z)=\pi\mathbf p_\psi(Z),
\label{eq:reader-equivariance}
\end{equation}

Candidate permutation was tested by permuting the input candidate planes and
undoing that permutation at the output.  The maximum numerical difference was
zero in the unit tests, including the pointwise and soft-spatial
variants.  Thus the implementation realizes
Eq.~\eqref{eq:reader-equivariance}, rather than learning that an early tensor
index means an old generation.

\paragraph{Training data and selection.}
The deployed checkpoint was trained with a frozen writer and decoder on 160
distinct COCO val2017 sources: 64 for training, 32 for development, and 64 for
sealed evaluation.  The 2,048-example training cache contains 32 independently
generated field/channel instances per training source.  Generation fields use
connected Voronoi, brush/polygon, rectangle/ellipse, holed, and disconnected
same-label regions rather than independent cell labels.  The fixed channel
battery contains JPEG 80/95, resize factors 0.75/1.25, Gaussian blur, and light
noise.

Training uses AdamW \citep{loshchilov2019adamw}, learning rate $3\times10^{-3}$, betas $(0.9,0.95)$,
weight decay $5\times10^{-3}$, batch size 128, gradient clipping at one, and
random candidate permutation.  The loss is balanced cross-entropy plus Dice
loss \citep{milletari2016vnet} and a $0.5$-weighted correct-key/wrong-key energy margin.  Development
metrics are evaluated every 100 steps; training stops after three non-improving
checks.  The frozen development criterion selects the step-200 checkpoint;
the sealed split is not used for selection.

The reader-training sources are disjoint from the 480 evaluation roots.

\subsection{Different Spatial Solvers Share the Same Underlying Gain}

Table~\ref{tab:app-reader-family} compares three ways of resolving the same
candidate-likelihood field.  The amortized soft reader is the deployed
4,633-parameter model above.  A hard-partition reader learns affinities,
constructs a region tree, and performs exact coded Bayesian decoding within
each partition.  A recurrent soft reader iteratively transports candidate
evidence and additionally contains content and rejection heads.  These are
not three unrelated systems; they are alternative inference operators over the
same keyed generation field.

\begin{table}[!htbp]
\caption{Alternative spatial inference operators on the same 480-root issued
successors.  Values are median ambiguous-cell accuracies.  The first row is
the independent-cell likelihood rule.}
\label{tab:app-reader-family}
\centering
\small
\setlength{\tabcolsep}{3pt}
\begin{tabular}{lrrr}
\toprule
Inference operator & Depth-1 acc. $\uparrow$ & Depth-2 acc. $\uparrow$ &
Depth-3 acc. $\uparrow$ \\
\midrule
Independent local likelihood & 46.67 & 35.14 & 25.56 \\
Hard partition + exact region decoding & 84.21 & 82.02 & 80.23 \\
Amortized soft spatial reader & 88.64 & \textbf{89.23} & 88.27 \\
Recurrent soft spatial reader & \textbf{88.89} & 88.66 & \textbf{89.01} \\
\bottomrule
\end{tabular}
\end{table}

All spatial operators substantially outperform independent cells.  The
4,633-parameter amortized reader matches the larger recurrent solver, while
hard partitioning is consistently lower.  This solver-family comparison
triangulates the role of spatial evidence; the controlled study below isolates
the two mechanisms that carry the gain.

\subsection{Recalibration and Soft Coupling Carry the Gain}
\label{app:reader-mechanism-ablation}

For the controlled mechanism study, all soft variants share a disjoint
80-source cache (56 train, 8 calibration, 16 development), a warm-started
encoder, a region-balanced generation NLL, and 400 training steps.  The 80
sources are disjoint from both the 480 evaluation roots and a separate 120-root
reserve.  Results are aggregated over the three own-head issued successor
conditions (Table~\ref{tab:app-reader-mechanisms}), so they must not be
numerically merged with the depth-conditioned
medians in Table~\ref{tab:app-collision-depth}.

\begin{table}[!htbp]
\caption{Controlled mechanism ladder on issued successors.  ``No spatial
kernel'' center-masks the learned depthwise kernels; shared GroupNorm and
permutation-symmetric candidate context remain, so it isolates learned kernel
coupling rather than every possible cross-cell statistic.}
\label{tab:app-reader-mechanisms}
\centering
\small
\begin{tabular}{lrr}
\toprule
Mechanism & Ambiguous-cell accuracy (\%) $\uparrow$ & Increment $\uparrow$ \\
\midrule
Independent local likelihood & 36.97 & -- \\
Learned evidence recalibration, no spatial kernel & 67.48 & +30.51 \\
$+$ soft multiscale spatial kernels & 87.83 & +20.35 \\
\midrule
Recurrent soft coupling (three seeds) & 87.70--88.11 & $\leq+0.28$ \\
Hard partition + exact decoding & 80.78 & $-7.05$ \\
\bottomrule
\end{tabular}
\end{table}

\begin{figure}[!htbp]
\centering
\includegraphics[width=0.6\linewidth]{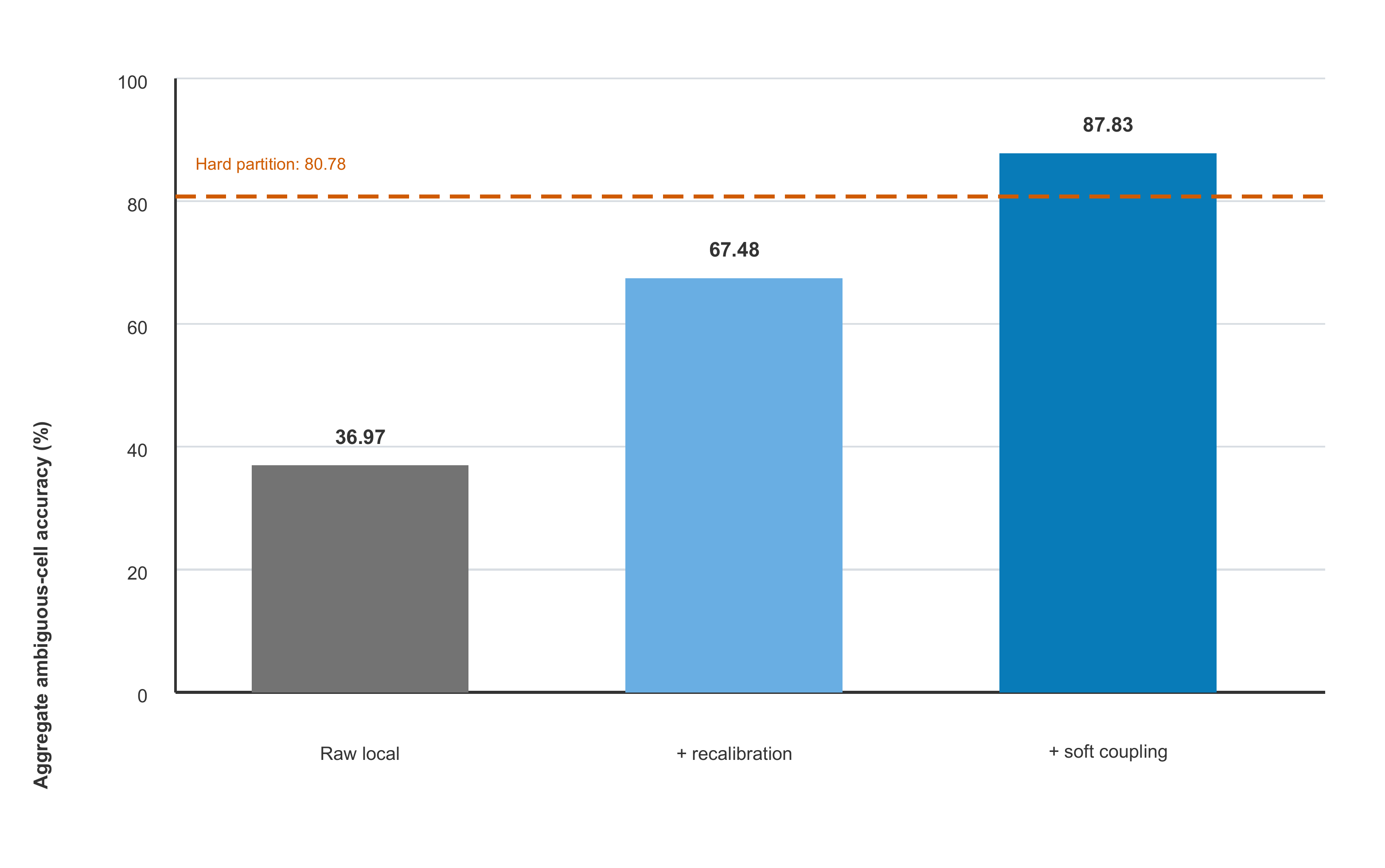}
\caption{Reader-mechanism attribution on the 480-root transfer set.  Shared
pointwise recalibration and soft spatial coupling provide the two principal
gains; the dashed line marks the hard-partition result.}
\label{fig:app-reader-mechanisms}
\end{figure}

The largest single improvement comes from learning how confidence, margin,
and bitwise evidence should recalibrate the raw candidate likelihood
(Figure~\ref{fig:app-reader-mechanisms}).  Spatial
coupling supplies the second indispensable improvement by allowing reliable
temporal anchors to support coherent neighboring cells.  Recurrence changes
the result by at most 0.28 points, showing that the compact feed-forward reader
already captures the useful evidence flow.  The independently trained solver
family in Table~\ref{tab:app-reader-family} shows the same ordering, supporting
soft evidence propagation over early hard partitioning.

\subsection{Editing Changes the Reader's Task}

Spatial inference is applied to issued fields whose carriers remain
observable.  Among unissued outputs whose generation carrier remains
readable, the controlled reader recovers ambiguous cells with 92.11\%
accuracy.  Where editing destroys that field, the joint verifier returns
$\bot$ regionally and uses lineage and disruption evidence to identify the
global \textsc{Open}/\textsc{Unknown} transition instead of inventing a local
history.

As an end-to-end consistency check, committed targets and decodable issued
fingerprints agree on 99.988\% of 133,373 cells.  All images are retained, and
the depth-conditioned conclusions are unchanged.  The result reinforces the
division of labor at the heart of GSF: the issuer's committed field defines
history, the image supplies evidence for it, and spatial inference reconstructs
that history while editing-induced destruction opens the next lifecycle
transition.

\section{One Field Answers When and Where}
\label{app:generation-map-details}

The generation field is more than a device for improving ambiguous-cell
accuracy.  It changes what can be asked of a watermarked issue.  Every label
slice answers \emph{when} the corresponding region was last authorized, while
the newest slice answers \emph{where} the current revision entered the
lineage.  Both answers come from the same recovered field and require neither
a second localization watermark nor a separate localization head.

\subsection{Repeated Reissuance Actually Produces Multi-Time Images}

The need for a generation map would be weak if each final image were dominated
by a single generation.  The opposite occurs.  Over the full $16\times16$
logical revision grid, 463 of 480 final issues contain all four authorized
labels, 16 contain three, and one contains two.  Table~\ref{tab:app-label-mass}
shows how much of the final field remains attributable to each time.  The
latest generation is largest on average, but every earlier generation remains
present in almost all final fields.

\begin{table}[!htbp]
\caption{Number of logical cells carrying each last-authorized generation in
the final ground-truth field.  Quartiles and extrema are over 480 roots on the
256-cell logical grid; ``Present'' counts fields with at least one such cell.}
\label{tab:app-label-mass}
\centering
\small
\begin{tabular}{crrrrrr}
\toprule
Label & Present & Minimum & Q1 & Median & Q3 & Maximum \\
\midrule
$g_0$ & 480 & 1  & 18 & 33 & 52 & 168 \\
$g_1$ & 466 & 0  & 9  & 19 & 33 & 102 \\
$g_2$ & 476 & 0  & 25 & 44 & 71 & 153 \\
$g_3$ & 480 & 64 & 126 & 152 & 170 & 220 \\
\bottomrule
\end{tabular}
\end{table}

This coexistence follows directly from inheritance.  A reissue assigns the
new label only on the issuer-committed revision scope and freshly re-embeds
the inherited labels elsewhere.  The committed scopes are substantial but do
not overwrite the whole field: their median sizes on the 256-cell logical grid
are 147, 147, and 152 cells at depths 1--3, with respective interquartile
ranges 120.5--168, 119--168, and 126--170.  Repeated revision therefore creates
a spatial mosaic of authorized times rather than a sequence of whole-image
version numbers.

The logical field is defined on all 256 cells.  The concrete carrier jointly
allocates 96 cells to lineage/issue evidence and 160 cells to generation
fingerprints.  Recovery metrics evaluate all 160 history-bearing cells.
Generation-specific IoU is reported whenever the corresponding label is
represented in that evaluated field.

\begin{revblock}
\paragraph{Overlap among committed scopes.}
With median sizes of 147--152 of the 256 cells, successive committed scopes
overlap.  Over the 160 field cells of the 480 roots, the median IoU
(interquartile range) is 45\% (30--61\%) between $A_1$ and $A_2$,
44\% (30--59\%) between $A_2$ and $A_3$, and 43\%
(30--55\%) between $A_1$ and $A_3$; for scopes of different roots,
which are independent, it is 38\% (27--47\%).
Appendix~\ref{app:generation-map-spatial} relates this overlap to the
accuracy of the map.
\end{revblock}

\subsection{Spatial Inference Recovers the Temporal Partition}
\label{app:generation-map-spatial}

\begin{figure}[!htbp]
\centering
\includegraphics[width=\linewidth]{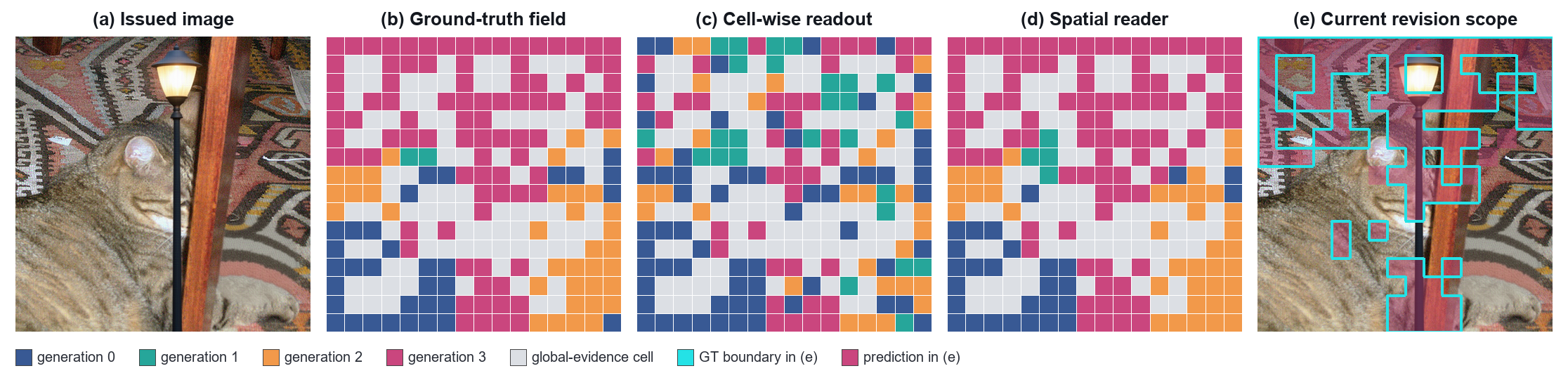}
\caption{Spatial inference recovers a representative four-generation field.
Independent cell-wise readout confuses historical labels, whereas the spatial
reader restores the field used for both temporal attribution and localization.
In (e), magenta is
$\widehat A_3=\{i:\widehat\tau_3(i)=3\}$ and cyan traces the committed
ground-truth $A_3$.  Gray cells carry global lineage and issue evidence and are
excluded from regional metrics.}
\label{fig:generation-map}
\end{figure}

\begin{figure}[!htbp]
\centering
\includegraphics[width=\linewidth]{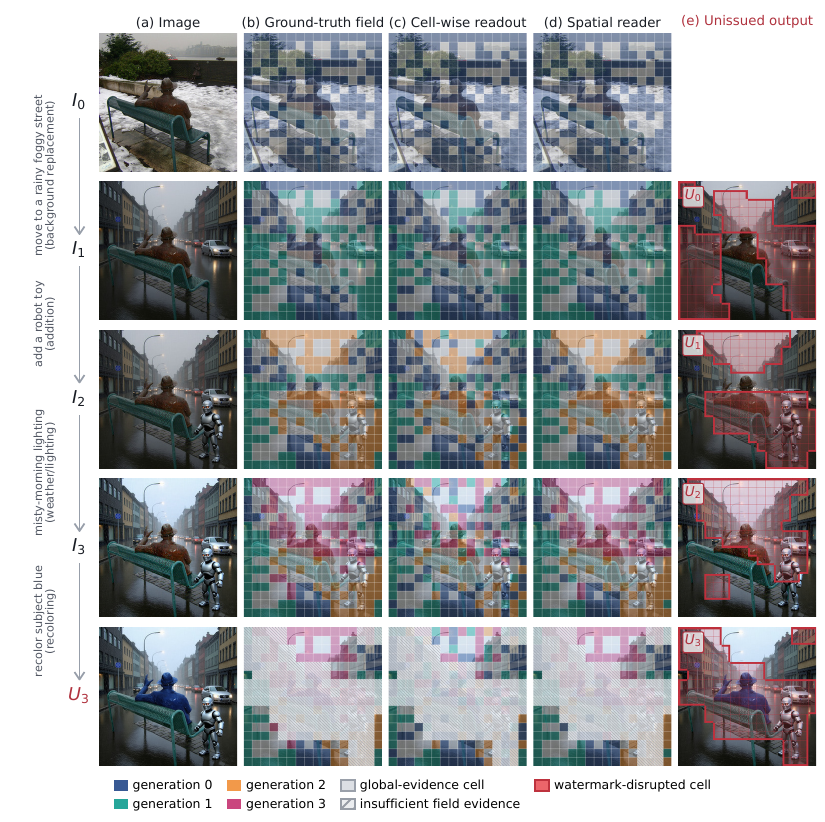}
\caption{A second lineage followed issue by issue.  Rows are the issues $I_0$--$I_3$ and the final unissued output $U_3$; the
label between two rows is the edit that produced the lower one.  Columns:
(a) the image; (b) the ground-truth field, for $U_3$ the field of its parent
$I_3$; (c) the cell-wise readout; (d) the spatial reader; (e) the unissued
output that the reissue in that row approved, with its watermark-disrupted
cells outlined, and in the last row $U_3$ itself.  Disrupted cells of $U_3$
carry insufficient field evidence and are hatched in (b)--(d).
Reissues by FLUX.1 Kontext throughout; $U_3$ by Qwen Image Edit.}
\label{fig:app-map-example-a}
\end{figure}

\begin{figure}[!htbp]
\centering
\includegraphics[width=\linewidth]{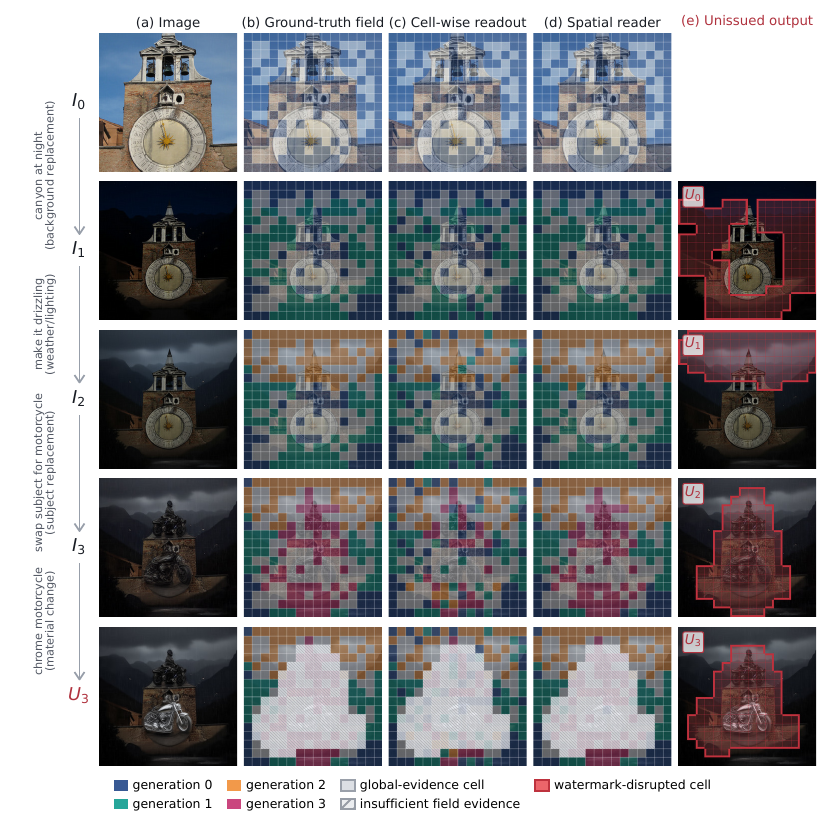}
\caption{A third lineage followed issue by issue.  Rows are the issues $I_0$--$I_3$ and the final unissued output $U_3$; the
label between two rows is the edit that produced the lower one.  Columns:
(a) the image; (b) the ground-truth field, for $U_3$ the field of its parent
$I_3$; (c) the cell-wise readout; (d) the spatial reader; (e) the unissued
output that the reissue in that row approved, with its watermark-disrupted
cells outlined, and in the last row $U_3$ itself.  Disrupted cells of $U_3$
carry insufficient field evidence and are hatched in (b)--(d).
Reissues by FLUX.1 Kontext, Qwen Image Edit, and InstructPix2Pix in turn;
$U_3$ by Qwen Image Edit.}
\label{fig:app-map-example-b}
\end{figure}

Table~\ref{tab:app-map-local-vs-spatial} moves from ambiguous cells to the
scientific object of interest: the complete regional-history map.  Independent
local likelihood deteriorates sharply as more historical labels coexist.
Spatial inference maintains 93.13\% full-field accuracy and 78.96\% mIoU even
at depth three, where a median 58.13\% of field cells are locally ambiguous.
Figure~\ref{fig:generation-map} shows the transition from independent to
spatial readout on a representative four-generation successor;
Figures~\ref{fig:app-map-example-a} and~\ref{fig:app-map-example-b} follow
two further lineages issue by issue, with the unissued output that each
reissue approved and the final unissued output.

\begin{table}[!htbp]
\caption{From collision recovery to generation-map recovery.  Map metrics use
all 160 generation cells.  Scope metrics use the newest-generation slice.
Entries are medians over 480 issued successors at each depth.}
\label{tab:app-map-local-vs-spatial}
\centering
\scriptsize
\resizebox{\linewidth}{!}{%
\begin{tabular}{c|rrrr|rrrr}
\toprule
& \multicolumn{4}{c|}{Independent local likelihood} &
  \multicolumn{4}{c}{Spatial generation-field reader} \\
\cmidrule(lr){2-5}\cmidrule(lr){6-9}
Depth & Map acc. $\uparrow$ & Map mIoU $\uparrow$ & Newest IoU $\uparrow$ &
Newest F1 $\uparrow$ & Map acc. $\uparrow$ & Map mIoU $\uparrow$ &
Newest IoU $\uparrow$ & Newest F1 $\uparrow$ \\
\midrule
1 & 86.88 & 75.82 & 75.00 & 85.71 & 97.19 & 93.87 & 94.92 & 97.39 \\
2 & 71.88 & 52.61 & 56.09 & 71.87 & 95.00 & 87.27 & 94.09 & 96.95 \\
3 & 57.50 & 36.76 & 42.19 & 59.34 & 93.13 & 78.96 & 92.86 & 96.30 \\
\bottomrule
\end{tabular}}
\end{table}

The aggregate decline is asymmetric rather than uniform across the field.
From depth one to depth three, generation-field mIoU falls from 93.87\% to
78.96\%, whereas newest-slice IoU moves only from 94.92\% to 92.86\%---an
end-to-end change of 2.06 percentage points.  Class-balanced overlap exposes
the harder part of the problem: older labels occupy smaller surviving regions,
so boundary errors weigh more heavily as those regions fragment and shrink.
Table~\ref{tab:app-generation-iou} localizes this loss.  At depth three, the
current $g_3$ slice retains 92.86\% IoU, while the historical $g_0$, $g_1$,
and $g_2$ slices reach 81.48\%, 72.57\%, and 81.32\%, respectively.  Within
the tested depths, the loss is therefore concentrated in the completeness of
fine-grained historical reconstruction rather than a comparable
destabilization of the current committed scope.

\begin{table}[!htbp]
\caption{Presence-conditioned IoU of every authorized-generation slice under
spatial inference.  A dash denotes a generation that does not yet exist at
that depth.  Entries are median percentages.}
\label{tab:app-generation-iou}
\centering
\small
\begin{tabular}{crrrr}
\toprule
& \multicolumn{4}{c}{Presence-conditioned IoU (\%) $\uparrow$} \\
\cmidrule(lr){2-5}
Depth & $g_0$ & $g_1$ & $g_2$ & $g_3$ \\
\midrule
1 & 93.99 & 94.92 & -- & -- \\
2 & 89.32 & 83.33 & 94.09 & -- \\
3 & 81.48 & 72.57 & 81.32 & 92.86 \\
\bottomrule
\end{tabular}
\end{table}

\begin{revblock}
\paragraph{Sensitivity to the observed overlap.}
Grouped into quartiles of the mean pairwise IoU of the three committed
scopes (13--80\%; 120 roots each), the root means of map
accuracy are 93.2, 92.9, 92.7, and 92.8\% and those of mIoU 80.8, 79.3,
78.2, and 76.8\%.  The decrease lies in $g_1$ and $g_2$, whose mean IoU
falls from 72.7 to 60.8\% and from 81.8 to 73.3\%; $g_0$ and $g_3$ do not
decrease.  With more overlap, later reissues assign their labels to more of
$A_1$ and $A_2$: the cells still carrying $g_1$ or $g_2$ fall from 61.5 to
32.6 per root, and the cells of $g_0$--$g_2$ lie more often in small
fragments, which are read less accurately (60.0, 70.9, 88.3, and 95.2\% in
fragments of 1, 2--4, 5--16, and at least 17 cells).  The accuracy of these
cells falls from 91.3 to 89.1\% across the quartiles; with the distribution
of fragment sizes held fixed, it varies within 0.8 points without a trend.

\paragraph{Controlled overlap.}
To vary the overlap alone, we kept the approved content $X_3$ of every root
and replaced its three committed scopes with bands of four of the grid's 16
lines (rows or columns, the direction set per root by a hash), each band
covering 0, 25, 50, 75, or 100\% of the previous one.  The frozen writer
embedded the resulting field, and the frozen verifier and reader read it;
with the original scopes, the same pipeline reproduces the $I_3$ of the
evaluation bit for bit on all 480 roots.

\begin{table}[!htbp]
\revtable{R9}
\caption{Controlled overlap of three committed bands written on the approved content of $I_3$ (480 roots, \%).}
\label{tab:rev-controlled-overlap}
\centering
\small
\revfit{%
\begin{tabular}{@{}p{5.4cm}ccccc@{}}
\toprule
Share of the previous band covered & 0\% & 25\% & 50\% & 75\% & 100\% \\
\midrule
Lines still carrying $g_1$ and $g_2$ (of 4) & 4 & 3 & 2 & 1 & 0 \\
Map acc. (mean) & 98.2 & 97.5 & 96.9 & \textbf{95.9} & 97.8 \\
mIoU (mean) & 96.5 & 94.4 & 91.2 & \textbf{80.9} & 95.8 \\
Newest slice read correctly & 98.4 & 98.4 & 98.5 & 98.6 & 99.1 \\
$g_1$ and $g_2$ read correctly & 98.4 & 96.7 & 93.4 & \textbf{80.8} & -- \\
\bottomrule
\end{tabular}}
\end{table}

As the surviving part of the first two bands narrows from four lines to one,
map accuracy falls from 98.2 to 95.9\%, mIoU from 96.5 to 80.9\%, and the
accuracy on the cells of $g_1$ and $g_2$ from 98.4 to 80.8\%: a surviving
region one line wide lies entirely on a boundary.  At 100\%, where none of
them survives, map accuracy and mIoU return to 97.8 and 95.8\%.  The newest
slice is read correctly in 98.4--99.1\% throughout, and writing the same
fields on the roots' unmarked sources, which carry no earlier field, changes
map accuracy by at most 0.1 points.  Map accuracy and mIoU exceed those of
Table~\ref{tab:generation-map-results} because the bands are contiguous.

In the evaluation, the cells of $g_3$ that were committed once, twice, and
three times are read correctly in 94.7, 95.0, and 95.3\%: repeated
commitment does not lower the accuracy of a cell, since every reissue
rewrites the whole field (Appendix~\ref{app:logical-physical-rewrite}).
Overlap therefore determines which label each cell carries and how much of
an older scope survives, and older labels lose accuracy as their surviving
regions shrink and fragment, as above and at larger depth
(Appendix~\ref{app:generation-map-depth7}).
\end{revblock}

\subsection{The Newest Slice Is the Current Authorized Scope}

For a committed successor $g$, the update law gives
\begin{equation}
  A_g=\{i:\tau_g(i)=g\},\qquad
  \widehat A_g=\{i:\widehat\tau_g(i)=g\}.
  \label{eq:app-newest-slice}
\end{equation}
Consequently, reading all slices yields the regional authorization history,
whereas selecting only the newest slice yields the current committed revision
scope.  Table~\ref{tab:app-map-local-vs-spatial} shows that its IoU remains
94.92\%, 94.09\%, and 92.86\% at depths 1--3, while its F1 remains above
96\% throughout.  This stability is the spatial half of the joint task: the
map does not merely say that a revision happened; it preserves the present
issuer-committed scope even as older authorization fragments accumulate, and
attributes every observable region to the authorization event that most
recently changed it.

In the evaluation, the proposed scope is constructed from blockwise
RGB difference between the issued parent and the editor's complete
return, using Eq.~\eqref{eq:app-scope-proposal}; the issuer then commits that
proposal.  Thus the newest slice localizes the \emph{committed,
pixel-difference-derived revision support}: the blockwise region through which
the revision entered authorized history.  This authorization target is
distinct from an editor's latent pixelwise semantic mask.  In another
deployment, spatial watermark disruption or authenticated edit records may
propose the scope; issuer commitment gives any such proposal its
authority.

Ordinary edit localization asks which pixels differ in one snapshot.  The
generation map instead records which authorized event currently governs each
region.  Its newest slice provides the current revision support, while its
older slices retain the temporal attribution that a binary edit mask discards.

\subsection{Status and Regional History Are Recovered Together}

Table~\ref{tab:app-joint-successors} pairs status and regional history on
the same issued successors and the same live head, showing that one verifier
supplies both answers from the same image.

\begin{table}[!htbp]
\caption{Paired global and regional results on the same own-head successor
queries.  Every depth is one query condition over 480 roots.  Regional entries are
the medians defined above; global accuracy is the fraction predicted
\textsc{Current}.}
\label{tab:app-joint-successors}
\centering
\small
\setlength{\tabcolsep}{3pt}
\begin{tabular}{crrrr}
\toprule
Depth & Predicted \textsc{Current} (\%) $\uparrow$ & Ambiguous-cell acc. (\%) $\uparrow$ &
Map accuracy (\%) $\uparrow$ & Newest IoU (\%) $\uparrow$ \\
\midrule
1 & 99.79 & 88.64 & 97.19 & 94.92 \\
2 & 99.79 & 89.23 & 95.00 & 94.09 \\
3 & 98.96 & 88.27 & 93.13 & 92.86 \\
\bottomrule
\end{tabular}
\end{table}

Across these $3\times480$ successors, 1,433 are globally recognized as the
live issue.  On the same queries, the regional field retains high full-map and
newest-scope accuracy despite increasing local collisions.  The resulting
object is therefore not a version number plus an unrelated mask.  It is a
single spatiotemporal authorization field: one field answers when and where,
and continuity is maintained jointly at the issue level and the
regional-history level.

\begin{revblock}
\subsection{Regional History at Depths 4--7}
\label{app:generation-map-depth7}

The evaluation of Section~\ref{sec:evaluation} covers depths 1--3.  Every
root was extended from $I_3$ to $I_7$ by 1,920 further reissues with the
same four editors (FLUX.1 Kontext for 468, InstructPix2Pix for 493, Qwen
Image Edit for 474, and SDXL-Turbo for 485), the same operation classes, and
the blockwise pixel-difference proposal of
Appendix~\ref{app:scope-construction}; writer, decoder, and reader are
frozen.  Table~\ref{tab:rev-eight-versions} reads the current
issue at depths 3--7.

\begin{table}[!htbp]
\revtable{R6}
\caption{Regional history at depths 3--7, that is, four to eight issues (medians, \%).  Upper bound: the map of a three-bit code whose logical regions are all read without error (Appendix~\ref{app:bitwidth-beyond-depth}).}
\label{tab:rev-eight-versions}
\centering
\footnotesize
\revfit{%
\begin{tabular}{@{}p{4.2cm}ccccc@{}}
\toprule
Depth & 3 (Sec.~\ref{sec:results-generation-map}) & 4 & 5 & 6 & 7 \\
\midrule
Map acc. & 93.1 & 91.3 & 90.0 & 88.1 & 86.9 \\
mIoU & 79.0 & 72.5 & 67.1 & 62.8 & 59.0 \\
Ambiguous cells, observed / Eq.~\eqref{eq:collision-law} & 58.1 / 57.8 & 68.8 / 68.4 & 76.3 / 76.3 & 82.5 / 82.2 & 86.9 / 86.7 \\
Ambiguous-cell accuracy & \textbf{88.3} & \textbf{87.4} & \textbf{86.9} & \textbf{85.7} & \textbf{84.8} \\
Three-bit code, error-free upper bound: map acc. / mIoU & 90.6 / 76.0 & 90.3 / 73.8 & 90.3 / 72.1 & 89.9 / 70.4 & 89.9 / 69.4 \\
\bottomrule
\end{tabular}}
\end{table}

The share of ambiguous cells rises from 58.1\% at depth 3 to 86.9\% at
depth 7, and the spatial reader recovers them with 88.3 and 84.8\%
accuracy.  Slices at least four generations older than the head, which a
direct two-bit code aliases with later labels
(Appendix~\ref{app:same-rate-counterfactual}), keep a median IoU of 75.0,
61.5, 50.0, and 50.0\% at depths 4--7.  As at depths 1--3
(Section~\ref{sec:results-generation-map}), mIoU falls because older slices
become small and fragmented.  At depths 4--7, cells of older slices are read
correctly in 42.9\% when isolated and in 62.2\% in fragments of 2--4 cells
(together 15.5\% of these cells), against 85.3 and 92.4\% in fragments of
5--16 and at least 17 cells; among older slices of 5--16 cells, those split
into two or more fragments reach a lower median IoU than contiguous ones
(60.0 against 71.4\%).  Appendix~\ref{app:bitwidth-beyond-depth} compares the field with
Eq.~\eqref{eq:collision-law} and with wider codes at the same depths.
\end{revblock}

\section{Canonicalized Verification Preserves Multi-Resolution Delivery}
\label{app:resolution}

The writer is trained and the keyed $16\times16$ field is defined on a
canonical $512\times512$ canvas.  Images may nevertheless circulate at
different delivery resolutions: verification restores the canonical canvas
before reading the field.  Across intermediate sizes from 768 to 2048 pixels,
this path preserves every tested issue and can even make fixed-pixel
delivery operations milder relative to the carrier.

\subsection{Resolution Round Trips Preserve the Carrier}

We write at 512, resize to an intermediate square canvas, and restore 512
before verification.  The supporting set contains 60 issued images.  Twelve
round-trip settings cover intermediate sizes 768, 1024, 1536, and 2048 and
PIL bilinear, bicubic, and Lanczos kernels, together with area downsampling
(Table~\ref{tab:app-resolution-roundtrip}).
No network is retrained and the verifier is unchanged.

\begin{table}[!htbp]
\caption{The physical carrier survives ordinary resize round trips.  The
carrier score is the median absolute two-bit decoder margin aggregated over
the field; the ratio is paired to the clean score.  Every tested resampling
pair in a row obtains the stated result on all 60 issues.}
\label{tab:app-resolution-roundtrip}
\centering
\small
\begin{tabular}{lcrrc}
\toprule
Intermediate canvas & Kernel pairs & Carrier score $\uparrow$ &
Clean ratio $\uparrow$ & \textsc{Current} $\uparrow$ \\
\midrule
None (clean) & 1 & 17.999 & 1.000 & $60/60$ \\
768          & 2 & 18.013--18.042 & 1.002--1.003 & $60/60$ \\
1024         & 6 & 17.929--18.040 & 0.996--1.002 & $60/60$ \\
1536         & 2 & 18.016--18.030 & 1.001         & $60/60$ \\
2048         & 2 & 18.030--18.048 & 1.001--1.002 & $60/60$ \\
\bottomrule
\end{tabular}
\end{table}

The median number of disrupted cells is zero in every setting.  This result
rules out the intuitive but incorrect explanation that the carrier lives near
the pixel Nyquist limit.  Its semantic lattice is $16\times16$, so its
characteristic scale is approximately 32 pixels on the canonical image.
The tested linear resamplers preserve structure at that scale even when they
alter individual pixels.  This also explains why benign resampling and AI
regeneration must not be conflated: the former filters an existing block-scale
signal, whereas the latter can replace the image-wide carrier evidence while
producing a new complete output.

\subsection{Canonical Coordinates Preserve Keyed Cell Identity}

Canonicalization is part of the spatial interface rather than an image-quality
repair.  It restores the same $16\times16$ lattice on which every keyed cell
index was written, so resizing changes the delivery canvas without changing
the semantic identity of a region.  The zero disrupted-cell median in
Table~\ref{tab:app-resolution-roundtrip} confirms that ordinary resampling
preserves the underlying block-scale evidence.

\subsection{A Canonical Hub Can Improve Delivery Robustness}

We next compare a native-512 delivery path with a path that
distributes the same issued image at 1024 and canonicalizes it to 512 for
verification.  Both arms use the same 60 roots and the same six benign
operations.  Table~\ref{tab:app-resolution-deployment} shows that the added
round trip introduces no clean penalty and improves the weakest fixed-pixel
transformations in this set.

\begin{table}[!htbp]
\caption{Matched benign verification at native 512 and through a
1024-distribution/512-verification hub.  Entries are \textsc{Current} rates
over 60 issues per condition.}
\label{tab:app-resolution-deployment}
\centering
\small
\begin{tabular}{lrrr}
\toprule
Condition & Native 512 $\uparrow$ & Canonical hub $\uparrow$ &
Difference $\uparrow$ \\
\midrule
Clean                         & 1.0000 & 1.0000 & 0.0000 \\
JPEG 75                       & 1.0000 & 1.0000 & 0.0000 \\
Downsample/upsample           & 1.0000 & 1.0000 & 0.0000 \\
Crop 90\% and resize         & 0.8000 & 0.9333 & $+0.1333$ \\
Rotate $2^\circ$              & 1.0000 & 1.0000 & 0.0000 \\
Screenshot simulation        & 0.9667 & 1.0000 & $+0.0333$ \\
\midrule
Pooled                        & 0.9611 & 0.9889 & $+0.0278$ \\
\bottomrule
\end{tabular}
\end{table}

The crop improvement is not explained by a different recovered scale: the
median alignment scale is 1.1087 in both arms.  A consistent explanation is
relative operator scale.  Distribution at 1024 doubles the carrier block to
approximately 64 pixels, while JPEG blocks and interpolation kernels remain
fixed in pixel units; those operations are therefore milder relative to the
carrier.  The pure geometric rotation, for which this argument predicts
little change, is unchanged.

\subsection{Alignment Lives in the Same Canonical Coordinates}

The blind alignment search in Appendix~\ref{app:frozen-instantiation} handles
isotropic scale, rotation, and local translation before field decoding.  Both
alignment and resolution normalization therefore operate in the same
canonical coordinates.  Across all tested delivery resolutions, the physical
evidence survives resampling and canonicalization restores the semantic
identity of every keyed spatial cell.

\section{Complete Outputs Change the Localization Endpoint}
\label{app:complete-output-controls}

The complete-output workflow changes which image is retained rather than the
size of the inpainting mask or the strength of a post-processing attack.  A
paste-back endpoint preserves the watermarked container
pixel for pixel outside the edit mask; a retained-output endpoint keeps the
editor's complete RGB return, including its reconstruction of the nominally
unmodified background.  This appendix isolates that single endpoint choice
and shows why snapshot localization and status verification answer
different questions under it.

The controlled comparison first verifies each snapshot localizer in its native
paste-back protocol, then changes only the retained endpoint.  It thereby
isolates the physical premise removed by complete-output editing.

\subsection{Matched Endpoint Construction}

The matched control uses all 1,000 images and masks in valAGE
\citep{zhang2024editguard}.  The median
mask area is 0.0918.  For each carrier, source, mask, prompt, and seed, one
inpainting call produces an edited image $Y$.  We then form
\begin{align}
 X_{\mathrm{paste}} &=(1-M)\odot X_{\mathrm{wm}}+M\odot Y, \\
 X_{\mathrm{full}}  &=Y,
 \label{eq:app-endpoint-control}
\end{align}
where $X_{\mathrm{wm}}$ is that method's watermarked container and $M$ is the
provided mask.  Thus the two endpoints within each method have identical
generated content and differ only in whether the protected background is
restored.  Active methods require their own containers, so pixels are not
shared across methods; the image set, masks, editor, prompts, and seed schedule
are shared.

All arms use the same SD-1.5 inpainting editor \citep{rombach2022ldm} with
DDIM \citep{song2021ddim}, 50 steps, guidance
scale 7.5, and the safety checker disabled.  Sharing the editor makes retained
endpoint the controlled variable; the native-endpoint results first anchor
both released method implementations
(Table~\ref{tab:app-external-configurations}).

\begin{table}[!htbp]
\caption{External localization configurations in the matched endpoint
control.  EditGuard uses its released mask rule; OmniGuard uses the standard
0.5 decision point for its sigmoid map.}
\label{tab:app-external-configurations}
\centering
\small
\begin{tabular}{llll}
\toprule
Method & Released weight & Localization signal & Native threshold \\
\midrule
EditGuard & clean & revealed blue residual & 0.2 \\
EditGuard & degradation-trained & revealed blue residual & 0.4 \\
OmniGuard & DTE checkpoint 1500 & sigmoid tamper probability & 0.5 \\
\bottomrule
\end{tabular}
\end{table}

EditGuard is loaded strictly with its released clean and degradation-trained
weights; both recover the embedded 64-bit message with zero BER on unedited
containers.  Its published mask rule, thresholding whether any revealed
channel differs from the blue reference, is used without learned
post-processing.  OmniGuard \citep{zhang2025omniguard} is loaded with complete checkpoint coverage and
uses its released fixed secret image, padding/normalization pipeline, and DTE
output.  Its clean-container artifact reconstruction and negligible false
positive area provide an independent forward/reverse-path check.

\subsection{Native Success Isolates the Endpoint Effect}

Table~\ref{tab:app-endpoint-localization} first evaluates the paste-back image,
for which the supplied mask remains the exact composition support.  Both
external localizers reproduce strong native behavior: EditGuard reaches
0.9718/0.9519 F1 with its two released weights, and OmniGuard reaches 0.9847.
Their average precisions are 0.9968, 0.9949, and 0.9994, respectively.  This
strong native performance anchors the checkpoints, mask convention, and
operating rules before the retained-output endpoint is evaluated.

\begin{table}[!htbp]
\caption{Changing only the retained endpoint changes the output of snapshot
localizers.  F1 and predicted area are medians over 1,000 images.  The mask
prior is 0.0918, so an all-positive prediction has F1 0.1681.}
\label{tab:app-endpoint-localization}
\centering
\small
\begin{tabular}{lrrrr}
\toprule
& \multicolumn{2}{c}{Native paste-back} &
  \multicolumn{2}{c}{Retained complete output} \\
\cmidrule(lr){2-3}\cmidrule(lr){4-5}
Method & F1 $\uparrow$ & Pred. area & F1 $\uparrow$ & Pred. area \\
\midrule
EditGuard, clean weight       & 0.9718 & 0.0962 & 0.1681 & 1.0000 \\
EditGuard, degradation weight & 0.9519 & 0.0850 & 0.1681 & 1.0000 \\
OmniGuard                     & 0.9847 & 0.0935 & 0.0000 & 0.0027 \\
\bottomrule
\end{tabular}
\end{table}

Once the full output is retained, the two localizers move in opposite
directions.  EditGuard predicts the entire image: its F1 equals the
all-positive identity $2|M|/(1+|M|)$ rather than adding localization
information.  OmniGuard predicts almost nothing.  The pattern holds across a
wide fixed-threshold sweep: on the retained outputs EditGuard's revealed
residual stays above threshold over the whole image while OmniGuard's soft
map has negligible support, and the carrier diagnostics in
Table~\ref{tab:app-endpoint-carriers} give the physical cause.

\begin{table}[!htbp]
\caption{Carrier evidence under clean, paste-back, and retained-complete
endpoints.  BER is better when lower; artifact PSNR is relative to
OmniGuard's secret localization image; GSF disruption is the fraction of
unreadable carrier cells.}
\label{tab:app-endpoint-carriers}
\centering
\small
\begin{tabular}{lrrr}
\toprule
Diagnostic & Clean & Paste-back & Complete output \\
\midrule
EditGuard clean-weight message BER $\downarrow$       & 0.0000 & 0.0000 & 0.5000 \\
EditGuard degradation-weight message BER $\downarrow$ & 0.0000 & 0.0000 & 0.5156 \\
OmniGuard artifact PSNR (dB) $\uparrow$                & 38.66  & 17.35  & 6.67 \\
GSF disrupted-cell fraction               & 0.0000 & 0.0837 & 0.8778 \\
\bottomrule
\end{tabular}
\end{table}

Paste-back leaves enough predecessor carrier outside $M$ for the native
localizers to compare intact and disrupted evidence.  Complete-output
generation instead randomizes EditGuard's decoded message and collapses
OmniGuard's recovered secret image from 38.66 to 6.67 dB.  The localization
outputs therefore reflect predecessor-evidence loss, not evidence that the
retained image is unedited.

\subsection{The Original Mask No Longer Describes Every Changed Pixel}

For $X_{\mathrm{paste}}$, pixels outside $M$ are copied from the container by
construction, so $M$ is a valid pixel-composition target.  For
$X_{\mathrm{full}}$, the editor has returned new RGB values outside $M$ as
well.  The original mask still identifies the requested inpainting support,
but it no longer describes every changed output pixel.  Consequently, the
retained-output F1 values above diagnose how a snapshot localizer behaves when
its native background-retention premise is removed; they are not a semantic
ranking of complete-output localization methods.

GSF addresses the reissuance that follows rather than reusing the
inpainting mask as its target.  The event graph identifies the complete return
as an unissued transition; after review, the issuer commits an observed
revision scope and writes a fresh successor carrying the inherited history.

\subsection{The Effect Persists Across Complete-Output Editors}

A supporting control replaces the inpainting editor with the four
complete-output editor families used in the reissuance evaluation.  It covers
12 roots (two FLUX, five InstructPix2Pix, three Qwen Image Edit, and two
SDXL-Turbo instances) under matched prompts and seeds, reproducing the endpoint
mechanism in every editor family (Table~\ref{tab:app-four-editor-control}).

\begin{table}[!htbp]
\caption{Supporting complete-output control over 12 roots and four editor
families.  The trivial F1 differs slightly across carriers because each
method writes its own container and therefore induces its own output.}
\label{tab:app-four-editor-control}
\centering
\small
\setlength{\tabcolsep}{3pt}
\begin{tabular}{lrrrl}
\toprule
Method & Carrier diagnostic & Pred. area & F1 $\uparrow$ & Output form \\
\midrule
EditGuard, clean weight & BER $\downarrow$: 0.5391 & 1.0000 & 0.7495 & all positive \\
EditGuard, degradation weight & BER $\downarrow$: 0.4844 & 1.0000 & 0.7532 & all positive \\
OmniGuard & artifact PSNR $\uparrow$: 6.57 dB & 0.0039 & 0.0096 & nearly empty \\
\bottomrule
\end{tabular}
\end{table}

The matched 1,000-image experiment supplies the causal endpoint comparison;
the four-editor control shows that its physical explanation is not peculiar
to a single inpainting implementation.  The consequence is that watermark
destruction has acquired a meaning in the protocol: it marks an unresolved,
unissued transition, and authorized reissuance converts that transition into
a new readable issue carrying forward the regional history.

\section{Fidelity, Statistical Design, and System Cost}
\label{app:fidelity-statistics}

This appendix quantifies image fidelity, identifies the independent
experimental unit, and separates constant image-resident rate from registry
growth.  Together, these measurements show that status and regional-history
results are obtained with a high-fidelity, compact carrier and root-level
statistical accounting.

\subsection{Initial Reissuable Issues Remain High Fidelity}

For every one of the 480 evaluation roots, the initial issue is written from its
canonical source and compared with that source in 8-bit RGB.  Table
~\ref{tab:app-fidelity} reports the complete PSNR distribution.  The median is
46.694 dB and the tenth percentile remains 42.504 dB.  Thus the
results do not rely on a visibly aggressive initial carrier.

\begin{table}[!htbp]
\caption{Initial-issue fidelity over 480 roots.  PSNR is computed between the
canonical source and its newly written issue.  Quantiles use linear
interpolation between order statistics.}
\label{tab:app-fidelity}
\centering
\small
\begin{tabular}{crrrrrr}
\toprule
& \multicolumn{6}{c}{PSNR (dB) $\uparrow$} \\
\cmidrule(lr){2-7}
$n$ & Minimum & P10 & Mean & Median & P90 & Maximum \\
\midrule
480 & 36.850 & 42.504 & 45.973 & \textbf{46.694} & 48.141 & 48.722 \\
\bottomrule
\end{tabular}
\end{table}

The interquartile range is 45.082--47.598 dB.  A 10,000-replicate root
bootstrap (seed 20260824) gives a 95\% interval of
$[46.498,46.816]$ dB for the median and $[45.766,46.175]$ dB for the mean.
The separate 24-image physical-channel validation set in
Table~\ref{tab:app-carrier-gate} additionally records median LPIPS 0.003431 at
45.646 dB PSNR, confirming that the learned carrier remains visually light
while supporting the complete reissuance protocol.

\subsection{The Independent Experimental Unit Is the Root}

The generation lattice yields many cells per image, and the same root is
queried at several live heads.  Neither repetition creates new independent
images.  Table~\ref{tab:app-statistical-units} records the unit and aggregation
behind each headline result.  Exact count tables are reported when available;
when an uncertainty interval is used, resampling is paired or clustered at
the root, never at the cell.

\begin{table}[!htbp]
\caption{Statistical interpretation of the main evidence.  Query conditions
and post-processing settings are repeated measurements within a root, not
additional independent image draws.}
\label{tab:app-statistical-units}
\centering
\scriptsize
\begin{tabular}{p{2.5cm}p{1.7cm}p{3.7cm}p{4.4cm}}
\toprule
Result & Unit & Denominator & Aggregation and dependence \\
\midrule
Status of issued images & root--head query &
$4\times480$ own-head and $6\times480$ later-head queries &
Pooled fixed-condition accuracy; roots cluster repeated heads \\
\addlinespace
Status of unissued outputs & root--head query &
$3\times480$ pre-reissue, $1\times480$ final-unissued, and four earlier-issue conditions
($249+249+231+480$) &
Pooled recall; event truth fixed before evidence; roots cluster repeated queries \\
\addlinespace
Post-processing robustness & root--condition query &
The clean images and 31 prespecified transformations on reused roots &
Per-condition rates followed by median/minimum stress summaries; conditions are not iid draws \\
\addlinespace
Collision recovery & successor root &
480 successors independently at each tested depth &
Ambiguous-cell accuracy is computed within image, then summarized across roots \\
\addlinespace
Generation maps & successor root &
480 successors at each tested depth &
Per-image accuracy/IoU, then median; absent labels do not create artificial zero-IoU classes \\
\addlinespace
Matched endpoint control & image &
1,000 valAGE images &
Paired by source, mask, editor configuration, and seed within each carrier \\
\bottomrule
\end{tabular}
\end{table}

This accounting is especially important for the reader result.  Reporting a
confidence interval over tens of thousands of spatial cells would make the
interval spuriously narrow because neighboring cells share content, edit
support, lineage, and decoder noise.  The paper instead reports per-successor
medians at each depth and keeps the 480-root aggregate mechanism ablation
separate from those depth-conditioned values.  Similarly, the joint global
and regional result uses the same successor/head pairs but does not condition
the regional metric on a correct global prediction.

\subsection{Fresh Authority, Offline Regional Readout, and Registry Cost}
\label{app:offline-readout}

Two facts must be fetched live: whether the lineage is still active and
which issue is current.  Appendix~\ref{app:formal-semantics} establishes why
this is structural.  A fixed bitmap and entirely fixed verifier state cannot
express a later change from \textsc{Current} to \textsc{Superseded}; the
transition requires a trusted mutable coordinate, and the live head supplies
it.  The registry's verification interface is exactly that head and the
lineage's label set $\mathcal G_L$ (Section~\ref{sec:problem}).  The label
set is append-only and changes only at issuance, so a verifier can cache it;
the registry is needed live only for $h$.

Regional history is read from the copy.  The spatial reader needs the image,
the key $K$, the lineage identifier decoded from the image, and the label set
that defines its candidate planes (Section~\ref{sec:spatial-inference}); it
does not consult the head, and it does not consult the committed scopes
$A_g$.  Those scopes remain in the registry as the authorization records that
a recovered label points to, but the readout does not depend on their being
retained, available, or shared, so a partner institution or a forensic unit
entrusted with $K$ reads each region's last authorized revision from the
pixels alone.  For an unissued output there is no registry record at all: its
inherited history and the disruption that marks the edit exist only in the
copy, which is the evidence Sections~\ref{sec:results-lifecycle}
and~\ref{sec:results-robustness} read.

The costs follow this placement (Table~\ref{tab:app-system-cost}).  GSF
keeps every issued image at a fixed
512-slot payload, independent of lineage depth.  Once a copy has circulated,
later issuance never requires locating, enlarging, or rewriting it.  Each new
issue instead appends one authenticated issue record and advances the live
head, which places growing authorization history in the component designed
to authenticate, audit, and update it: the registry.

\begin{table}[!htbp]
\caption{Placement of model, payload, authorization-state, and query costs in the
evaluated instantiation.  $n$ denotes the number of issues in a
lineage.}
\label{tab:app-system-cost}
\centering
\scriptsize
\begin{tabular}{p{3.1cm}p{2.7cm}p{5.7cm}}
\toprule
Component & Evaluated cost & Scaling or role \\
\midrule
\multicolumn{3}{l}{\textit{Fixed distributed representation}} \\
Writer & 596,355 parameters & one full rewrite per issuance \\
Decoder & 2,329,155 parameters & one canonical evidence field \\
Spatial reader & 4,633 parameters & shared over candidate and spatial axes \\
\cmidrule(lr){1-3}
Learned total & 2,930,143 parameters & independent of lineage depth \\
Image-resident payload & 512 raw binary slots & 2 slots per cell; depth independent \\
\addlinespace
\multicolumn{3}{l}{\textit{Authenticated authorization state}} \\
Mutable live head & one authenticated issue label or inactive marker & constant record count; value uses $O(\log n)$ bits \\
Issue registry & one authenticated record per issue & $O(n)$ records preserving the complete issue-level audit history \\
\addlinespace
\multicolumn{3}{l}{\textit{Current verification work}} \\
Candidate evaluation & $n$ candidate checks & query-time cost of the exhaustive scan in the evaluated verifier \\
\bottomrule
\end{tabular}
\end{table}

The evaluation uses four legal generations, while shared candidate
weights provide a variable-set interface.  Each appended issue record extends
the authenticated audit trail without changing any circulated image payload or
learned model.  GSF therefore concentrates lineage growth in an updateable
registry rather than distributing it across immutable image copies: the
representation carried by every copy remains fixed, while authorization
history remains centrally auditable.

The evaluated verifier exhaustively scores the candidate set $\mathcal G_L$,
giving $O(n)$ query work.  Table~\ref{tab:app-system-cost} reports this linear
scan as an implementation property, separately from the $O(n)$ persistent
audit history.  Candidate retrieval remains an implementation dimension rather
than a consequence of the constant-rate field.

\subsection{Evidence Map}

Table~\ref{tab:app-scope-of-evidence} connects each experimental component to
the claim it establishes.

\begin{table}[!htbp]
\caption{How the experimental evidence supports the paper's claims.}
\label{tab:app-scope-of-evidence}
\centering
\small
\begin{tabular}{p{2.7cm}p{4.7cm}p{5.3cm}}
\toprule
Axis & Evidence & Established result \\
\midrule
Status semantics &
Current/superseded/open behavior across repeated complete-output editing,
branching, and head changes on 480 roots &
A live head expires old issues, while reissuance separates unissued edits from
issued successors \\
\addlinespace
Regional history &
Four coexisting labels, depth-conditioned collisions, and generation-map
recovery &
One constant-rate field recovers temporal attribution and current revision
support \\
\addlinespace
Resolution &
Canonicalized delivery from 768 through 2048 pixels &
Resolution changes preserve keyed cell identity and all tested
issues \\
\addlinespace
Post-processing &
Paired issued/unissued queries under 31 frozen transformations &
Routine delivery preserves both status decisions; heavier stress reveals
their distinct evidence curves \\
\addlinespace
Authorized scope &
Issuer commitment of a blockwise pixel-difference proposal derived from each
complete editor return &
The newest generation slice localizes where the current revision entered
authorized history \\
\addlinespace
External controls &
Matched native-paste-back versus retained-output endpoint under a shared
SD-1.5 inpainting setup &
Complete outputs remove the predecessor-residual premise used by snapshot
localizers \\
\bottomrule
\end{tabular}
\end{table}

The interface also extends beyond AI editing: conventional layer- or
pixel-based operations can propose a revision scope and enter the same
issuer-commitment workflow.  Complete-output AI editors provide the demanding
instantiation evaluated here.

\subsection{Artifact Traceability and Evidence Chain}

The anonymous manifest records each artifact's digest, version, producing
script, and table/figure dependencies.  Together with the frozen component
inventory in Appendix~\ref{app:formal-semantics}, it traces every aggregate
back to its root, issuance event, and verifier output.

Taken together, the evidence forms one chain rather than a collection of
independent benchmarks: complete-output editing opens an unresolved
transition; authenticated reissuance restores a readable issue without
requiring pixelwise survival of its predecessor; and the same constant-rate
field recovers where each region last entered the authorized history.
Continuity is thus maintained jointly at the status and regional-history
levels.

\section{Two Same-Rate Codes for the Field: Direct Labels or Keyed Compatibility}
\label{app:same-rate-counterfactual}

The generation field of Section~\ref{sec:writer} can be written at the same
two-bit rate with either of two codes, both implemented and evaluated here on
the same 480 roots.  A direct code gives each generation label its own
two-bit symbol: for a fixed alphabet of at most four labels the assignment is
a bijection, and an intact symbol maps back to its label without any
candidate collision.  The keyed field instead truncates a keyed fingerprint of
$(K,L,a,i)$ to two bits, which deliberately permits position-dependent local
collisions among candidates.

The two codes answer different requirements.  Where the label alphabet is
fixed in advance and holds at most four generations, the direct code is the
better local code, and Appendix~\ref{app:same-rate-closed} measures by how
much.  GSF is designed for a candidate set that grows after the carrier is
frozen, for a code-level signal that a readable symbol matches no issued
candidate, and for spatial anchors that survive local ambiguity.  A direct
code spends the entire two-bit alphabet on the closed lookup: its fifth label
is not occasionally ambiguous but indistinguishable from an earlier label
everywhere, and a damaged symbol is still read as one of the four labels.
The choice is therefore not between collisions and no collisions but between
deterministic global aliasing and stochastic local ambiguity.  The former
erases a distinction from the field; the latter is detectable and can leave
distinguishing evidence at other positions.

\subsection{One Rate, Two Different Representational Commitments}

Let $a$ denote an issued generation, $i$ a spatial cell, and
$q_{L,i}=\operatorname{Trunc}_2F_K(L\Vert i)$.  We compare
\begin{align}
c^{\mathrm{key}}_{a,i}
  &=\operatorname{Trunc}_2F_K(L\Vert a\Vert i), \\
c^{\mathrm{dir}}_{a,i}
  &=\operatorname{bin}_2(a\bmod 4), \\
c^{\mathrm{pdir}}_{a,i}
  &=c^{\mathrm{dir}}_{a,i}\oplus q_{L,i}.
\label{eq:app-same-rate-codes}
\end{align}
The first is GSF's candidate-compatibility fingerprint.  The second is a
position-invariant two-bit version tag; on the alphabet $\{0,1,2,3\}$ used
throughout, $a\bmod4$ is a bijection, and any label outside that alphabet
receives the symbol of a label inside it at every cell.  The third is a
diagnostic keyed-permuted direct control: an XOR pad hides the visible
label-to-symbol mapping while preserving the direct code's generation
aliases.  In particular, adding such a pad does not distinguish $a$ from
$a+4$.

All three assignments use the same 320 physical binary slots and the same 160
two-bit logical cells.  Given a hard symbol $z_i$, exact candidate matching
returns a label only when $\{a:c_{a,i}=z_i\}$ is a singleton.  A soft local
rule instead maximizes the candidate likelihood in
Eq.~\eqref{eq:local-likelihood}; a coverage-matched selective rule abstains at
a threshold chosen to equal the keyed exact rule's issued-cell answer rate.
The GSF spatial reader receives the same candidate-likelihood field but may
use evidence outside $i$.  Abstentions count as errors in every unconditional
accuracy reported below.

Table~\ref{tab:app-same-rate-semantics} summarizes why equal code rate does
not imply equal field semantics.  The direct code is optimized for the first
row; the keyed field is designed around the remaining rows as well.

\begin{table}[!htbp]
\caption{Representational commitments at the same two-bit rate.  Candidate
growth means adding issued generations while retaining the existing
carrier interface.  The rows state properties of the codes themselves; a
readout rule may add a confidence threshold to either code
(Appendix~\ref{app:same-rate-closed}).}
\label{tab:app-same-rate-semantics}
\centering
\footnotesize
\setlength{\tabcolsep}{3pt}
\begin{tabular}{p{0.19\linewidth}p{0.34\linewidth}p{0.34\linewidth}}
\toprule
Property & Direct two-bit tag & GSF keyed field \\
\midrule
Intact labels $0$--$3$ & Collision-free bijection & Predictable local stochastic ambiguity \\
Candidate growth & $a$ and $a+4$ alias at every cell & Same interface; unique anchors decrease with candidate count \\
Encoded relation & Generation modulo four & Compatibility for $(K,L,a,i)$ \\
Damaged symbol & Every two-bit value remains a legal label & Exact matching can return no match or multiple matches \\
Spatial evidence & A global alias leaves no distinguishing position & A pair may collide here and differ elsewhere \\
Wider code & Postpones alias but reduces fixed-slot spatial resolution & Retains the two-bit, 160-cell operating point \\
\bottomrule
\end{tabular}
\end{table}

\subsection{A Code-Only Counterfactual on Fixed Physical Evidence}

To isolate this representational difference, the code-only comparison holds
the decoded physical evidence fixed.  If slot $j$ was written with sign
$s_j\in\{-1,+1\}$ and has decoder logit $\ell_j$, its truth-aligned evidence
is $e_j=s_j\ell_j$.  For an alternative assignment with sign $s'_j$, the
counterfactual logit is
\begin{equation}
\ell'_j=s'_j e_j.
\label{eq:app-slot-counterfactual}
\end{equation}
Equation~\eqref{eq:app-slot-counterfactual} preserves evidence magnitude,
revision field, slot location, and image-level query while changing only the
code assignment.  The construction favors the direct code in one respect: it
does not charge that code for the carrier's co-adaptation to the keyed code
and its alignment objective, which the physical-recoding control of
Appendix~\ref{app:same-rate-physical} measures separately.  Reinstating the
native keyed signs reproduces the original measurements exactly.

The counterfactual covers all 17 issued, unissued, and branch query
conditions over the same 480 roots.  Complete per-condition results are
provided in the machine-readable artifacts; here we retain the comparisons
needed to separate a closed-set lookup from a reissuable field.

\subsection{Direct Labels Are the Better Local Code on a Fixed Four-Label Alphabet}
\label{app:same-rate-closed}

On issued images, the keyed exact rule answers 57.8876\% of cells and
obtains 57.8806\% unconditional accuracy; 42.1124\% of cells admit more than
one candidate.  The direct assignment is injective over the four labels in
this experiment and reaches 0.99983 unconditional issued-cell accuracy.  Its
soft local rule reaches 0.99990.  When the direct rule is thresholded to match
the keyed exact rule's answer rate, its selective error is
$4.2\times10^{-5}$, compared with $1.33\times10^{-4}$ for keyed exact
matching; a confidence threshold thus makes the direct code selective as
well.

The keyed ambiguity is not unexplained decoder noise.  With one, two, and
three other candidates, its cell-pooled ambiguous fractions are 0.248698,
0.436289, and 0.578138, closely matching
$1-(3/4)^d=0.25,0.4375,0.578125$; the largest deviation is 0.0013.  The
24.38\%, 43.75\%, and 58.13\% values in
Table~\ref{tab:generation-map-results} are per-image medians and therefore use
a different aggregation.

The direct code is thus the better local code for this closed, intact,
four-label question, as a bijection over a fully known alphabet should be.
The comparison rests on two assumptions that the following sections release
in turn: that the alphabet is fixed in advance, and that the observed symbol
remains meaningful.

\subsection{Candidate Growth and Local Rejection}
\label{app:same-rate-analysis}

\paragraph{The depth failure is global, not local.}
For direct $b$-bit coding,
$c^{\mathrm{dir}}_a=c^{\mathrm{dir}}_{a'}$ whenever
$a\equiv a'\pmod{2^b}$.  At two bits,
\begin{equation}
c^{\mathrm{dir}}_{a,i}=c^{\mathrm{dir}}_{a+4,i}
\qquad \text{for every cell } i.
\end{equation}
Labels 0 and 4 therefore do not merely collide in an unlucky subset of cells;
the field contains no position at which they differ.  No local decoder,
larger neighborhood, or spatial reader can recover a distinction that was
never encoded.

The keyed field moves this failure mode.  For an ideal keyed assignment, a
candidate pair collides with probability $1/4$ at any particular cell, but a
collision at one position does not force a collision at another.  Adding an
issued candidate leaves every existing fingerprint definition unchanged;
the price is the explicit, quantifiable reduction in locally unique anchors
given by Eq.~\eqref{eq:collision-law}, rather than a deterministic whole-field
alias.  GSF does not claim collision-free operation at arbitrary depth.  It
uses a fixed interface for a changing finite candidate set and exposes where
that interface is locally insufficient.

\paragraph{Rejection happens at two levels.}
At the level of the code, four active labels fill all four two-bit symbols,
so a hard readout under the direct code always names a label, whereas keyed
candidate matching can return no match or several matches.  At the level of
the readout rule, either code can abstain on low decoder confidence; the
coverage-matched selective rule of Appendix~\ref{app:same-rate-closed} is
such a rule for the direct code.  What a confidence threshold cannot supply is
the code-level signal: a symbol read with a high margin that matches no
issued candidate is, under the keyed code, evidence that the cell no
longer carries any issued generation, and the verifier counts such cells
as positive \textsc{Open} evidence ($C_{\mathrm{nokey}}$ in
Appendix~\ref{app:frozen-instantiation}); under the direct code the same
symbol is another legal label.  Table~\ref{tab:app-same-rate-failure-semantics}
isolates the code-level behavior on carrier-destroyed cells, read as hard
symbols without the readability mask that the verifier applies.

\begin{table}[!htbp]
\caption{Local outcome decomposition on carrier-destroyed cells, read as hard
symbols without the verifier's readability mask.  This is not a total-system
ranking: abstention counts as an error in unconditional accuracy, but
preserves the distinction between unsupported and asserted history.}
\label{tab:app-same-rate-failure-semantics}
\centering
\small
\begin{tabular}{lrr}
\toprule
Local outcome & Direct lookup & Keyed exact \\
\midrule
Correct generation $\uparrow$ & 42.04\% & 17.60\% \\
Wrong generation $\downarrow$ & 57.96\% & 27.52\% \\
Abstain / ambiguous & 0.00\% & 54.88\% \\
\bottomrule
\end{tabular}
\end{table}

The direct rule is correct more often in this subset, 42.04\% versus 17.60\%,
when abstention is scored as error.  That advantage and its failure mode are
two sides of the same compulsory decision: it also asserts a wrong generation
on the majority of these cells and never rejects at the code level.  For
reissuable provenance, these outcomes are not interchangeable.  A no-match does not
solve the cell, but it leaves an honest uncertainty state for spatial
inference or an \textsc{Unknown} decision; a confident wrong generation turns
damaged evidence into asserted history.  The keyed field's no-match and
multi-match outcomes are therefore not merely lost accuracy; they also form
an explicit uncertainty channel.

\paragraph{The field encodes compatibility, not a public version number.}
The direct symbol expresses $a\bmod4$ and repeats that convention across
positions and lineages.  The GSF symbol evaluates compatibility for the full
tuple $(K,L,a,i)$: changing the key, lineage, generation, or cell changes the
expected field.  The keyed-permuted direct control shows that a direct lookup can be
masked by a key and position, but masking does not make generations $a$ and
$a+4$ distinct.  Both still receive the same padded symbol at every cell.
This is why a keyed candidate-compatibility field is not equivalent to hiding
a two-bit version tag.

\subsection{Width Trades Against Spatial Granularity}

Increasing the direct code width postpones deterministic aliasing from four
to $2^b$ labels, but it spends the same fixed carrier budget in a different
place.  With 320 binary slots, $b=2,3,4,8$ provide 160, 106, 80, and 40
logical regions, respectively.  Coarser groups are also more likely to cross
a true revision boundary; at depth three, 58.05\% of the $b=8$ groups are
mixed.  A wider direct tag can therefore buy a longer closed label window only
by reducing the resolution of the regional history map.  It does not jointly
obtain collision-free arbitrary candidate depth, the same two-bit cell rate, and the
same spatial granularity.  Appendix~\ref{app:bitwidth-ladder} gives the full
fixed-slot width law, break-even depths, and mixed-group analysis.

\subsection{Editing Creates a Different Information Loss}

The preceding argument concerns what the representation can express.  Editing
adds a second failure mode by changing the observed symbol itself.  Define a
decoded-code-flip cell as one whose hard two-bit observation differs from the
symbol written for its true generation.  Conditional on that event, neither
keyed exact matching nor direct lookup can recover the true label from the
damaged cell alone: both are exactly zero in every depth-three edit condition
in Table~\ref{tab:app-same-rate-damage}.

\begin{table}[!htbp]
\caption{Recovery on decoded-code-flip cells in the five depth-three edit
conditions.  Cells/image is the mean number of affected cells.  Local
accuracies are structurally zero; the spatial reader uses evidence outside the
damaged cell.}
\label{tab:app-same-rate-damage}
\centering
\small
\begin{tabular}{lrrrr}
\toprule
& & \multicolumn{3}{c}{Generation accuracy $\uparrow$} \\
\cmidrule(lr){3-5}
Query & Cells/image & Keyed exact & Direct lookup & GSF spatial \\
\midrule
$U_0$ at head 3 & 31.8 & 0.0000 & 0.0000 & 0.2248 \\
$U_1$ at head 3 & 32.7 & 0.0000 & 0.0000 & 0.2188 \\
$U_2$ at head 3 & 31.0 & 0.0000 & 0.0000 & 0.2355 \\
$U_3$ at head 3 & 31.4 & 0.0000 & 0.0000 & 0.2303 \\
$U_p$ at head 3 & 31.6 & 0.0000 & 0.0000 & 0.2377 \\
\bottomrule
\end{tabular}
\end{table}

The GSF reader remains nonzero, at 0.2188--0.2377, because it can consult
surviving temporal anchors and coherent neighboring regions.  This experiment
establishes a deliberately narrow point: nonlocal evidence is needed after a
physical symbol flip even when the local code is collision-free over the
active labels.  It is not presented as a ranking against every possible
direct-code spatial reader.  Rather, it shows that removing keyed local
collisions would not remove the inference problem created by editing.

\subsection{Carrier and Code Are Jointly Optimized}
\label{app:same-rate-physical}

The code-only construction isolates representation, but a codebook is not a
free plug-in to the physical system.  A complementary control writes each
assignment through the same frozen carrier, decoder, and alignment stack
on 12 source images and ten post-processing conditions.  On the clean
generation-zero issue, exact two-bit readback is 1.0000 for the native keyed
assignment, 0.9299 for direct assignment, and 0.9053 for keyed-permuted direct
assignment.  Across the ten conditions, decoded-symbol change rates range
from 3.6--10.3\% for the native code, 13.3--25.9\% for direct recoding, and
13.1--21.9\% for keyed-permuted direct recoding.

Table~\ref{tab:app-physical-recoding} compares physical recoding with the
slot-level counterfactual on non-saturated cells.  Their agreement is
0.72--0.87 rather than near identity, whereas the largest difference between
positive- and negative-written-slot change probabilities is only 0.0095.
Thus a simple bit-sign asymmetry does not explain the discrepancy.

\begin{table}[!htbp]
\caption{Agreement between physical recoding and the code-only slot-level
counterfactual of Eq.~\eqref{eq:app-slot-counterfactual} on non-saturated
generation slots.  The last column
is the absolute positive/negative written-sign asymmetry.}
\label{tab:app-physical-recoding}
\centering
\small
\setlength{\tabcolsep}{3pt}
\begin{tabular}{lrrr}
\toprule
Condition & Direct agreement $\uparrow$ & Permuted-direct agreement $\uparrow$ &
Sign asymmetry $\downarrow$ \\
\midrule
Clean                 & 0.8499 & 0.8471 & 0.0058 \\
Blur $\sigma=1.0$     & 0.7677 & 0.8191 & 0.0037 \\
Brightness $1.15$     & 0.8567 & 0.8535 & 0.0047 \\
Brightness $1.30$     & 0.8544 & 0.8271 & 0.0069 \\
Contrast $1.50$       & 0.8673 & 0.8465 & 0.0088 \\
JPEG 50               & 0.7247 & 0.7711 & 0.0014 \\
JPEG 75               & 0.8028 & 0.8300 & 0.0059 \\
Platform-style pipeline & 0.7196 & 0.8120 & 0.0036 \\
Resize 256            & 0.7863 & 0.8008 & 0.0095 \\
Resize 384            & 0.8353 & 0.8451 & 0.0093 \\
\bottomrule
\end{tabular}
\end{table}

The coupling has a concrete source: the translation stage of alignment scores
candidate translations against the native keyed expected signs, whereas
rotation and scale are chosen by the key-independent decoder margin.
Replacing the code therefore changes both the semantic lookup after decoding
and the translation objective before decoding.  The counterfactual should
consequently be read as a clean comparison of representation geometry that
favors the direct code.  The physical control shows only that a complete
direct alternative would require joint optimization and fresh evaluation of
the writer, decoder, and alignment rule; it is not an intrinsic upper bound on
such a system.

\subsection{Why the Keyed Operating Point}

On a fixed alphabet of four labels the direct code is the better local code,
and the price of the keyed field can be stated in numbers: on intact issued
cells the code-only counterfactual gives the direct code 0.99983
unconditional accuracy against 0.5788 for keyed exact matching, a gap the
spatial reader closes only in part (Table~\ref{tab:generation-map-results}),
and direct recoding on the same carrier still reads back 0.9299 of its
symbols.  GSF pays that price for four properties the direct code does not
have.  When the candidate set grows, the direct code aliases the fifth label
with an earlier one at every position, whereas a keyed candidate leaves every
existing fingerprint unchanged and costs only the predictable loss of unique
anchors of Eq.~\eqref{eq:collision-law}.  A readable symbol that matches no
issued candidate is evidence under the keyed code and a legal label under
the direct code, so the field's rejections feed the verifier instead of being
absorbed into a wrong generation.  Local ambiguity varies by position, so a
spatial reader can recover part of the regional history from other cells,
whereas a global alias leaves no such cell.  And the frozen writer, decoder,
and alignment rule are co-adapted to the keyed code, so a direct alternative
is a different system to be optimized and evaluated, not a drop-in
replacement.  A deployment whose label set never exceeds four
could adopt the direct code outright; the keyed field is the choice once the
candidate set is allowed to grow.

All representation-level results use the same 480 roots, candidate
sets, revision fields, 320 physical slots, and 17 query conditions.
Alternative assignments are generated deterministically from the issued
label, cell coordinate, lineage key, and fixed slot evidence.  Reinstating the
native keyed assignment reproduces every original local and spatial input,
and the two-bit rows independently match the fixed-slot width experiment.
These reconstruction checks keep the two claims separate: the direct code's
four-label advantage is measured rather than assumed, and the keyed field is
selected for the representation that a reissuable field requires.

\section{A Fixed-Slot Design Law for Temporal Fingerprints}
\label{app:bitwidth-ladder}

Wider fingerprints reduce the probability that two candidates predict the
same local symbol, but a fixed carrier cannot increase width without reducing
the number of spatial regions.  We derive the resulting anchor law, verify it
at five widths, and show that two bits provide the strongest worst-case balance
over the three evaluated successor depths.  The same law also predicts how a
deeper lineage gradually shifts the preferred width.

\subsection{A Fixed-Slot Anchor Lemma}

Let a writer provide $S_{\mathrm{slot}}$ physical binary generation slots.
Grouping $b$ consecutive slots gives
$M_b=\lfloor S_{\mathrm{slot}}/b\rfloor$ logical regions.  Assume that, at each
logical region, the $b$-bit fingerprints of different candidates
are independent and uniform.  If the true label must be separated from $d$
other candidates, a region is a unique temporal anchor with probability
$(1-2^{-b})^d$.  Linearity of expectation therefore gives
\begin{equation}
N_S(b,d)=\left\lfloor\frac{S_{\mathrm{slot}}}{b}\right\rfloor
          (1-2^{-b})^d.
\label{eq:app-anchor-count}
\end{equation}
For a declared depth set $D$, comparison to the best width separately at each
depth yields
\begin{equation}
\eta_D(b)=\min_{d\in D}
\frac{N_S(b,d)}{\max_{k\geq1}N_S(k,d)}.
\label{eq:app-normalized-maximin}
\end{equation}

Code width is frozen when the carrier is instantiated, whereas the realized
candidate depth within the declared range $D$ is not known at that point.  A
single width must therefore serve every $d\in D$.  Normalizing by the best
allocation at each depth places these operating points on a common relative-
efficiency scale, and the outer minimum selects the width that preserves the
largest worst-case fraction of the depth-wise optimum.  This criterion follows
from the ex-ante frozen-width decision and is independent of the reader
outcomes reported below.

\paragraph{Normalized-maximin result.}
Ignoring only the finite grouping remainder, the common factor
$S_{\mathrm{slot}}$ cancels and the comparison becomes
$b^{-1}(1-2^{-b})^d$.  At $D=\{1,2,3\}$, the per-depth maximizing widths are
$1,2,3$.  Width two has relative efficiency $3/4$ at $d=1$, one at $d=2$,
and greater than $3/4$ at $d=3$.  Width one falls to
$192/343\approx0.560$ at $d=3$.  For every integer $b\geq3$, its relative
efficiency at $d=1$ is at most
\begin{equation}
\frac{2(1-2^{-b})}{b}\leq\frac{7}{12}<\frac{3}{4}.
\label{eq:app-maximin-bound}
\end{equation}
Thus $b=2$ is the unique normalized-maximin
allocation under the fixed-slot, keyed-iid assumptions and this declared depth
range.

Our writer has $S_{\mathrm{slot}}=320$.  The main slot order lists the 160
generation cells in row-major grid order and places their two channel slots
consecutively; width-$b$ groups are consecutive blocks in this 320-slot list.
The $b=3$ case discards the final two slots, the largest remainder among the
tested widths: $2/320=0.625\%$.  Table~\ref{tab:app-anchor-ladder} shows that
finite grouping does not change the winner or any relevant ordering.

\begin{table}[!htbp]
\caption{Expected unique anchors per image from
Eq.~\eqref{eq:app-anchor-count}.  The final row is finite-$S$ normalized
maximin efficiency over $D=\{1,2,3\}$.}
\label{tab:app-anchor-ladder}
\centering
\small
\begin{tabular}{crrrrr}
\toprule
& \multicolumn{5}{c}{Expected unique anchors $\uparrow$} \\
\cmidrule(lr){2-6}
$d$ & $b=1$ & $b=2$ & $b=3$ & $b=4$ & $b=8$ \\
\midrule
1 & 160.00 & 120.00 & 92.75 & 75.00 & 39.84 \\
2 &  80.00 &  90.00 & 81.16 & 70.31 & 39.69 \\
3 &  40.00 &  67.50 & 71.01 & 65.92 & 39.53 \\
4 &  20.00 &  50.62 & 62.14 & 61.80 & 39.38 \\
5 &  10.00 &  37.97 & 54.37 & 57.94 & 39.22 \\
6 &   5.00 &  28.48 & 47.57 & 54.31 & 39.07 \\
7 &   2.50 &  21.36 & 41.63 & 50.92 & 38.92 \\
\midrule
$\eta_{\{1,2,3\}}(b)$ $\uparrow$ & 0.5633 & \textbf{0.7500} & 0.5797 & 0.4688 & 0.2490 \\
\bottomrule
\end{tabular}
\end{table}

\begin{figure}[!htbp]
\centering
\includegraphics[width=0.7\linewidth]{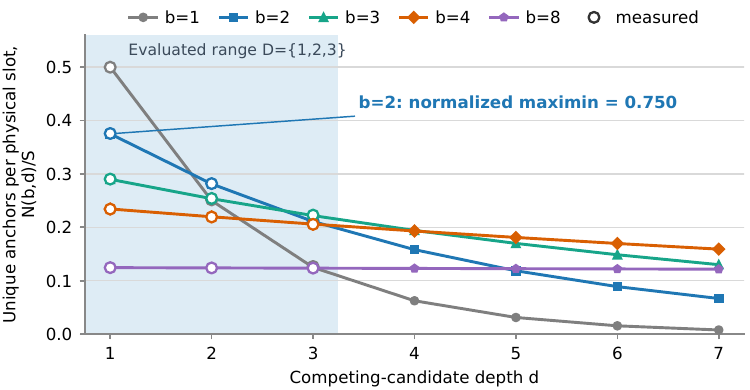}
\caption{Fixed-slot anchor yield across candidate depth.  Curves show
$N_S(b,d)/S_{\mathrm{slot}}$; hollow markers are measurements at the three
evaluated depths.  The shaded region is $D=\{1,2,3\}$, where
$\eta_D(2)=0.750$ is the worst relative efficiency, not the plotted anchor
yield.}
\label{fig:app-bitwidth-tradeoff}
\end{figure}

Figure~\ref{fig:app-bitwidth-tradeoff} makes the trade-off visible: no width
dominates at every depth, but two bits preserve the strongest worst-case
fraction of the depth-wise optimum over the evaluated range.

\subsection{When Does an Additional Bit Pay for Itself?}

In the continuous-slot comparison, the first depth at which width $b+1$
provides more total anchors than width $b$ is
\begin{equation}
d_b^*=\min\left\{d\in\mathbb N:
\frac{(1-2^{-(b+1)})^d}{b+1}>
\frac{(1-2^{-b})^d}{b}\right\}.
\label{eq:app-width-break-even}
\end{equation}
The full ladder is shown in Table~\ref{tab:app-bit-break-even}.  The second bit
already pays for itself once there are two competing candidates; the third at
three.  Further bits pay progressively later.  In particular, the eighth bit
does not beat the seventh until $d=34$.  The preferred width therefore moves
gradually with lineage depth rather than being a depth-independent constant.

\begin{table}[!htbp]
\caption{First competing-candidate depth at which each additional bit
increases total unique-anchor count under a fixed continuous slot budget.}
\label{tab:app-bit-break-even}
\centering
\small
\begin{tabular}{lrrrrrrr}
\toprule
Added bit & 2nd & 3rd & 4th & 5th & 6th & 7th & 8th \\
\midrule
Break-even depth $d_b^*$ & 2 & 3 & 5 & 7 & 12 & 20 & 34 \\
\bottomrule
\end{tabular}
\end{table}

\subsection{The Collision Identity Holds Across Widths}

For keyed fingerprints the blind fraction is
$1-(1-2^{-b})^d$, so the uniquely readable fraction is
$(1-2^{-b})^d$.  Table~\ref{tab:app-width-identity} evaluates all 15
width--depth points using intact issued fields.  The largest absolute
theory--measurement difference is 0.002857, at $b=3,d=3$.  Bit width therefore
buys collision resistance exactly as predicted; the design question is how
many spatial anchors remain after that width is charged against a fixed slot
budget.

\begin{table}[!htbp]
\caption{Closed-form and observed uniquely readable fractions for keyed
fingerprints.  Observations pool logical regions across 480 roots; values in
the main Results table are instead per-image medians.}
\label{tab:app-width-identity}
\centering
\small
\setlength{\tabcolsep}{3.5pt}
\begin{tabular}{crrrrrr}
\toprule
& \multicolumn{2}{c}{$d=1$} & \multicolumn{2}{c}{$d=2$} & \multicolumn{2}{c}{$d=3$} \\
\cmidrule(lr){2-3}\cmidrule(lr){4-5}\cmidrule(lr){6-7}
$b$ & Theory $\uparrow$ & Observed $\uparrow$ & Theory $\uparrow$ &
Observed $\uparrow$ & Theory $\uparrow$ & Observed $\uparrow$ \\
\midrule
1 & 0.5000 & 0.4997 & 0.2500 & 0.2514 & 0.1250 & 0.1267 \\
2 & 0.7500 & 0.7513 & 0.5625 & 0.5637 & 0.4219 & 0.4234 \\
3 & 0.8750 & 0.8763 & 0.7656 & 0.7670 & 0.6699 & 0.6728 \\
4 & 0.9375 & 0.9365 & 0.8789 & 0.8775 & 0.8240 & 0.8224 \\
8 & 0.9961 & 0.9962 & 0.9922 & 0.9918 & 0.9883 & 0.9892 \\
\bottomrule
\end{tabular}
\end{table}

\subsection{Width Is Invisible to the Physical-Slot Law}

Let $s_{a,j}^{(b)}$ be the expected sign of candidate $a$ at physical slot
$j$ under width $b$, and let the true candidate be $a^*$.  The truth-aligned
slot evidence is
$E_j=s_{a^*,j}^{(b)}\ell_j$.  Under iid keyed codewords and the fixed-slot
counterfactual of Eq.~\eqref{eq:app-slot-counterfactual}, the distribution of
$E_j$ is independent of $b$: changing width neither adds observations nor
changes their marginal signed-evidence law.  Width changes which slots form a
logical region, how often another candidate matches the entire group, and how
that region intersects a revision boundary.

We test this invariance with independently trained keyed two- and eight-bit
readers under the same frozen data and optimization rules.  As
Table~\ref{tab:app-width-reader-law} shows, the eight-bit reader is numerically
higher in all three reported domains: by 0.0060 over all valid generation
slots and by 0.0250--0.0288 on the two smaller damaged subdomains.  Because the
readers were trained independently, these values provide a descriptive reader
comparison rather than a causal decomposition of the difference.  The
physical-slot result above has a narrower scope: it concerns the marginal
truth-aligned evidence law, not equality between independently trained and
differently grouped readers.  Table~\ref{tab:app-width-reader-law} therefore
pairs the observed reader differences with the spatial product delivered under
the same physical-slot budget.

\begin{table}[!htbp]
\caption{Local reader accuracy and regional-history cost for separately
trained keyed readers under the same 320-slot budget.  ``All'' is the pooled
17-condition generation-slot domain; ``decoded flip'' and
``carrier-destroyed'' are nested diagnostic domains.}
\label{tab:app-width-reader-law}
\centering
\small
\begin{tabular}{lrrp{3.0cm}}
\toprule
Metric & Keyed 2-bit & Keyed 8-bit & Eight-bit change \\
\midrule
\multicolumn{4}{l}{\textit{Conditional reader accuracy}} \\
All valid generation slots, accuracy $\uparrow$ & 0.8539 & 0.8599 & $+0.0060$ \\
Decoded-code-flip slots, accuracy $\uparrow$     & 0.2743 & 0.3031 & $+0.0288$ \\
Carrier-destroyed slots, accuracy $\uparrow$     & 0.4685 & 0.4935 & $+0.0250$ \\
\addlinespace
\multicolumn{4}{l}{\textit{Regional-history product under the same 320-slot budget}} \\
Logical regions $\uparrow$ & 160 & 40 & one quarter as many \\
Depth-3 mixed groups $\downarrow$ & 0.0000 & 0.5805 & $+58.05$ percentage points \\
$\eta_{\{1,2,3\}}(b)$ $\uparrow$ & 0.7500 & 0.2490 & $-0.5010$ \\
\bottomrule
\end{tabular}
\end{table}

The eight-bit reader therefore obtains an observed local conditional advantage
of at most 0.0288.  Under the same 320-slot budget, that advantage comes with
one quarter as many logical regions---40 rather than 160---while 58.05\% of
depth-three groups cross a generation boundary and normalized-maximin
efficiency is 0.249 rather than 0.750.  Eight bits improve a local readout
statistic by producing a substantially coarser spatial product; they do not
dominate the two-bit allocation on the same regional-history objective.

\subsection{Fully Damaged Windows Contain No Clean Local Evidence}

A fully damaged logical window contains no constituent physical slot whose
hard observation retains its written symbol.  Head-zero queries contain only
one candidate, so correctness there is vacuous and is excluded
from this diagnostic.  For heads with at least one competing candidate, keyed
and direct one- and two-bit local decisions are exactly zero over 160,101
one-bit windows and 79,312 two-bit windows.  No clean local evidence in those
windows points to the original code.

Widths three and above show residual local accuracy of 0.024--0.048, but this
does not reverse the information statement.  Their coarser groups can cross a
ground-truth revision boundary; a minority subset of slots may then agree with
the label used to score the mixed group by chance.  The nonzero value is a
boundary-mixing effect, not recovery of an erased homogeneous codeword.

\subsection{Coarse Grouping Changes the Spatial Product}

Table~\ref{tab:app-mixed-groups} expands the boundary-mixing row of
Table~\ref{tab:app-width-reader-law} across all five widths and three issued
successor depths.  It measures the fraction of logical groups that contain
physical slots from more than one ground-truth generation.  One- and two-bit
groups preserve the original cell support.  At eight bits, 58.05\% of groups
in the deepest field cross a generation boundary.  This statistic measures
grouping granularity rather than pixelwise localization error.

\begin{table}[!htbp]
\caption{Fraction of logical groups crossing a ground-truth generation
boundary at each issued successor depth.}
\label{tab:app-mixed-groups}
\centering
\small
\begin{tabular}{crrrr}
\toprule
& \multicolumn{3}{c}{Mixed-group fraction $\downarrow$} & \\
\cmidrule(lr){2-4}
Width & Depth 1 & Depth 2 & Depth 3 & Logical regions $\uparrow$ \\
\midrule
1 & 0.0000 & 0.0000 & 0.0000 & 320 \\
2 & 0.0000 & 0.0000 & 0.0000 & 160 \\
3 & 0.1961 & 0.2589 & 0.2808 & 106 \\
4 & 0.1871 & 0.2459 & 0.2661 & 80 \\
8 & 0.4547 & 0.5488 & 0.5805 & 40 \\
\bottomrule
\end{tabular}
\end{table}

The main grouping follows row-major generation-cell order.  Repeating the
analysis with column-major order changes aggregate keyed accuracy by 0.0027 at
$b=4$ but by 0.0194 at $b=8$.  Coarse groups are consequently more sensitive
to an otherwise arbitrary grouping geometry.  Together with the normalized
anchor efficiencies in Table~\ref{tab:app-anchor-ladder}, this explains the
three spatial costs of very wide codes: fewer logical regions, more mixed
regions, and greater grouping sensitivity.

\subsection{Keyed and Direct Codes Allocate Ambiguity Differently}

The keyed lemma describes a variable-candidate fingerprint family.  A
position-invariant direct code instead eliminates stochastic collisions inside
its finite injective window and aliases deterministically when the alphabet is
exhausted.  Appendix~\ref{app:same-rate-counterfactual} gives the matched
two-bit comparison and shows that either family still loses local information
after a decoded-symbol change.  At one bit, the keyed field is ambiguous on
87.5\% of regions at $d=3$, while the direct field already aliases labels 0
with 2 and 1 with 3.

\subsection{Local Aggregation and Spatial Cost Move Together}

Wider groups can aggregate more weak local evidence, and direct assignments
can sharpen decisions inside a finite injective window.  Under a fixed carrier,
however, those gains arrive together with the spatial costs measured above.
At eight bits the field contains only 40 logical regions, 58.05\% of deepest
groups cross a generation boundary, and normalized maximin efficiency falls to
0.249 over the evaluated depth range.  The design law therefore compares
collision resistance and spatial evidence jointly rather than optimizing a
single local readout statistic.

\subsection{Complete Results and Reproducibility}

The full width-by-readout matrices cover all 17 query conditions for keyed
and direct codes at $b\in\{1,2,3,4,8\}$.  The accompanying artifacts provide
these matrices in machine-readable form together with the code that reproduces
the anchor counts, collision identities, grouping statistics, and decision
rules reported above.

\paragraph{Validation.}
Keyed width two uses the native GSF code; the other widths use disjoint PRF
domains over the same 320 ordered slots.  Independent executions agree on all
8,160 width-two query rows.  The theoretical and empirical results therefore
close the same loop: wider fingerprints reduce local collision by spending
spatial regions, while two bits uniquely maximize worst-case relative anchor
efficiency over the evaluated depths.

\begin{revblock}
\subsection{Wider Codes beyond the Declared Depth Range}
\label{app:bitwidth-beyond-depth}

Appendix~\ref{app:generation-map-depth7} reads the field beyond the declared
depth range $D=\{1,2,3\}$, on roots extended to depth 7.  The observed
ambiguity follows Eq.~\eqref{eq:collision-law} at every depth
(Table~\ref{tab:rev-eight-versions}, third row: 58.1 against 57.8\% at depth
3 and 86.9 against 86.7\% at depth 7).  Table~\ref{tab:rev-wider-ceilings}
compares the measured two-bit field with wider codes on the same 320 ordered
slots, which give 106 logical regions at three bits, 80 at four, and 40 at
eight, instead of 160.  For each width it reports an error-free upper bound:
the map obtained when every logical region is read without error and
assigned the most frequent ground-truth generation of its slots.  At depths
1--3 the two-bit field exceeds the upper bound of every wider code.  At
depths 4--7 the three-bit bound exceeds it, in mIoU from depth 4 and in
accuracy from depth 5, while the four- and eight-bit bounds stay below it at
every depth.  Both results agree with Section~\ref{sec:writer}: two bits
maximize the worst-case expected number of uniquely identifying cells over
$d\in\{1,2,3\}$, and deeper histories gradually favor wider codes.

\begin{table}[!htbp]
\revtable{R10}
\caption{The measured two-bit field against error-free upper bounds of wider codes on the same 320 ordered slots (medians, \%).  The declared depth range is $D=\{1,2,3\}$.}
\label{tab:rev-wider-ceilings}
\centering
\small
\setlength{\tabcolsep}{4pt}
\revfit{%
\begin{tabular}{@{}lccccccc@{}}
\toprule
Depth & 1 & 2 & 3 & 4 & 5 & 6 & 7 \\
\midrule
Map acc., measured two-bit field & 97.2 & 95.0 & 93.1 & 91.3 & 90.0 & 88.1 & 86.9 \\
Map acc., 3-bit upper bound (106 regions) & 93.4 & 91.5 & 90.6 & 90.3 & 90.3 & 89.9 & 89.9 \\
Map acc., 4-bit upper bound (80 regions) & 90.6 & 88.1 & 86.9 & 85.6 & 86.3 & 85.6 & 85.3 \\
Map acc., 8-bit upper bound (40 regions) & 84.4 & 80.6 & 78.8 & 78.1 & 78.1 & 77.5 & 76.9 \\
\midrule
mIoU, measured two-bit field & 93.9 & 87.3 & 79.0 & 72.5 & 67.1 & 62.8 & 59.0 \\
mIoU, 3-bit upper bound (106 regions) & 86.4 & 79.8 & 76.0 & 73.8 & 72.1 & 70.4 & 69.4 \\
mIoU, 4-bit upper bound (80 regions) & 81.2 & 71.2 & 63.9 & 59.0 & 55.9 & 53.7 & 50.5 \\
mIoU, 8-bit upper bound (40 regions) & 70.1 & 57.4 & 49.4 & 44.4 & 40.8 & 38.7 & 36.7 \\
\bottomrule
\end{tabular}}
\end{table}
\end{revblock}

\section{Transferring the Reissuance Protocol to Released Watermarks}
\label{app:cross-method-comparison}

Section~\ref{sec:results-cross-method} shows that the reissuance protocol
transfers to released watermarks
\citep{lu2025vine,bui2025trustmark,zhang2025omniguard,sander2025wam} and
asks which parts of the joint task transfer with it.  This appendix states
what is shared, what is method-specific, and what each number in Table~\ref{tab:cross-method-lifecycle} and
Figure~\ref{fig:delivery-robustness}(e) means.  All results are on the 480
roots of Section~\ref{sec:evaluation}.

\subsection{A Common Task with Method-Specific Constructions}
\label{app:cross-method-construction}

A released watermark has no vocabulary for reissuance.  In a separate
24-root check with released decoders and a constant message, under
inpainting and super-resolution edits rather than the editors of
Section~\ref{sec:evaluation}, the current issue and its archived predecessor
both read ``present'' on every root for all four marks tested (VINE-R,
TrustMark-Q, EditGuard, WAM), with no difference that could express
\textsc{Superseded}, and edited outputs read ``present'' on 62.5--100\% of
roots depending on the mark.  The comparison therefore gives each released
watermark the reissuance protocol of Section~\ref{sec:method}, and asks
which parts of the task it can then support.  The transferred carriers are
post-hoc marks in the encoder--decoder tradition \citep{zhu2018hidden};
in-generation watermarks \citep{fernandez2023stablesignature,wen2023treering,yang2024gaussianshading}
embed at sampling time and cannot be written into an already rendered
image, so they are not transferred.

\paragraph{Shared task.}
Every system follows the same 480 roots through four issues $I_0$--$I_3$
under the same edit-instruction table: the 2,400 rows of backbone, prompt,
seed, operation, and parent that produced GSF's outputs are matched row by
row.  Each system generates its own lineage from the source image, so its
issued and unissued pixels differ from GSF's while the instructions
coincide; this costs 12,000 editor calls.  The queries per system are
$4\times480$ under an issue's own head (truth \textsc{Current}),
$6\times480$ under a later head (truth \textsc{Superseded}), and six
unissued query conditions (truth \textsc{Open}): the outputs before
issues 1--3 at their own heads, the output after $I_3$ at head 3, and the
branch at its parent's head and at head 3.  The seventh condition of
Table~\ref{tab:app-open-legs}, a branch from $I_1$ queried at head 2,
exists only for GSF and is excluded here.
Regional history is read on the same current issues at depths 1--3.

\paragraph{Shared protocol.}
For the global-message watermarks every reissue writes a
generation-conditioned 100-bit issue message: a lineage identifier and a
keyed content descriptor, protected by the same error-correcting format and
masked by a keystream conditioned on lineage and generation.  The live head
and the registry are those of Section~\ref{sec:method}.  At query time
every system is decided by the routing chain of
Table~\ref{tab:app-verifier-parameters} at the operating point of the main
results, applied to the evidence its carrier provides.  For a global message
that evidence is the message alone: a decoded message that names the head
is \textsc{Current}, a decoded predecessor is \textsc{Superseded}, and an
undecodable message is \textsc{Open}, the route that also carries three
quarters of GSF's own \textsc{Open} decisions
(Table~\ref{tab:app-open-evidence-routes}); with no field to lose, the
chain's abstention route never fires.  WAM's adapted field takes the field
routes.  \rev{The full-field-refresh baseline of
Appendix~\ref{app:cross-method-refresh} is decided by the same chain at
thresholds tuned in its favor.}  The issuance semantics and live-head rule are shared; each construction maps
its available evidence to the task outputs as specified in
Table~\ref{tab:app-cross-method-adapters}.\footnote{TrustMark-RM applies the
released remover, which the ICCV version of TrustMark names ReMark
\citep{bui2025trustmark}; we keep the released-code name.}  No baseline network is
retrained.  What transfers with the protocol is the
issue decision against the live head; what the comparison then measures is
how much of the remaining task each carrier's edit evidence supports.

\begin{table*}[!htbp]
\caption{Method-specific construction of the compared systems.  A
constant field is a deliberately non-spatial reference: it labels
every cell with the decoded current generation and cannot recover inherited
regional history.}
\label{tab:app-cross-method-adapters}
\centering
\scriptsize
\begin{tabular}{@{}p{0.15\textwidth}p{0.28\textwidth}p{0.23\textwidth}p{0.25\textwidth}@{}}
\toprule
Watermark & Reissuance & Evidence for an unissued output & Regional-history readout \\
\midrule
VINE-R
& Rewrite the generation-conditioned issue message over the complete image.
& Message survival yields an issue match; message loss yields \textsc{Open}.
& Head label at every cell. \\
TrustMark-Q
& Rewrite the same issue message with the released quality-oriented TrustMark model.
& Message survival yields an issue match; message loss yields \textsc{Open}.
& Head label at every cell. \\
TrustMark-RM
& Apply the released remover before rewriting the next generation-conditioned TrustMark message.
& The rewritten message has the same two outcomes.
& Head label at every cell. \\
OmniGuard
& Rewrite the issue message with the released robust message watermark.
& The released tamper-mask output supports \textsc{Open}; message loss yields \textsc{Open}.
& Head label at every cell; the snapshot mask carries no earlier generations. \\
WAM
& A contiguous generation block and a 160-cell field in GSF's layout; reissue rewrites the block and the committed cells.
& An unreadable generation block or a disrupted cluster supports \textsc{Open}; a field lost entirely or disrupted beyond one cluster abstains.
& Generation code decoded on cells with readable WAM evidence. \\
GSF
& Lineage evidence, generation-conditioned issue evidence, live head, and inherited keyed field.
& Once the image is attributed to the lineage, undecodable issue evidence or a disrupted cluster supports \textsc{Open} without asserting an issue.
& Spatial reader recovers the last authorized generation at every field cell. \\
\bottomrule
\end{tabular}
\end{table*}

\paragraph{Readability on clean images.}
On the clean images at head 0, the issue message is recovered on 479/480
roots for VINE-R and on 480/480 for both TrustMark variants; OmniGuard
recovers it on 480/480 but reads 30 clean issues as \textsc{Open} because
its own tamper mask reports tampering (90 such cases over the four issues
at their own heads).  WAM's generation block, the
contiguous cells that carry its generation code, is readable on 442/480
clean roots and 391 read \textsc{Current}.  No root is excluded from any reported number; the WAM row
is a diagnostic of a GSF-like layout realized through WAM, whose native
localized-message task is a separate product.

\paragraph{WAM's layout.}
WAM writes one message per masked region and composes the regions in the
frame, whereas GSF's field is addressable per cell
(Eq.~\ref{eq:app-prf-implementation}), so one write carries every
inherited label at once.  Coexisting WAM messages interfere: with three
(lineage header, generation block, field), the header decodes on only
185/480 clean roots; with one message per field cell, every cell is
detected as marked yet none of the 160 returns its own code.  The adapted
layout therefore keeps coexisting messages few: it writes the generation
code as one block and rewrites only the committed cells at each reissue,
leaving inherited cells as previously written;
Table~\ref{tab:app-cross-method-maps} measures what survives of them.
\rev{Appendix~\ref{app:cross-method-refresh} evaluates the opposite choice, rewriting
every cell at each reissue.}

\subsection{Currency Transfers}
\label{app:cross-method-currency}

Table~\ref{tab:app-cross-method-registered} expands the issued columns of
Table~\ref{tab:cross-method-lifecycle} by live head.  With a
generation-conditioned message and a live head, VINE-R, both TrustMark
variants, and OmniGuard separate \textsc{Current} from \textsc{Superseded}
at every head, and the two TrustMark variants differ by at most 0.4 points
under any condition.  Currency transfers with the protocol wherever the
carrier's rewritten message stays readable, and the released robust
watermarks support it as accurately as GSF or slightly better.  WAM is
the exception, and its failure has a structure.  It reads its generation
block on the $I_0$ images (81.5\% \textsc{Current} at head 0) but loses
issue identity once a real edit has been reissued (0.2\%, 0.0\%, and 0.0\%
at heads 1--3, where repeated editing and re-embedding leave its field
disrupted and the shared chain returns \textsc{Open} or abstains); its
40.8\% \textsc{Superseded} is the same set of readable
$I_0$ images correctly expired under later heads, not a separate ability.

\begin{table}[!htbp]
\caption{Issued images by live head (\% \textsc{Current}/\textsc{Superseded}
accuracy).  At head $h$ the \textsc{Current} condition has 480 queries and
the \textsc{Superseded} conditions $480h$; overall columns contain
$4\times480$ and $6\times480$ queries.  ``$+$'' as in
Table~\ref{tab:cross-method-lifecycle}; the GSF row is
Table~\ref{tab:app-registered-legs}.}
\label{tab:app-cross-method-registered}
\centering
\scriptsize
\setlength{\tabcolsep}{3.4pt}
\begin{tabular}{@{}lccccc@{}}
\toprule
& \multicolumn{5}{c}{\textsc{Current}/\textsc{Superseded} accuracy (\%) $\uparrow$} \\
\cmidrule(lr){2-6}
System & $h=0$ & $h=1$ & $h=2$ & $h=3$ & Overall \\
\midrule
$+$ VINE-R                    & 99.8/-- & 98.3/99.8 & 97.1/99.1 & 97.7/98.4 & 98.2/98.9 \\
$+$ TrustMark-Q               & \textbf{100.0}/-- & \textbf{100.0/100.0} & \textbf{100.0/100.0} & \textbf{99.8/100.0} & \textbf{99.9/100.0} \\
$+$ TrustMark-RM              & \textbf{100.0}/-- & 99.6/100.0 & 99.6/99.8 & 99.4/99.7 & 99.6/99.8 \\
$+$ OmniGuard $+$ tamper mask & 93.8/-- & 95.2/93.8 & 96.5/94.5 & 95.8/95.1 & 95.3/94.7 \\
$+$ WAM in GSF layout$^\dagger$ & 81.5/-- & 0.2/81.5 & 0.0/40.8 & 0.0/27.2 & 20.4/40.8 \\
\textbf{Full GSF (ours)}      & 99.6/-- & 99.8/99.6 & 99.8/99.8 & 99.0/99.8 & 99.5/99.8 \\
\bottomrule
\end{tabular}
\end{table}

\subsection{Unissued Recognition Needs Evidence beyond the Message}
\label{app:cross-method-open}

Table~\ref{tab:app-cross-method-open} gives the six unissued query
conditions with 95\% Clopper--Pearson intervals; the GSF row is the sum of
the corresponding six rows of Table~\ref{tab:app-open-legs}.  For the
three message-only watermarks (VINE-R, TrustMark-Q, TrustMark-RM) the
editor decides the outcome.  A destroyed message reads as \textsc{Open}:
63.5\%, 70.1\%, and 70.5\%.  A surviving message names the issue it was
written for, which is a false \textsc{Current} at that issue's own head
and a false \textsc{Superseded} after the head advances (VINE-R 879 and
172 of its 1,051 false issues; TrustMark-Q 722 and 140; TrustMark-RM 710
and 139).  SDXL-Turbo destroys every message of every watermark (0 of 730
readable), FLUX.1 Kontext preserves the most (521 of 720 for VINE-R), with
InstructPix2Pix and Qwen Image Edit in between.  An unissued output whose
message survives thus reads as the issue it names; GSF's field adds a
second observation, a disrupted cluster, which catches most such outputs
and halves false acceptance (Tables~\ref{tab:lifecycle-evidence}
and~\ref{tab:app-open-evidence-routes}).  On these conditions GSF's
94.0\% is 75.1 points of undecodable issue evidence, the route it shares
with the message-only watermarks, and 18.9 points of field evidence, 17.1
of them from a disrupted cluster.
GSF's own recall on the four sequential conditions is 96.04\%, 94.17\%,
92.92\%, and 91.67\% (outputs before issues 1--3 and after $I_3$), a slow
decline with depth.

OmniGuard's 61.3\% splits the same way: 35.0 points of undecodable message
and 26.3 from its tamper mask, a second observation; its robust message
nevertheless survives more edits than any other watermark, so it also has
the highest false-issue rate, 38.7\%.  WAM's 0.4\% false issue must be read
with its 48.8\% \textsc{Unknown}: after the first real edit its generation
block is unreadable, and its field is lost entirely on 1,164 of the 2,880
outputs and disrupted beyond one cluster on 242 more, the two routes that
abstain under the shared chain; its 50.8\% \textsc{Open} is decoding
failure rather than discrimination.

\begin{table}[!htbp]
\caption{Unissued outputs on the six query conditions shared by all systems
($6\times480=2{,}880$ queries; \% with 95\% Clopper--Pearson intervals).
False issue is a prediction of \textsc{Current} or \textsc{Superseded};
\textsc{Unknown} is never credited as \textsc{Open}; --: no abstention
route.}
\label{tab:app-cross-method-open}
\centering
\scriptsize
\setlength{\tabcolsep}{3.5pt}
\begin{tabular}{@{}lccc@{}}
\toprule
System & \textsc{Open} $\uparrow$ & False issue $\downarrow$ & \textsc{Unknown} $\downarrow$ \\
\midrule
$+$ VINE-R & 63.5 [61.7, 65.3] & 36.5 [34.7, 38.3] & -- \\
$+$ TrustMark-Q & 70.1 [68.4, 71.7] & 29.9 [28.3, 31.6] & -- \\
$+$ TrustMark-RM & 70.5 [68.8, 72.2] & 29.5 [27.8, 31.2] & -- \\
$+$ OmniGuard $+$ tamper mask & 61.3 [59.5, 63.1] & 38.7 [36.9, 40.5] & -- \\
$+$ WAM in GSF layout$^\dagger$ & 50.8 [48.9, 52.6] & 0.4 [0.2, 0.7] & 48.8 [47.0, 50.7] \\
\textbf{Full GSF (ours)} & \textbf{94.0 [93.0, 94.8]} & 2.1 [1.6, 2.7] & 3.9 [3.2, 4.7] \\
\bottomrule
\end{tabular}
\end{table}

\subsection{A Content Commitment Trades Currency for Caution}
\label{app:cross-method-commitment}

The message's content descriptor does separate unissued outputs from issued
images: as a scalar distance it reaches an AUROC of 0.92--0.94 for every
watermark (0.945 for GSF's own commitment).  Thresholding it, however, only
moves outcomes between \textsc{Current} and withheld
(Table~\ref{tab:app-cross-method-commit}).  At the strict threshold
$\tau_{\mathrm{id}}=1/48$, \textsc{Current} accuracy under an issue's own
head falls to 67.4--79.9\% while \textsc{Open} recall rises by only
1.8--5.3 points for the message-only watermarks and 4.5 for OmniGuard;
false issue drops to 7.1--7.8\% because surviving messages
are now withheld rather than accepted.  Relaxing to $2/48$ restores 90.1--97.6\% \textsc{Current} at the
same \textsc{Open} recall.  At the thresholds and routing tested, a content
commitment trades currency for caution along one scalar and adds little
\textsc{Open} recall.

\begin{table}[!htbp]
\caption{Effect of a strict content commitment on the global-message
watermarks (\%).  Native: the protocol as evaluated in
Table~\ref{tab:cross-method-lifecycle}; $\tau_{\mathrm{id}}$: the message's
content-descriptor mismatch threshold in units of $1/48$.  Own/later:
\textsc{Current}/\textsc{Superseded} accuracy under the issue's own head and
a later head; \textsc{Open} and false issue on the six unissued query
conditions.}
\label{tab:app-cross-method-commit}
\centering
\scriptsize
\setlength{\tabcolsep}{3pt}
\begin{tabular}{@{}lccc@{}}
\toprule
& Native & $\tau_{\mathrm{id}}=1/48$ & $\tau_{\mathrm{id}}=2/48$ \\
System & own / later / Open / false issue & own / later / Open / false issue & own / later / Open / false issue \\
\midrule
$+$ VINE-R      & 98.2 / 98.9 / 63.5 / 36.5 & 75.7 / 98.9 / 68.9 / 7.8 & 94.2 / 98.9 / 68.9 / 10.8 \\
$+$ TrustMark-Q & 99.9 / 100.0 / 70.1 / 29.9 & 79.8 / 100.0 / 72.0 / 7.1 & 97.6 / 100.0 / 72.0 / 10.6 \\
$+$ TrustMark-RM & 99.6 / 99.8 / 70.5 / 29.5 & 79.9 / 99.8 / 72.4 / 7.1 & 97.6 / 99.8 / 72.4 / 10.2 \\
$+$ OmniGuard $+$ tamper mask & 95.3 / 94.7 / 61.3 / 38.7 & 67.4 / 94.7 / 65.8 / 7.7 & 90.1 / 94.7 / 65.8 / 11.3 \\
\bottomrule
\end{tabular}
\end{table}

\subsection{Regional History}
\label{app:cross-method-maps}

Table~\ref{tab:app-cross-method-maps} reports the map metrics of
Table~\ref{tab:generation-map-results} for every system, adding the
newest-slice IoU used by Figure~\ref{fig:delivery-robustness}(e).  A
global message produces no field, so its map is a constant field: every
cell labeled with the decoded head.  Its accuracy is the area of the latest
committed scope and its newest-slice IoU coincides with that accuracy; its
mIoU falls with depth because every inherited generation is missed.  The
constant field was scored over all 256 cells of the grid, whereas WAM and
GSF are scored on the 160 field cells; on GSF's own fields it gives
27.50\%, 18.12\%, and 14.30\% mIoU on the 160 cells against 28.71\%,
19.14\%, and 14.89\% on all 256, so the difference is about one point and
does not affect any comparison.  WAM's adapted field
is the closest spatial counterpart.  On the cells where its evidence survives
it is accurate (91.3--94.7\% cell accuracy), but 26.9--36.9\% of cells are
unreadable after repeated editing and re-embedding and count as errors, so
its mIoU falls from 49.05\% to 24.30\%; its high newest-slice IoU reflects
that the most recently rewritten cells are exactly the ones still readable.  Localizing a single
edit is not where the systems differ: on unissued outputs, the average
precision of OmniGuard's mask, WAM's evidence map, and GSF's raw
disruption map against the proposed scope is 0.625, 0.642, and 0.634.
The separation lies in inherited history, which only the keyed field
carries.

\subsection{Delivery Range}
\label{app:cross-method-delivery}

Table~\ref{tab:app-cross-method-delivery} reports \textsc{Current}
retention of the current issue at head 3 under the 22 conditions of
Appendix~\ref{app:rendition-robustness} that were run for every system
(the ten finer levels of that appendix were run for GSF only).  The
message-only watermarks hold 95.6--98.1\% (VINE-R), 84.4--99.8\%
(TrustMark-Q), and 81.2--99.4\% (TrustMark-RM) across all 22 conditions,
OmniGuard 52.9--96.2\%, whereas GSF falls to 0--65\% under JPEG at quality
50 or below, resizing to 192 pixels, the platform pipelines, contrast gain,
and brightness $\times1.30$, and to 72.5\% at brightness $\times1.15$.  WAM
is omitted: its current issue reads \textsc{Current} in 0.0\% of clean
roots at head 3 (Table~\ref{tab:app-cross-method-registered}), so delivery
adds nothing.  A robust static message thus survives a wider range of
delivery transformations; the same survival keeps naming the old issue on
an unissued output and carries no history.  Over the same 22 conditions,
GSF's median map accuracy stays at or above 85.00\%
(Table~\ref{tab:app-map-delivery}).

\begin{table}[!htbp]
\begin{minipage}[t]{0.45\textwidth}
\caption{Regional history at depths 1--3 (median \% over 480 roots).  Acc.\
and mIoU as in Table~\ref{tab:generation-map-results}; Newest IoU is
the IoU of the newest slice with the committed scope.  Global-message rows
are the constant field scored over all 256 cells (see text);
WAM and GSF are scored on the 160 field cells.  Baseline rows carry the
protocol of Table~\ref{tab:app-cross-method-adapters}.}
\label{tab:app-cross-method-maps}
\centering
\scriptsize
\setlength{\tabcolsep}{3pt}
\begin{tabular}{@{}clccc@{}}
\toprule
Depth & System & Acc.\ $\uparrow$ & mIoU $\uparrow$ & Newest IoU $\uparrow$ \\
\midrule
1 & $+$ VINE-R & 57.81 & 28.91 & 57.81 \\
1 & $+$ TrustMark-Q & 57.03 & 28.52 & 57.03 \\
1 & $+$ TrustMark-RM & 57.03 & 28.52 & 57.03 \\
1 & $+$ OmniGuard & 56.64 & 28.32 & 56.64 \\
1 & $+$ WAM$^\dagger$ & 60.00 & 49.05 & 90.97 \\
1 & \textbf{GSF (ours)} & \textbf{97.19} & \textbf{93.87} & \textbf{94.92} \\
\addlinespace[2pt]
2 & $+$ VINE-R & 57.81 & 19.40 & 57.81 \\
2 & $+$ TrustMark-Q & 57.81 & 19.27 & 57.81 \\
2 & $+$ TrustMark-RM & 57.42 & 19.21 & 57.42 \\
2 & $+$ OmniGuard & 57.03 & 19.08 & 57.03 \\
2 & $+$ WAM$^\dagger$ & 52.50 & 32.38 & 93.92 \\
2 & \textbf{GSF (ours)} & \textbf{95.00} & \textbf{87.27} & \textbf{94.09} \\
\addlinespace[2pt]
3 & $+$ VINE-R & 58.20 & 14.70 & 58.20 \\
3 & $+$ TrustMark-Q & 57.03 & 14.40 & 57.03 \\
3 & $+$ TrustMark-RM & 57.42 & 14.65 & 57.42 \\
3 & $+$ OmniGuard & 58.20 & 14.65 & 58.20 \\
3 & $+$ WAM$^\dagger$ & 53.75 & 24.30 & 93.44 \\
3 & \textbf{GSF (ours)} & \textbf{93.13} & \textbf{78.96} & \textbf{92.86} \\
\bottomrule
\end{tabular}
\end{minipage}\hfill
\begin{minipage}[t]{0.54\textwidth}
\caption{\textsc{Current} retention of the current issue $I_3$ at head 3
under the 22 shared delivery conditions (\% of 480 roots).  Operators as in
Appendix~\ref{app:robustness-setup}; the GSF column is the $I_3$ column of
Table~\ref{tab:app-robust-matrix}; baseline columns carry the protocol (TM: TrustMark).}
\label{tab:app-cross-method-delivery}
\centering
\scriptsize
\setlength{\tabcolsep}{2.2pt}
\begin{tabular}{@{}lrrrrr@{}}
\toprule
Condition & $+$VINE-R & $+$TM-Q & $+$TM-RM & $+$OmniGuard & GSF \\
\midrule
Clean & 97.7 & 99.8 & 99.4 & 95.8 & 99.0 \\
\addlinespace[2pt]
JPEG 75 & 97.5 & 99.8 & 98.8 & 87.1 & 95.6 \\
JPEG 50 & 96.9 & 98.5 & 97.5 & 89.2 & 35.0 \\
JPEG 40 & 96.7 & 97.3 & 96.9 & 84.0 & 1.7 \\
JPEG 30 & 96.5 & 94.6 & 93.8 & 70.0 & 0.0 \\
Double JPEG 75 & 97.7 & 99.8 & 98.5 & 89.0 & 95.2 \\
\addlinespace[2pt]
Resize 384 & 97.9 & 99.8 & 99.4 & 83.5 & 99.4 \\
Resize 320 & 97.9 & 99.8 & 99.4 & 79.0 & 99.0 \\
Resize 256 & 97.7 & 99.8 & 99.4 & 76.0 & 99.4 \\
Resize 192 & 97.1 & 99.8 & 99.4 & 72.7 & 1.0 \\
\addlinespace[2pt]
Blur $0.5$ & 97.9 & 99.8 & 99.4 & 89.6 & 99.0 \\
Blur $1.0$ & 98.1 & 99.6 & 99.4 & 57.7 & 98.1 \\
\addlinespace[2pt]
Brightness $\times0.85$ & 97.7 & 99.8 & 99.4 & 94.0 & 99.4 \\
Brightness $\times1.15$ & 97.3 & 98.5 & 98.3 & 77.3 & 72.5 \\
Brightness $\times1.30$ & 95.6 & 94.0 & 94.2 & 63.5 & 39.4 \\
\addlinespace[2pt]
Contrast $\times0.80$ & 97.3 & 99.8 & 99.4 & 95.0 & 99.2 \\
Contrast $\times1.25$ & 97.3 & 93.3 & 93.3 & 71.7 & 64.6 \\
Contrast $\times1.50$ & 96.7 & 84.4 & 81.2 & 52.9 & 31.3 \\
\addlinespace[2pt]
Gamma $0.80$ & 97.9 & 99.8 & 99.4 & 96.2 & 75.0 \\
Gamma $1.25$ & 96.7 & 99.6 & 99.0 & 77.7 & 66.7 \\
\addlinespace[2pt]
Platform A & 97.1 & 98.8 & 97.5 & 92.3 & 58.1 \\
Platform B & 97.1 & 99.4 & 98.3 & 92.7 & 2.5 \\
\bottomrule
\end{tabular}
\end{minipage}
\end{table}

\subsection{A Registry Lookup Reproduces the Map of an Issued Image}
\label{app:cross-method-lookup}

The registry of Section~\ref{sec:problem} records every committed scope
$A_g$ and parent link, so Eq.~\eqref{eq:field-update} reconstructs $\tau_g$
exactly from a decoded label.  A deployment that consults the registry at
every verification therefore needs no field on issued images: a robust
global message carrying the protocol returns the exact map whenever its
label decodes, at the carrier's status accuracy---for TrustMark-Q,
99.9/100.0 on clean images and 84.4--99.8\% \textsc{Current} across the 22
delivery conditions of Table~\ref{tab:app-cross-method-delivery}---on clean
images above the 93.13\% that GSF reads from the pixels at depth 3.  That
design keeps the history in the registry and reads it online.  GSF keeps it
in the copy (Appendix~\ref{app:offline-readout}): a partner entrusted with
$K$ and the list of issues reads each region's last authorized revision
without the registry's records, and an unissued output, on which a
surviving message names the old issue
(Table~\ref{tab:app-cross-method-open}), shows in its field where it
departs from its parent (Figure~\ref{fig:lifecycle-sequence}).  Both
designs run the same reissuance protocol---generation-conditioned issue
message, live head, registry, and reissuance transaction---and it is the
protocol that gives every row of Table~\ref{tab:cross-method-lifecycle}
supersession without retraining.

\begin{revblock}
\subsection{A Registry Lookup with the Issuer's Complete Record}
\label{app:cross-method-registry-record}

Appendix~\ref{app:cross-method-lookup} treats a registry lookup as a
different design, and the two designs give the verifier different inputs.
Our task gives it a query image, the issuer's key, and a small trusted
reference to the lineage's current authorization state
(Section~\ref{sec:introduction}), and regional history is read from the copy
with the key and the list of issues, without the issuer's edit records
(Appendix~\ref{app:cross-method-lookup}).  A registry lookup reads more: at
every query, the scope and parent link that the registry records for every
issue (Section~\ref{sec:problem}) and, if the lookup is keyed by content, a
stored content hash of every issue.  This subsection measures that design
without a watermark, and what the field adds beyond it.

\paragraph{Setup.}
For each of the 1,920 issues the registry stores an exact hash and a 256-bit
perceptual hash.  A query is matched to the nearest issue of its own root
within $t$ bits (no match: not an issue), and the head then gives
\textsc{Current} or \textsc{Superseded}.  The lookup is given the root of
every query and its complete record, which favors it.  Three tolerances were
fixed in advance, without reference to GSF: $t=0$; $t=27$, the 99.9th
percentile of the distances under benign delivery on calibration roots over
the 22 conditions of Table~\ref{tab:app-cross-method-delivery}; and $t=12$,
at which 2.1\% of the delivered current issues are lost over the same
conditions, which extend beyond the envelope of
Appendix~\ref{app:robustness-envelope}.

\begin{table}[!htbp]
\revtable{R1}
\caption{Status with a registry keyed by a content hash (\%).  Clean: the queries of Table~\ref{tab:lifecycle-main}.  Delivered: the nine conditions of Table~\ref{tab:app-cross-method-delivery} within the envelope of Appendix~\ref{app:robustness-envelope}.}
\label{tab:rev-registry-status}
\centering
\small
\revfit{%
\begin{tabular}{@{}lccc@{}}
\toprule
Registry key & \shortstack{Clean:\\\textsc{Current} /\\\textsc{Superseded} $\uparrow$} & \shortstack{Clean:\\false issue $\downarrow$} & \shortstack{Delivered:\\\textsc{Current}\\retention $\uparrow$} \\
\midrule
Exact hash & 100.0 / 100.0 & 0.0 & 0.0 \\
Perceptual hash, $t=0$ & 100.0 / 99.7 & 0.5 & 83.1 \\
Perceptual hash, $t=12$ & 100.0 / 99.7 & 14.8 & 100.0 \\
Perceptual hash, $t=27$ & 100.0 / 99.7 & 31.1 & 100.0 \\
Perceptual hash (oracle) & 100.0 / 99.7 & 2.6 & 99.4 \\
GSF (reads no per-issue record) & \textbf{99.5 / 99.8} & \textbf{2.2} & \textbf{98.2} \\
\bottomrule
\end{tabular}}
\end{table}

On clean images the exact hash identifies every issue, and under delivery it
matches none of them.  Delivery requires a tolerance, and the tolerance
admits edits: of the tolerances fixed in advance, those that keep the
delivered current issues ($t=12$ and $t=27$) accept 14.8 and 31.1\% of the
clean unissued outputs as issues, as the global-message watermarks of
Table~\ref{tab:cross-method-lifecycle} do (29.5--38.7\%).  GSF does not
decide from one content distance: its issue evidence keeps decoding under
ordinary delivery and does not decode on 74.94\% of unissued outputs, and a
disrupted cluster catches most outputs whose issue evidence survives the
edit (Appendix~\ref{app:lifecycle-evidence-ablation}).  Its thresholds were
placed before the evaluation (Table~\ref{tab:app-verifier-parameters}), and
Appendix~\ref{app:robustness-sweep} shows how moving one of them trades
retention against false acceptance under harsher delivery.

\paragraph{Regional history of an issued image.}
A lookup identifies which issue an image is.  Which revision each region
comes from is not in the hash; it is in the issuer's records of committed
scopes and parent links.

\begin{table}[!htbp]
\revtable{R2}
\caption{Regional history of the clean current issue $I_3$ (480 roots; medians, \%).}
\label{tab:rev-registry-history}
\centering
\small
\revfit{%
\begin{tabular}{@{}p{9.6cm}cc@{}}
\toprule
Where the map comes from & Map acc. & mIoU \\
\midrule
Lookup alone, no edit record read: every cell gets the identified issue & 56.3 & 14.3 \\
Lookup $+$ replay of every issue's committed scope and parent link & 100.0 & 100.0 \\
GSF: read from the copy, with the key and the list of issues & \textbf{93.1} & \textbf{79.0} \\
\bottomrule
\end{tabular}}
\end{table}

The lookup's map is thus a replay of the issuer's records: exact where they
are read, as Appendix~\ref{app:cross-method-lookup} states, and absent where
they are not.

\paragraph{Paste-back edits.}
For an unissued output the target is its parent's field
(Section~\ref{sec:problem}).  A lookup that identifies the parent and
replays its records therefore returns the right history; what an
identification and a replayed record do not provide by themselves is where
the image departs from its parent.  The unissued outputs of
Section~\ref{sec:evaluation} are complete returns, in which the median edit
changes every cell (Table~\ref{tab:app-complete-change}), so no region of
such an output can be assumed to be the parent's.  We therefore formed
paste-back endpoints as in Appendix~\ref{app:complete-output-controls}: in
the current issue $I_3$ of every root, the largest object (a median of
21.1\% of the image) was regenerated with SD-1.5 inpainting
\citep{rombach2022ldm} and with SDXL-Turbo \citep{sauer2024adversarial} and
pasted back, giving 960 unissued edits that keep the pixels of $I_3$, and
with them its watermark, outside the object.  At $t=0$, 12, and 27 the
lookup matches 0.1, 11.1, and 42.0\% of them to an issue.  For the edits it
matches, it replays a history and, by the same match, accepts an unissued
edit as an issue; for the others it has no history to return, so no edit
receives both the right status and a history.  GSF reads no edit record.
It recognizes 99.8\% of these edits as unissued, accepts none as an issue,
and flags 79.3\% of the edited cells as disrupted (1.2\% of the untouched
ones).  In the edited region it does not recover the parent's history; on
the regions the edit left untouched, it reads the history as accurately as
on $I_3$ itself (Table~\ref{tab:rev-local-edit-history}).

\begin{table}[!htbp]
\revtable{R3}
\caption{Regional history read by GSF on the untouched regions of the paste-back edits (952 edits with untouched cells; medians, \%).}
\label{tab:rev-local-edit-history}
\centering
\small
\revfit{%
\begin{tabular}{@{}lcc@{}}
\toprule
Read from & Map acc. & mIoU \\
\midrule
$I_3$, before the edit & 93.6 & 79.7 \\
Paste-back edit of $I_3$ & \textbf{93.7} & \textbf{79.4} \\
\bottomrule
\end{tabular}}
\end{table}

\paragraph{What the field adds.}
GSF's contribution is thus not that a database could not, in principle,
provide status and regional history.  It is that a verifier without access
to the issuer's per-issue edit records can read the regional history from
the copy (Table~\ref{tab:rev-registry-history}), including on the untouched
regions of a paste-back edit (Table~\ref{tab:rev-local-edit-history});
recognize unissued edits (Tables~\ref{tab:cross-method-lifecycle}
and~\ref{tab:rev-registry-status}); and obtain issue evidence bound to the
image content (Table~\ref{tab:rev-new-experiments}).  A system with similar
functions could be built on a database that holds the issuer's complete
record, by identifying the parent issue, retrieving its recorded history,
and comparing image content; but that takes additional inputs, a
content-comparison step, and deployment assumptions, and is no longer a
plain ID lookup.  GSF embeds this evidence in the image, so that it travels
with the copy.
\end{revblock}

\begin{revblock}
\subsection{A Full-Field-Refresh Regional Baseline}
\label{app:cross-method-refresh}

The adapted WAM of Appendix~\ref{app:cross-method-construction} rewrites only
the committed cells at each reissue and leaves inherited cells as previously
written, whereas GSF rewrites every field cell (Section~\ref{sec:writer}).
No regional watermark we know of defines a reissue step, so we built a
full-field-refresh baseline from the closest released one, WAM, with each
choice made in its favor.

\paragraph{Construction.}
The initial issue is written as in
Appendix~\ref{app:cross-method-construction}.  At each reissue the writer
rewrites all 256 cells: the lineage header, the generation block of the new
issue, and the 160 field cells, which the field $\tau_g$ partitions into one
region per generation.  Each region carries the 32-bit keyed generation code
of its generation, so that every cell carries the label it carries in GSF;
WAM composes the regions in the frame as in
Appendix~\ref{app:cross-method-construction}, and the network is not
retrained.  Two strengths bracket the distortion budget: the released
strength scaled by 0.3472, at which the median initial-issue PSNR matches
GSF's 46.7~dB (Table~\ref{tab:app-fidelity}), and WAM's released strength,
at 37.8~dB, which spends about 9~dB more distortion.  The verifier knows
every candidate code, an informed decoding that favors the baseline over
WAM's own blind decoder.  A cell is read by the bitwise majority of its
pixels detected as marked and accepted within a Hamming tolerance of a
candidate code.

\paragraph{Operating point.}
The readout tolerance and the thresholds $C_{\mathrm{nokey}}$,
$C_{\mathrm{cc}}$, and $B_{\mathrm{bg}}$ of
Table~\ref{tab:app-verifier-parameters} were tuned jointly on a 2,268-point
grid over the 480 evaluation roots; $G_{\mathrm{dead}}=1$, which routes a
field lost entirely to \textsc{Unknown}, is GSF's and was not tuned.  The
reader of the map was selected on the same roots as the one most favorable
to the baseline: it assigns every cell with readable WAM evidence to its
nearest candidate.  At each strength, the rows of
Table~\ref{tab:cross-method-lifecycle} report the grid point with the highest
\textsc{Open} recall among those that keep both issued columns at or above
90\%, with that reader, so the baseline enters the comparison at its most
favorable point under this constraint.  Table~\ref{tab:rev-refresh-operating}
gives it also under the frozen rules, with the readout of the committed-cells
WAM row and the thresholds of Table~\ref{tab:app-verifier-parameters}.

\begin{table}[!htbp]
\revtable{R5}
\caption{Inherited regional history read from the copy (mIoU, medians, \%).  Full-refresh rows: the reader most favorable to the baseline.}
\label{tab:rev-refresh-history}
\centering
\small
\revfit{%
\begin{tabular}{@{}lccc@{}}
\toprule
Carrier (initial-issue PSNR) & $d{=}1$ & $d{=}2$ & $d{=}3$ \\
\midrule
WAM in GSF layout, committed cells only (Table~\ref{tab:cross-method-lifecycle}; 37.8~dB) & 49.05 & 32.38 & 24.30 \\
WAM in GSF layout, full refresh (46.7~dB) & 63.61 & 43.56 & 34.22 \\
WAM in GSF layout, full refresh (37.8~dB) & 80.40 & 55.68 & 43.15 \\
GSF (46.7~dB) & \textbf{93.87} & \textbf{87.27} & \textbf{78.96} \\
\bottomrule
\end{tabular}}
\end{table}

\begin{table}[!htbp]
\revtable{R11}
\caption{The full-field-refresh WAM baseline under the frozen rules and at its tuned point, beside GSF (\%).  Frozen rules: the readout of the committed-cells WAM row of Table~\ref{tab:cross-method-lifecycle} and the thresholds of Table~\ref{tab:app-verifier-parameters}, with the map read at the same tolerance.  Tuned: the rows of Table~\ref{tab:cross-method-lifecycle}, with the nearest-candidate reader.  GSF's map is read by its spatial reader.  Status rows as the columns of Table~\ref{tab:cross-method-lifecycle}.}
\label{tab:rev-refresh-operating}
\centering
\footnotesize
\setlength{\tabcolsep}{4pt}
\revfit{%
\begin{tabular}{@{}p{3.4cm}ccccc@{}}
\toprule
& \multicolumn{2}{c}{Full refresh, 46.7~dB} & \multicolumn{2}{c}{Full refresh, 37.8~dB} & GSF \\
\cmidrule(lr){2-3}\cmidrule(lr){4-5}
& Frozen rules & Tuned & Frozen rules & Tuned & \\
\midrule
Readout tolerance (bits) & 4 & 10 & 4 & 8 & -- \\
$C_{\mathrm{nokey}}$ & 1 & 64 & 1 & 64 & 1 \\
$C_{\mathrm{cc}}$ & 1 & 256 & 1 & 1 & 1 \\
$B_{\mathrm{bg}}$ & 1/4 & 1/7 & 1/4 & 1 & 1/4 \\
\midrule
\textsc{Current} $\uparrow$ & 6.9 & 90.7 & 21.0 & 90.9 & 99.5 \\
\textsc{Superseded} $\uparrow$ & 13.6 & 95.7 & 41.3 & 95.7 & 99.8 \\
\textsc{Open} $\uparrow$ & 41.6 & 19.1 & 61.4 & 58.5 & 94.0 \\
False issue $\downarrow$ & 0.1 & 23.2 & 0.4 & 3.8 & 2.1 \\
\textsc{Unknown} $\downarrow$ & 58.3 & 57.7 & 38.3 & 37.7 & 3.9 \\
\midrule
mIoU, $d{=}1$ $\uparrow$ & 45.37 & 63.61 & 68.41 & 80.40 & 93.87 \\
mIoU, $d{=}2$ $\uparrow$ & 28.55 & 43.56 & 43.93 & 55.68 & 87.27 \\
mIoU, $d{=}3$ $\uparrow$ & 18.98 & 34.22 & 31.12 & 43.15 & 78.96 \\
\bottomrule
\end{tabular}}
\end{table}

\paragraph{Results.}
Under the frozen rules at its released strength
(Table~\ref{tab:rev-refresh-operating}), rewriting every cell raises the
baseline's \textsc{Open} recall from 50.8 to 61.4\% and its depth-3 mIoU
from 24.30 to 31.12\% over the committed-cells row of
Table~\ref{tab:cross-method-lifecycle}, and leaves its issued images read
\textsc{Current} and \textsc{Superseded} in 21.0 and 41.3\%, against 20.4
and 40.8\%.  At its tuned point it reads them in 90.9 and 95.7\%,
recognizes 58.5\% of unissued outputs as \textsc{Open}, and accepts 3.8\% as
issues (GSF: 99.5, 99.8, 94.0, and 2.1\%); at GSF's PSNR it reads them in
90.7 and 95.7\%, recognizes 19.1\%, and accepts 23.2\%.  The other 37.7\%
of its unissued outputs at the released strength, and 57.7\% at GSF's
PSNR, read \textsc{Unknown}: no field cell is readable in them, and
$G_{\mathrm{dead}}$ routes a field lost entirely to \textsc{Unknown}.  Its inherited
history at depth 3 reaches 43.15\% mIoU at the released strength (34.22\%
at GSF's PSNR), and stays 30.3, 43.7, and 44.7 points below GSF at depths 1--3 at
GSF's PSNR and 13.5, 31.6, and 35.8 points below at its released strength
(Table~\ref{tab:rev-refresh-history}).
On older slices of at most 16 cells its median IoU is 0 at both strengths
(GSF: 40.0 and 66.7\% for slices of 1--4 and 5--16 cells), also on control
images written on the roots' unmarked sources, where no earlier watermark
can interfere.  This is the layout constraint of
Appendix~\ref{app:cross-method-construction}: a WAM message is decoded per
region, and small regions do not decode, whereas GSF's field is addressable
per cell.

\paragraph{With the registry's full history.}
Any carrier replays the map of an issued image exactly from the records
(Appendix~\ref{app:cross-method-lookup}), so the comparison reduces to
identifying the issue.  On its 4,800 issued-image queries, under the
conditions of Section~\ref{sec:results-lifecycle}, the baseline identifies
the issue in 94.1\% at GSF's PSNR and 97.1\% at its released strength, with
its readout tolerance widened in its favor from 4 to 12 bits on the same
roots; GSF's issue evidence matches in 99.67\% of its own
(Appendix~\ref{app:lifecycle-evidence-ablation}).  For an unissued output
the target is its parent's field (Section~\ref{sec:problem}), so a replayed
record is right whenever the parent is identified;
Appendix~\ref{app:cross-method-registry-record} measures what that takes.
\end{revblock}

\subsection{Reproducibility Notes}
\label{app:cross-method-reproducibility}

The five baseline runs (480 lineages each, generated once per system) took
about 124 GPU-hours in total; GSF's own run of the 480 lineages took about
23 GPU-hours including the final map decoding.  \rev{The full-field-refresh baseline of
Appendix~\ref{app:cross-method-refresh} took about 47 GPU-hours, the extension to depth 7 of
Appendix~\ref{app:generation-map-depth7} about 23, the paste-back edits, further editors, and attacks of
Appendices~\ref{app:mark-preserving} and~\ref{app:active-attacks} about 13, and the TrustMark-Q runs of
Appendix~\ref{app:robustness-sweep} about 10.}  Qwen Image Edit is not deterministic across runs:
regenerating six roots changed 91 of 240 editor outputs at the bit level
and 36 of 570 verifier decisions, and every divergence originated at a Qwen
step.  We did not force deterministic kernels, which would change the
numerics; comparability rests on the matched instruction table rather than
on pixel identity, and GSF's outputs were generated once under the same
editors.  The verifier is deterministic: the delivery run of
Appendix~\ref{app:rendition-robustness} reproduced all 1,920 earlier clean
decisions.

\section{Negative Queries: Images outside the Named Lineage}
\label{app:negative-queries}

Every query of Section~\ref{sec:evaluation} presents an image of the
lineage it names, so the verifier there decides among the four statuses of
an attributed image.  This appendix tests the attribution itself: the query
names a lineage, the image is unmarked or issued by another lineage, and the
verifier evaluates the lineage term $\lambda_L$ of
Eq.~\eqref{eq:lifecycle-semantics} before any other evidence, so the
correct answer is \textsc{Unknown}.  The negatives are a proxy for the open
world constructed for this test: three classes of clean image at one head,
chosen to exercise the attribution rule.  A deployed verifier meets the same
question on a wider input---edited and degraded images
\citep{zhao2024removable,saberi2024robustness}, forged marks
\citep{muller2025forgery}, and keys and lineages beyond this study.

Read with its own key at head 3, every $I_3$ of the evaluation decodes its lineage
identifier and its issue label, 3 (480/480 each), and receives the decisions
of Table~\ref{tab:app-registered-legs}.  Read with the keys of nine other
roots each, none of the 4,320 pairs decodes either, and unmarked images
decode neither: 0/480 on COCO val2017 images disjoint from the roots under
one root's key each, and 0/480 on the roots' own unmarked sources under
their own keys.  Over these 5,280 reads the per-candidate false-match rate
is at most 0.07\% (95\% Clopper--Pearson); analytically, a non-member read
passes the identifier's BCH check on \mbox{4--7\%} of decodes and must then
match 36 bits, about $6\%\times2^{-36}<10^{-12}$ per candidate.
Association is thus supported for the lineage's own images alone; every
image outside it returns \textsc{Unknown}, and none is accepted as an issue
(Table~\ref{tab:app-negative-queries}).

\begin{table}[!htbp]
\caption{Queries outside the named lineage on clean images at head 3 (truth
\textsc{Unknown}) and the lineage's own $I_3$ as control (truth
\textsc{Current}): the number of queries whose issue evidence decodes and
whose lineage identifier matches, and the decision taken with the lineage
term evaluated first.  Wrong keys are nine per root; control decisions are
those of the main evaluation.}
\label{tab:app-negative-queries}
\centering
\small
\setlength{\tabcolsep}{3pt}
\begin{tabular}{@{}lrrrl@{}}
\toprule
Query & $n$ & Issue evidence & Identifier & Decision \\
\midrule
Unmarked image, other content & 480 & 0 & 0 & \textsc{Unknown} 480 \\
Unmarked source of the named lineage & 480 & 0 & 0 & \textsc{Unknown} 480 \\
Issued $I_3$ of another lineage & 4,320 & 0 & 0 & \textsc{Unknown} 4,320 \\
\midrule
$I_3$ of the named lineage (control) & 480 & 480 & 480 & \textsc{Current} 475, \textsc{Open} 2, \textsc{Unk.} 3 \\
\bottomrule
\end{tabular}
\end{table}

\begin{revblock}
\section{Mark-Preserving Edits and Further Editors}
\label{app:mark-preserving}

The four editors of Section~\ref{sec:evaluation} return complete outputs, in
which the median edit changes every cell (Table~\ref{tab:app-complete-change});
yet a global-message watermark is still accepted as an issue on 29.5--38.7\%
of their outputs (Table~\ref{tab:cross-method-lifecycle}), and 18.9 of GSF's
94.0 points of \textsc{Open} recall come from field evidence rather than
from undecodable issue evidence (Appendix~\ref{app:cross-method-open}).  This
appendix adds edits that preserve the watermark outside the edited region,
as paste-back endpoints (Appendix~\ref{app:complete-output-controls}), and
complete outputs of five editors not used in the evaluation.  Each
experiment uses all 480 roots, with the verifier and its frozen rules
unchanged.

\paragraph{Object regenerated and pasted back.}
In the 960 paste-back edits of
Appendix~\ref{app:cross-method-registry-record}, everything outside the
regenerated object is $I_3$ itself, watermark included.  GSF recognizes
99.8\% of them as unissued and accepts none as an issue; TrustMark-Q,
carrying our reissuance protocol, accepts 38.8\% of the same edits, and
4.4\% with the strict content commitment of
Appendix~\ref{app:cross-method-commitment}.

\paragraph{Unissued output pasted into the issue.}
A harder case keeps the issue evidence intact.  The field cells in which
$U_3$ differs most from $I_3$ are pasted into $I_3$ over 5, 10, 20, or 40\%
of the field, and the 96 cells of lineage and issue evidence are left intact
(1,920 edits).  GSF recognizes 81.5--90.6\% of them as unissued and accepts
3.3--5.8\% as issues; TrustMark-Q accepts 68.5--98.5\%.  With the strict
content commitment it accepts 5.6--13.5\%, and its \textsc{Current}
accuracy on the clean $I_3$ falls from 99.8 to 77.3\% (GSF:
99.0\%).  OmniGuard's tamper mask, which localizes paste-back edits
(Appendix~\ref{app:complete-output-controls}), recognizes 87.3--93.8\% of
these pastes as unissued and accepts 6.3--12.7\%.  A paste confined to
one field cell (0.4\% of the image; 480 edits) is accepted in 34.4\% and
read \textsc{Unknown} in 63.3\%: a disruption confined to one cell supplies
no \textsc{Open} evidence (Table~\ref{tab:app-verifier-parameters}), and the
current-issue safeguard withholds most of the rest.

\paragraph{Further editors.}
Five editors not used in the evaluation edit the current issue of every root
and return complete outputs: SD-1.5 inpainting, returning the whole image
\citep{rombach2022ldm}; SD-Turbo \citep{sauer2024adversarial}; LCM
\citep{luo2023lcm}; SD-1.5 image-to-image \citep{meng2022sdedit}; and
PeRFlow \citep{yan2024perflow}, its SD-1.5 network with the scheduler
shipped with it (2,400 unissued outputs).  GSF recognizes 92.3--100.0\% of
them as unissued, depending on the editor, and accepts none as an issue.

\begin{table}[!htbp]
\revtable{R4}
\caption{False issue acceptance in the experiments of Appendices~\ref{app:cross-method-registry-record}, \ref{app:mark-preserving}, and~\ref{app:active-attacks} (\%).  GSF's ranges are over the four pasted areas or the four kinds of unissued edit; the last reference range is over three watermarks.}
\label{tab:rev-new-experiments}
\centering
\footnotesize
\revfit{%
\begin{tabular}{@{}p{4.3cm}rcp{4.6cm}@{}}
\toprule
Experiment & $n$ & GSF $\downarrow$ & Reference $\downarrow$ \\
\midrule
Object regenerated and pasted back & 960 & \textbf{0.0} & $+$ TrustMark-Q: 38.8 \\
Unissued output pasted into the issue & 1,920 & \textbf{3.3--5.8} & $+$ TrustMark-Q: 68.5--98.5 \\
Further editors & 2,400 & \textbf{0.0} & -- \\
Removal & 2,400 & \textbf{0.0} & -- \\
Transplanted issue evidence & 1,920 & \textbf{0.0--3.3} & -- \\
Copied watermark & 1,920 & \textbf{0.4--8.3} & global-message watermarks: 96.5--99.8 \\
\bottomrule
\end{tabular}}
\end{table}

Table~\ref{tab:rev-new-experiments} collects false issue acceptance in these
experiments and in the attacks of Appendix~\ref{app:active-attacks}.  The
route to \textsc{Open} differs: in most edits that regenerate an object, the
issue evidence does not decode, whereas the pastes over 5--40\% of the field
leave it intact, and a disrupted cluster supplies most of their
\textsc{Open} evidence.
\end{revblock}

\begin{revblock}
\section{Attacks on the Watermark}
\label{app:active-attacks}

Appendix~\ref{app:negative-queries} names forged marks among the inputs a
deployed verifier meets.  Under the assumptions of
Appendix~\ref{app:security-assumptions}, the issuer controls the key and the
registry, and a bitmap cannot become a new issue unless the issuer
authenticates the issuance transaction.  This appendix measures how often
the verifier's decision agrees with this when a party without the key
removes, disrupts, or forges the watermark.

\subsection{Threat Model}
The attacking party has the public writer and decoder, the current issue
$I_3$, and unissued outputs, but not the key.  The verifier and its frozen
rules are unchanged.  All 480 roots are attacked, about 13,000 images in
all, and two outcomes are counted separately.  In the first, an issue loses
its status, and the prediction then authorizes nothing
(Section~\ref{sec:problem}).  In the second, an unissued edit is accepted as
an issue, that is, read \textsc{Current} or \textsc{Superseded}, which is
what the assumption excludes.

\subsection{Attacks}
Removal applies VAE re-encoding, diffusion purification at three strengths
\citep{zhao2024removable,saberi2024robustness}, and a bilateral filter to
$I_3$ and to an editor's complete output of $I_3$; a $3\times3$ median filter
is a mild control.  Disruption embeds a released watermark (TrustMark-Q,
VINE-R, or WAM) or a second GSF watermark under another key over $I_3$.
Forgery transplants from $I_3$ the blocks of the 96 cells of lineage and
issue evidence, or copies the whole watermark of $I_3$ (all 256 cells
decoded and re-embedded), onto four kinds of unissued edit: an editor's
complete output of $I_3$ or of an earlier issue; $I_3$ with 40\% of its 256
cells pasted from $U_3$ without sparing the cells of lineage and issue
evidence; and $I_3$ with one object regenerated by SD-1.5 inpainting and
pasted back (Appendix~\ref{app:cross-method-registry-record}).

\subsection{Results}

\begin{table}[!htbp]
\revtable{R8}
\caption{The two outcomes under attack (480 roots).}
\label{tab:rev-attacks}
\centering
\small
\revfit{%
\begin{tabular}{@{}p{5.0cm}cp{4.2cm}@{}}
\toprule
Attack & \shortstack{$I_3$ read\\\textsc{Current}} & Unissued edits accepted as issues \\
\midrule
No attack & 475 of 480 & 34 of 1,920 (0.0--2.5\% by kind) \\
Removal (five operators) & 1 of 2,400 & \textbf{0 of 2,400 editor outputs} \\
3$\times$3 median filter & 96.3\% & -- \\
Released watermark over $I_3$ & 90.6--98.8\% & -- \\
Second GSF watermark under another key & 0 of 480 & -- \\
Transplanted issue evidence & -- & \textbf{32 of 1,920 (0.0--3.3\%)} \\
Copied watermark & -- & \textbf{112 of 1,920 (0.4--8.3\%)} \\
Copied message: VINE-R / TrustMark-Q / OmniGuard & -- & 97.9 / 99.8 / 96.5\% \\
\bottomrule
\end{tabular}}
\end{table}

Removal and a second GSF watermark leave 1 of 2,400 and 0 of 480 issues
\textsc{Current}, a released watermark 90.6--98.8\%, and the median filter
96.3\%; an issue that loses its status reads \textsc{Open} or
\textsc{Unknown}, and none is read \textsc{Superseded}.

An issued status requires a matched issue
(Eq.~\eqref{eq:lifecycle-semantics}), so destroying evidence takes a status
away and gives none: removal makes none of the 2,400 attacked editor outputs
an issue, against 12 of the 480 unattacked ones.  A transplant does not
raise acceptance overall: 32 of the 1,920 edits are accepted,
against 34 without attack (0.0--3.3\% against 0.0--2.5\% by kind),
because the field stays as disrupted as the edit left it (\textsc{Open} in
86.7--97.5\%).  A copied watermark raises acceptance to 0.4--8.3\%
(\textsc{Unknown} in 51.3--73.3\%, \textsc{Open} in 18.3--48.3\%), and every
accepted forgery reads \textsc{Current}.  The issue evidence is bound to a
descriptor of the issued content (Section~\ref{sec:writer}): a descriptor
mismatch above $\tau_{\mathrm{issue}}$ supplies \textsc{Open} evidence, and
one above $\tau_{\mathrm{id}}$ withholds \textsc{Current}
(Table~\ref{tab:app-verifier-parameters}), so a copied watermark is accepted
only where the edit leaves the descriptor within these thresholds.

For the global-message watermarks of
Appendix~\ref{app:cross-method-comparison}, the same attack reads the
message of each one's current issue and re-embeds it into its own editor
outputs (480 each).  It is accepted in 97.9\% for VINE-R, 99.8\% for
TrustMark-Q, and 96.5\% for OmniGuard, against 7.5\% for GSF on the same
kind of edit, since their native protocol does not threshold their content
descriptor.  With the strict content commitment of
Appendix~\ref{app:cross-method-commitment}, their acceptance falls to
2.9--5.4\%, at the cost that appendix reports: \textsc{Current} accuracy
under an issue's own head falls to 67.4--79.9\% (GSF: 99.5\%,
Table~\ref{tab:cross-method-lifecycle}).

\subsection{Limitations}
A party without the key can remove the watermark by regeneration through a
VAE or a diffusion model, as with invisible watermarks in general
\citep{zhao2024removable,saberi2024robustness}, or by a bilateral filter,
and can disrupt it with a second GSF watermark.  The attacked issue then
produces the evidence of an unissued edit and loses its status
(Appendix~\ref{app:robustness-abstention}), while the registry and the
other copies are unaffected.  The issue-consistency check thresholds a
descriptor mismatch so that ordinary delivery passes, and carries no
cryptographic guarantee: a copied watermark passes it on 0.4--8.3\% of
edits (Table~\ref{tab:rev-attacks}).  A disruption confined to one field
cell supplies no \textsc{Open} evidence
(Table~\ref{tab:app-verifier-parameters}), and a paste of that size is
accepted in 34.4\% (Appendix~\ref{app:mark-preserving}).
\end{revblock}

\end{document}